\documentclass[longauth]{aaEC}

\usepackage{graphicx}
\usepackage{natbib}
\usepackage{scalerel}
\usepackage{lastpage}

\usepackage[table]{xcolor}

\usepackage[version=4]{mhchem}

\usepackage{txfonts}
\usepackage[pdfencoding=auto,psdextra]{hyperref}
\hypersetup{
    colorlinks=true,
    linkcolor=blue,
    filecolor=magenta,      
    urlcolor=blue,
    citecolor=blue
}
\makeatletter
\renewcommand*\aa@pageof{, page \thepage{} of \pageref*{LastPage}}
\makeatother

\usepackage[utf8]{inputenc}

\usepackage[switch, modulo]{lineno}

\usepackage[nameinlink,capitalise]{cleveref}
\crefname{section}{Sect.}{Sects.}
\Crefname{section}{Section}{Sections}
\crefname{figure}{Fig.}{Figs.}
\Crefname{figure}{Figure}{Figures}
\crefname{equation}{Eq.}{Eqs.}
\Crefname{equation}{Equation}{Equations}
\crefname{table}{Table}{Tables}
\crefname{appendix}{Appendix}{Appendices}

\usepackage{euclid}

\usepackage[nopostdot, nonumberlist, acronym]{glossaries}
\glsdisablehyper

\newcommand{\pz}{\phantom{0}}
\newcommand{\ps}{\phantom{$-$}}

\usepackage{orcidlink} 

\usepackage[nolist,nohyperlinks]{acronym}

\begin{document}

\acrodef{2PCF}{two-point correlation function}
\acrodef{2dFGRS}{2-degree Field Galaxy Redshift Survey}
\acrodef{3PCF}{three-point correlation function}
\acrodef{AA}{azimuth or $\alpha$ angle}
\acrodef{ACT}{Atacama Cosmology Telescope}
\acrodef{ADC}{analogue-to-digital converter}
\acrodef{ADQL}{Astronomical Data Query Language}
\acrodef{ADS}{Airbus Defence and Space}
\acrodef{ADU}{analogue-to-digital unit}
\acrodef{AGN}{active galactic nuclei}
\acrodef{ALMA}{Atacama Large Millimeter/Submillimeter Array}
\acrodef{AM}{abundance matching}
\acrodef{AOCS}{attitude and orbit-control system}
\acrodef{AOI}{angle of incidence}
\acrodef{APE}{absolute pointing error}
\acrodef{AP}{Alcock-Paczynski}
\acrodef{ASIC}{application specific integrated circuit}
\acrodef{BAO}{baryon acoustic oscillation}
\acrodef{BA}{beta angle}
\acrodef{BBN}{big bang nucleosynthesis}
\acrodef{BFE}{brighter-fatter effect}
\acrodef{BGS}{Bright Galaxy Sample}
\acrodef{BG}{blue grism}
\acrodef{BNT}{Bernardeau--Nishimichi--Taruya}
\acrodef{BOSS}{Baryon Oscillation Spectroscopic Survey}
\acrodef{BPM}{bad-pixel mask}
\acrodef{BPT}{Baldwin--Phillips--Terlevich}
\acrodef{CAMB}{Code for Anisotropies in the Microwave Background}
\acrodef{CAS}{concentration, asymmetry and smoothness}
\acrodef{CCD}{charge-coupled device}
\acrodef{CDFS}{Chandra Deep Field South}
\acrodef{CDPU}{Control and Data Processing Unit}
\acrodef{CDSdawn}[CDaS]{Cosmic Dawn Survey}
\acrodef{CDSdata}[CDS]{Centre de Donn\'ees astronomiques de Strasbourg}
\acrodef{CFC}{cryo-flex cable}
\acrodef{CFHT}{Canada-France-Hawaii Telescope}
\acrodef{CFIS}{Canada-France Imaging Survey}
\acrodef{CFRP}{carbon-fibre reinforced plastic}
\acrodef{CGH}{computer-generated hologram}
\acrodef{CIB}{cosmic infrared background}
\acrodef{CLASS}{Cosmic Linear Anisotropy Solving System}
\acrodef{CLOE}{Cosmology Likelihood for Observables in Euclid}
\acrodef{CMB}{cosmic microwave background}
\acrodef{CMD}{colour-magnitude diagram}
\acrodef{CME}{coronal mass ejection}
\acrodef{CNES}{Centre National d'Etude Spacial}
\acrodef{COSEBIs}{Complete Orthogonal Sets of E/B-Integrals}
\acrodef{CPC}{completeness and purity calibration}
\acrodef{CPL}{Chevallier--Polarski--Linder}
\acrodef{CPPM}{Centre de Physique des Particules de Marseille}
\acrodef{CPU}{central processing unit}
\acrodef{CR}{cosmic ray}
\acrodef{CTE}{coefficient of thermal expansion}
\acrodef{CTI}{charge-transfer inefficiency}
\acrodef{CTIO}{Cerro Tololo Inter-American Observatory}
\acrodef{CU}{calibration unit}
\acrodef{CXB}{cosmic X-ray background}
\acrodef{CaLA}{camera-lens assembly}
\acrodef{CoLA}{corrector-lens assembly}
\acrodef{DCU}{Detector Control Unit}
\acrodef{DECam}{Dark Energy Camera}
\acrodef{DESI}{Dark Energy Spectroscopic Instrument experiment}
\acrodef{DES}{Dark Energy Survey}
\acrodef{DI}{direct integration}
\acrodef{DNL}{differential nonlinearity}
\acrodef{DPDD}{Data Product Description Document}
\acrodef{DPS}{data-processing system}
\acrodef{DPU}{Data Processing Unit}
\acrodef{DP}{Data Product}
\acrodef{DQC}{data quality control}
\acrodef{DQ}{data quality}
\acrodef{DR1}{first data release}
\acrodef{DR}{data release}
\acrodef{DS}{Detector System}
\acrodef{DSPL}{double-source-plane lense}
\acrodef{DTCP}{daily telemetry communication period}
\acrodef{DUNE}{Dark Universe Explorer}
\acrodef{EAF}{Euclid Auxiliary Field}
\acrodef{EAS}{Euclid Archive System}
\acrodef{ECSS}{European Cooperation for Space Standardization}
\acrodef{EC}{Euclid Consortium}
\acrodef{EDF-F}{Euclid Deep Field Fornax}
\acrodef{EDF-N}{Euclid Deep Field North}
\acrodef{EDF-S}{Euclid Deep Field South}
\acrodef{EDF}{Euclid Deep Field}
\acrodef{EDS}{Euclid Deep Survey}
\acrodef{EE}{encircled energy}
\acrodef{EFS}{Euclid Flagship Simulation}
\acrodef{EFT}{effective field theory}
\acrodef{EGC}{extragalactic globular cluster}
\acrodef{ELG}{emission line galaxy}
\acrodef{EMC}{electromagnetic compatibility}
\acrodef{EMDS}{Euclid Medium-Deep Survey}
\acrodef{EOR}{epoch of reionisation}
\acrodef{EPER}{extended pixel-edge response}
\acrodef{ERO}{Early Release Observations}
\acrodef{ESAC}{European Space Astronomy Centre}
\acrodef{ESA}{European Space Agency}
\acrodef{ESOC}{European Science Operations Centre}
\acrodef{ESOP}{early science operations}
\acrodef{ESP}{Emission of Solar Protons}
\acrodef{EUDF}{Euclid Ultra-Deep Field}
\acrodef{EWS}{Euclid Wide Survey}
\acrodef{EW}{equivalent width}
\acrodef{EXT-PF}{EXT processing function}
\acrodef{EXT}{EXTernal data}
\acrodef{FDIR}{Fault Detection, Isolation and Recovery}
\acrodef{FFP}{free-floating planet}
\acrodef{FFT}{fast Fourier transform}
\acrodef{FGS}{fine-guidance sensor}
\acrodef{FITS}{flexible image transport system}
\acrodef{FKP}{Feldman--Kaiser--Peacock}
\acrodef{FOM}{figure of merit}
\acrodef{FOV}{field of view}
\acrodef{FPA}{focal-plane array}
\acrodef{FPR}{false positive rate}
\acrodef{FWA}{filter-wheel assembly}
\acrodef{FWC}{full-well capacity}
\acrodef{FWHM}{full width at half maximum}
\acrodef{G3L}{galaxy-galaxy-galaxy lensing}
\acrodef{GBTDS}{Galactic Bulge Time Domain Survey}
\acrodef{GCR}{Galactic cosmic ray}
\acrodef{GC}{globular cluster}
\acrodef{GF}{Gaussian fit}
\acrodef{GGL}{galaxy-galaxy lensing}
\acrodef{GOES}{Geostationary Operational Environmental Satellites}
\acrodef{GP}{Gaussian process}
\acrodef{GR}{general relativity}
\acrodef{GWA}{grism-wheel assembly}
\acrodef{H2RG}{HAWAII-2RG}
\acrodef{HDU}{header data unit}
\acrodef{HGA}{high-gain antenna}
\acrodef{HKTM}{housekeeping telemetry}
\acrodef{HIERO}{HST-to-IRAC extremely red object}
\acrodef{HK}{housekeeping}
\acrodef{HOD}{halo-occupation distribution}
\acrodef{HOPS}{Herschel Orion Protostar Survey}
\acrodef{HOS}{higher-order statistics}
\acrodef{HSC-SSP}{Hyper Suprime-Cam Subaru Strategic Program}
\acrodef{HSC}{Subaru Hyper Suprime Camera}
\acrodef{HST}{\textit{Hubble} Space Telescope}
\acrodef{IAD}{ion-assisted deposition}
\acrodef{IA}{intrinsic alignment}
\acrodef{ICRS}{International Celestial Reference System}
\acrodef{ICU}{Instrument Control Unit}
\acrodef{IGM}{intergalactic medium}
\acrodef{IOT}{Instrument Operation Team}
\acrodef{IP2I}{Institut de Physique des 2 Infinis de Lyon}
\acrodef{IPC}{inter-pixel capacitance}
\acrodef{ISCS}{IRAC Shallow Cluster Survey}
\acrodef{ISES}{International Space Environmental Services}
\acrodef{ISW}{integrated Sachs--Wolfe}
\acrodef{IS}{inverse sensitivity}
\acrodef{JPL}{NASA Jet Propulsion Laboratory}
\acrodef{JWST}{\textit{James Webb} Space Telescope}
\acrodef{KSB}{Kaiser--Squires--Broadhurst}
\acrodef{KS}{Kaiser--Squires}
\acrodef{KiDS}{Kilo-Degree Survey}
\acrodef{LAM}{Laboratoire d'Astrophysique de Marseille}
\acrodef{LBT}{Large Binocular Telescope}
\acrodef{LDN}{Lynd's Catalogue of Dark Nebulae}
\acrodef{LE1}{Level 1}
\acrodef{LE2}{Level 2}
\acrodef{LE3}{Level 3}
\acrodef{LED}{light-emitting diode}
\acrodef{LEMON}{LEns MOdelling with Neural networks}
\acrodef{LF}{luminosity function}
\acrodef{LOS}{line of sight}
\acrodef{LRD}{little red dot}
\acrodef{LRG}{luminous red galaxy}
\acrodef{LSB}{low surface brightness}
\acrodef{LSST}{Legacy Survey of Space and Time}
\acrodef{LSS}{large-scale structure}
\acrodef{M2M}{M2 mechanism}
\acrodef{MACC}{multi-accumulate}
\acrodef{MAD}{median absolute deviation}
\acrodef{MAMBO}{Mocks with Abundance Matching in BOlogna}
\acrodef{MCMC}{Markov chain Monte Carlo}
\acrodef{MDB}{Mission Database}
\acrodef{MEF}{multi-extension FITS}
\acrodef{MER-PF}{merge-datasets processing function}
\acrodef{MER}{MERge datasets}
\acrodef{MLI}{multi-layer insulation}
\acrodef{MMU}{Mass Memory Unit}
\acrodef{MOC}{mission operation centre}
\acrodef{MPE}{Max-Planck-Institut für extraterrestrische Physik}
\acrodef{MPIA}{Max-Planck-Institut für Astronomie}
\acrodef{MS}{main sequence}
\acrodef{MZ-CGH}{multi-zonal computer-generated hologram}
\acrodef{NASA}{National Aeronautic and Space Administration}
\acrodef{NA}{numerical aperture}
\acrodef{NEP}{North Ecliptic Pole}
\acrodef{NFW}{Navarro--Frenk--White}
\acrodef{NI-CU}{NISP calibration unit}
\acrodef{NI-GWA}{NISP Grism Wheel Assembly}
\acrodef{NI-OA}{near-infrared optical assembly}
\acrodef{NIEL}{non-ionising energy loss}
\acrodef{NIRCAM}{Near-InfraRed Camera}
\acrodef{NIR}{near-infrared}
\acrodef{NIR-PF}{NISP processing function}
\acrodef{NISP}{Near-Infrared Spectrometer and Photometer}
\acrodef{NLA}{nonlinear linear alignment}
\acrodef{NMAD}{normalised absolute median deviation}
\acrodef{NOAA}{National Oceanic and Atmospheric Administration}
\acrodef{OU}{Organisation Unit}
\acrodef{PARMS}{plasma-assisted reactive magnetron sputtering}
\acrodef{PA}{position angle}
\acrodef{PDC}{phase-diversity calibration}
\acrodef{PDF}{probability density function}
\acrodef{PE}{processing element}
\acrodef{PF}{processing function}
\acrodef{SIR}{NISP Soectroscopy}
\acrodef{SPE}{NISP SPEctra analysis}
\acrodef{PF}{processing function}
\acrodef{PHZ}{PHotometric redshift}
\acrodef{PHZ-PF}{photometric-redshift processing function}
\acrodef{PLM}{payload module}
\acrodef{PPO}{pipeline processing order}
\acrodef{PRNU}{pixel-response non-uniformity}
\acrodef{PSF}{point-spread function}
\acrodef{PTC}{photon transfer curve}
\acrodef{PTFE}{polytetrafluoroethylene}
\acrodef{PV}{performance-verification}
\acrodef{PWM}{pulse-width modulation}
\acrodef{Pan-STARRS}{Panchromatic Survey Telescope and Rapid Response System}
\acrodef{QE}{quantum efficiency}
\acrodef{QF}{quality factor}
\acrodef{QSO}{quasi-stellar object}
\acrodef{RG}{red grism}
\acrodef{RMS}{root mean square}
\acrodef{ROE}{readout electronic block unit}
\acrodef{ROIC}{readout-integrated circuit}
\acrodef{ROI}{region of interest}
\acrodef{ROS}{reference observing sequence}
\acrodef{RPE}{relative pointing error}
\acrodef{RSD}{redshift-space distortion}
\acrodef{RSU}{readout shutter unit}
\acrodef{S/N}{signal-to-noise ratio}
\acrodef{SAA}{Solar aspect angle}
\acrodef{SBF}{surface brightness fluctuation}
\acrodef{SCA}{sensor-chip array}
\acrodef{SCE}{sensor-chip electronic}
\acrodef{SCS}{sensor-chip system}
\acrodef{SDC}{Science Data Centre}
\acrodef{SDSS}{Sloan Digital Sky Survey}
\acrodef{SED}{spectral energy distribution}
\acrodef{SEF}{single-epoch frame}
\acrodef{SEP}{South Ecliptic Pole}
\acrodef{SFE}{surface figure error}
\acrodef{SFMS}{star-forming main sequence}
\acrodef{SFR}{star-formation rate}
\acrodef{SGPS}{Solar and Galactic Proton Sensor}
\acrodef{SGS}{Science Ground Segment}
\acrodef{SHMR}{stellar mass-halo mass relation}
\acrodef{SHS}{Shack-Hartmann sensor}
\acrodef{SIR}{Spectroscopy InfraRed}
\acrodef{SLICS}{Scinet LIghtCone Simulations}
\acrodef{SMBH}{supermassive black hole}
\acrodef{SMF}{stellar mass function}
\acrodef{SNR}[S/N]{signal-to-noise ratio}
\acrodef{SNe}{supernovae}
\acrodef{SOC}{Science Operations Centre}
\acrodef{SOM}{self-organising map}
\acrodef{SO}{Simons Observatory}
\acrodef{SPACE}{Spectroscopic All-Sky Cosmic Explorer}
\acrodef{SPE}{SPEctroscopy}
\acrodef{SPT}{South Pole Telescope}
\acrodef{SPV}{Science Performance Verification}
\acrodef{SSN}{Sunspot number}
\acrodef{SSO}{Solar System object}
\acrodef{STIX}{Spectrometer/Telescope for Imaging X-rays}
\acrodef{STOP}{structural thermal optical performance}
\acrodef{STR}{star tracker}
\acrodef{SUDARE}{SUpernovae Diversity And Rate Evolution}
\acrodef{SUTR}{sample up-the-ramp}
\acrodef{SVM}{service module}
\acrodef{SZ}{Sunyaev--Zeldovich}
\acrodef{SiC}{silicon carbide}
\acrodef{SolO}{Solar Orbiter}
\acrodef{TCM}{transfer correction manoeuvre}
\acrodef{TDE}{tidal disruption event}
\acrodef{TPR}{true positive rate}
\acrodef{TP}{trap pumping}
\acrodef{UCD}{ultra-cool dwarf}
\acrodef{UMAP}[UMAP]{uniform manifold approximation and projection}
\acrodef{UNIONS}{Ultraviolet Near-Infrared Optical Northern Survey}
\acrodef{UTR}{up-the-ramp}
\acrodef{VIPERS}{VIMOS Public Extragalactic Redshift Survey}
\acrodef{VIS-PF}{VIS processing function}
\acrodef{VIS}{VISible instrument}  
\acrodef{VISTA}{Visible and Infrared Survey Telescope for Astronomy}
\acrodef{VLT}{Very Large Telescope}
\acrodef{WCS}{world coordinate system}
\acrodef{WD}{white dwarf}
\acrodef{WFC3}{Wide Field Camera 3}
\acrodef{WFE}{wavefront error}
\acrodef{WHIGS}{Waterloo-Hawaii-IfA $g$-band Survey}
\acrodef{WISHES}{Wide Imaging with Subaru-Hyper Suprime-Cam Euclid Sky}
\acrodef{WL}{weak lensing}
\acrodef{YSO}{young stellar object}
\acrodef{ZP}{zero point}
\acrodef{eROSITA}{extended ROentgen Survey with and Imaging Telescope Array}
\acrodef{kSZ}{kinetic Sunyaev--Zeldovich}
\acrodef{pRF}{probabilistic random forest}
\acrodef{tSZ}{thermal Sunyaev--Zeldovich}
\acrodef{zPDF}{redshift probability density function}

\title{\Euclid Quick Data Release (Q1)}
\subtitle{Data release overview}

\newcommand{\orcid}[1]{} 
\author{Euclid Collaboration: H.~Aussel\orcid{0000-0002-1371-5705}\thanks{\email{herve.aussel@cea.fr}}\inst{\ref{aff1}}
\and I.~Tereno\inst{\ref{aff2},\ref{aff3}}
\and M.~Schirmer\orcid{0000-0003-2568-9994}\inst{\ref{aff4}}
\and G.~Alguero\inst{\ref{aff5}}
\and B.~Altieri\orcid{0000-0003-3936-0284}\inst{\ref{aff6}}
\and E.~Balbinot\orcid{0000-0002-1322-3153}\inst{\ref{aff7},\ref{aff8}}
\and T.~de~Boer\orcid{0000-0001-5486-2747}\inst{\ref{aff9}}
\and P.~Casenove\orcid{0009-0006-6736-1670}\inst{\ref{aff10}}
\and P.~Corcho-Caballero\orcid{0000-0001-6327-7080}\inst{\ref{aff7}}
\and H.~Furusawa\inst{\ref{aff11},\ref{aff12}}
\and J.~Furusawa\orcid{0000-0002-1968-5762}\inst{\ref{aff11}}
\and M.~J.~Hudson\orcid{0000-0002-1437-3786}\inst{\ref{aff13},\ref{aff14},\ref{aff15}}
\and K.~Jahnke\orcid{0000-0003-3804-2137}\inst{\ref{aff4}}
\and G.~Libet\inst{\ref{aff10}}
\and J.~Macias-Perez\orcid{0000-0002-5385-2763}\inst{\ref{aff5}}
\and N.~Masoumzadeh\orcid{0000-0002-2664-8054}\inst{\ref{aff16}}
\and J.~J.~Mohr\orcid{0000-0002-6875-2087}\inst{\ref{aff17}}
\and J.~Odier\orcid{0000-0002-1650-2246}\inst{\ref{aff5}}
\and D.~Scott\orcid{0000-0002-6878-9840}\inst{\ref{aff18}}
\and T.~Vassallo\orcid{0000-0001-6512-6358}\inst{\ref{aff19},\ref{aff20}}
\and G.~Verdoes~Kleijn\orcid{0000-0001-5803-2580}\inst{\ref{aff7}}
\and A.~Zacchei\orcid{0000-0003-0396-1192}\inst{\ref{aff20},\ref{aff21}}
\and N.~Aghanim\orcid{0000-0002-6688-8992}\inst{\ref{aff22}}
\and A.~Amara\inst{\ref{aff23}}
\and S.~Andreon\orcid{0000-0002-2041-8784}\inst{\ref{aff24}}
\and N.~Auricchio\orcid{0000-0003-4444-8651}\inst{\ref{aff25}}
\and S.~Awan\inst{\ref{aff26}}
\and R.~Azzollini\orcid{0000-0002-0438-0886}\inst{\ref{aff26}}
\and C.~Baccigalupi\orcid{0000-0002-8211-1630}\inst{\ref{aff21},\ref{aff20},\ref{aff27},\ref{aff28}}
\and M.~Baldi\orcid{0000-0003-4145-1943}\inst{\ref{aff29},\ref{aff25},\ref{aff30}}
\and A.~Balestra\orcid{0000-0002-6967-261X}\inst{\ref{aff31}}
\and S.~Bardelli\orcid{0000-0002-8900-0298}\inst{\ref{aff25}}
\and A.~Basset\inst{\ref{aff10}}
\and P.~Battaglia\orcid{0000-0002-7337-5909}\inst{\ref{aff25}}
\and A.~N.~Belikov\inst{\ref{aff7},\ref{aff32}}
\and R.~Bender\orcid{0000-0001-7179-0626}\inst{\ref{aff16},\ref{aff19}}
\and A.~Biviano\orcid{0000-0002-0857-0732}\inst{\ref{aff20},\ref{aff21}}
\and A.~Bonchi\orcid{0000-0002-2667-5482}\inst{\ref{aff33}}
\and D.~Bonino\orcid{0000-0002-3336-9977}\inst{\ref{aff34}}
\and E.~Branchini\orcid{0000-0002-0808-6908}\inst{\ref{aff35},\ref{aff36},\ref{aff24}}
\and M.~Brescia\orcid{0000-0001-9506-5680}\inst{\ref{aff37},\ref{aff38}}
\and J.~Brinchmann\orcid{0000-0003-4359-8797}\inst{\ref{aff39},\ref{aff40}}
\and S.~Camera\orcid{0000-0003-3399-3574}\inst{\ref{aff41},\ref{aff42},\ref{aff34}}
\and G.~Ca\~nas-Herrera\orcid{0000-0003-2796-2149}\inst{\ref{aff43},\ref{aff44},\ref{aff8}}
\and V.~Capobianco\orcid{0000-0002-3309-7692}\inst{\ref{aff34}}
\and C.~Carbone\orcid{0000-0003-0125-3563}\inst{\ref{aff45}}
\and V.~F.~Cardone\inst{\ref{aff46},\ref{aff47}}
\and J.~Carretero\orcid{0000-0002-3130-0204}\inst{\ref{aff48},\ref{aff49}}
\and S.~Casas\orcid{0000-0002-4751-5138}\inst{\ref{aff50}}
\and F.~J.~Castander\orcid{0000-0001-7316-4573}\inst{\ref{aff51},\ref{aff52}}
\and M.~Castellano\orcid{0000-0001-9875-8263}\inst{\ref{aff46}}
\and G.~Castignani\orcid{0000-0001-6831-0687}\inst{\ref{aff25}}
\and S.~Cavuoti\orcid{0000-0002-3787-4196}\inst{\ref{aff38},\ref{aff53}}
\and K.~C.~Chambers\orcid{0000-0001-6965-7789}\inst{\ref{aff9}}
\and A.~Cimatti\inst{\ref{aff54}}
\and C.~Colodro-Conde\inst{\ref{aff55}}
\and G.~Congedo\orcid{0000-0003-2508-0046}\inst{\ref{aff56}}
\and C.~J.~Conselice\orcid{0000-0003-1949-7638}\inst{\ref{aff57}}
\and L.~Conversi\orcid{0000-0002-6710-8476}\inst{\ref{aff58},\ref{aff6}}
\and Y.~Copin\orcid{0000-0002-5317-7518}\inst{\ref{aff59}}
\and F.~Courbin\orcid{0000-0003-0758-6510}\inst{\ref{aff60},\ref{aff61}}
\and H.~M.~Courtois\orcid{0000-0003-0509-1776}\inst{\ref{aff62}}
\and M.~Cropper\orcid{0000-0003-4571-9468}\inst{\ref{aff26}}
\and J.-G.~Cuby\orcid{0000-0002-8767-1442}\inst{\ref{aff63},\ref{aff64}}
\and A.~Da~Silva\orcid{0000-0002-6385-1609}\inst{\ref{aff2},\ref{aff65}}
\and R.~da~Silva\orcid{0000-0003-4788-677X}\inst{\ref{aff46},\ref{aff33}}
\and H.~Degaudenzi\orcid{0000-0002-5887-6799}\inst{\ref{aff66}}
\and J.~T.~A.~de~Jong\orcid{0000-0002-9511-1357}\inst{\ref{aff7}}
\and G.~De~Lucia\orcid{0000-0002-6220-9104}\inst{\ref{aff20}}
\and A.~M.~Di~Giorgio\orcid{0000-0002-4767-2360}\inst{\ref{aff67}}
\and J.~Dinis\orcid{0000-0001-5075-1601}\inst{\ref{aff2},\ref{aff65}}
\and C.~Dolding\orcid{0009-0003-7199-6108}\inst{\ref{aff26}}
\and H.~Dole\orcid{0000-0002-9767-3839}\inst{\ref{aff22}}
\and M.~Douspis\orcid{0000-0003-4203-3954}\inst{\ref{aff22}}
\and F.~Dubath\orcid{0000-0002-6533-2810}\inst{\ref{aff66}}
\and C.~A.~J.~Duncan\orcid{0009-0003-3573-0791}\inst{\ref{aff56},\ref{aff57}}
\and X.~Dupac\inst{\ref{aff6}}
\and S.~Dusini\orcid{0000-0002-1128-0664}\inst{\ref{aff68}}
\and A.~Ealet\orcid{0000-0003-3070-014X}\inst{\ref{aff59}}
\and S.~Escoffier\orcid{0000-0002-2847-7498}\inst{\ref{aff69}}
\and M.~Fabricius\orcid{0000-0002-7025-6058}\inst{\ref{aff16},\ref{aff19}}
\and M.~Farina\orcid{0000-0002-3089-7846}\inst{\ref{aff67}}
\and R.~Farinelli\inst{\ref{aff25}}
\and F.~Faustini\orcid{0000-0001-6274-5145}\inst{\ref{aff46},\ref{aff33}}
\and S.~Ferriol\inst{\ref{aff59}}
\and S.~Fotopoulou\orcid{0000-0002-9686-254X}\inst{\ref{aff70}}
\and N.~Fourmanoit\orcid{0009-0005-6816-6925}\inst{\ref{aff69}}
\and M.~Frailis\orcid{0000-0002-7400-2135}\inst{\ref{aff20}}
\and E.~Franceschi\orcid{0000-0002-0585-6591}\inst{\ref{aff25}}
\and P.~Franzetti\inst{\ref{aff45}}
\and S.~Galeotta\orcid{0000-0002-3748-5115}\inst{\ref{aff20}}
\and K.~George\orcid{0000-0002-1734-8455}\inst{\ref{aff19}}
\and W.~Gillard\orcid{0000-0003-4744-9748}\inst{\ref{aff69}}
\and B.~Gillis\orcid{0000-0002-4478-1270}\inst{\ref{aff56}}
\and C.~Giocoli\orcid{0000-0002-9590-7961}\inst{\ref{aff25},\ref{aff30}}
\and P.~G\'omez-Alvarez\orcid{0000-0002-8594-5358}\inst{\ref{aff71},\ref{aff6}}
\and J.~Gracia-Carpio\inst{\ref{aff16}}
\and B.~R.~Granett\orcid{0000-0003-2694-9284}\inst{\ref{aff24}}
\and A.~Grazian\orcid{0000-0002-5688-0663}\inst{\ref{aff31}}
\and F.~Grupp\inst{\ref{aff16},\ref{aff19}}
\and L.~Guzzo\orcid{0000-0001-8264-5192}\inst{\ref{aff72},\ref{aff24},\ref{aff73}}
\and S.~Gwyn\orcid{0000-0001-8221-8406}\inst{\ref{aff74}}
\and S.~V.~H.~Haugan\orcid{0000-0001-9648-7260}\inst{\ref{aff75}}
\and O.~Herent\inst{\ref{aff76}}
\and J.~Hoar\inst{\ref{aff6}}
\and H.~Hoekstra\orcid{0000-0002-0641-3231}\inst{\ref{aff8}}
\and M.~S.~Holliman\inst{\ref{aff56}}
\and W.~Holmes\inst{\ref{aff77}}
\and I.~M.~Hook\orcid{0000-0002-2960-978X}\inst{\ref{aff78}}
\and F.~Hormuth\inst{\ref{aff79}}
\and A.~Hornstrup\orcid{0000-0002-3363-0936}\inst{\ref{aff80},\ref{aff81}}
\and P.~Hudelot\inst{\ref{aff76}}
\and S.~Ili\'c\orcid{0000-0003-4285-9086}\inst{\ref{aff82},\ref{aff83}}
\and M.~Jhabvala\inst{\ref{aff84}}
\and B.~Joachimi\orcid{0000-0001-7494-1303}\inst{\ref{aff85}}
\and E.~Keih\"anen\orcid{0000-0003-1804-7715}\inst{\ref{aff86}}
\and S.~Kermiche\orcid{0000-0002-0302-5735}\inst{\ref{aff69}}
\and A.~Kiessling\orcid{0000-0002-2590-1273}\inst{\ref{aff77}}
\and B.~Kubik\orcid{0009-0006-5823-4880}\inst{\ref{aff59}}
\and K.~Kuijken\orcid{0000-0002-3827-0175}\inst{\ref{aff8}}
\and M.~K\"ummel\orcid{0000-0003-2791-2117}\inst{\ref{aff19}}
\and M.~Kunz\orcid{0000-0002-3052-7394}\inst{\ref{aff87}}
\and H.~Kurki-Suonio\orcid{0000-0002-4618-3063}\inst{\ref{aff88},\ref{aff89}}
\and O.~Lahav\orcid{0000-0002-1134-9035}\inst{\ref{aff85}}
\and Q.~Le~Boulc'h\inst{\ref{aff90}}
\and A.~M.~C.~Le~Brun\orcid{0000-0002-0936-4594}\inst{\ref{aff91}}
\and D.~Le~Mignant\orcid{0000-0002-5339-5515}\inst{\ref{aff64}}
\and P.~Liebing\inst{\ref{aff26}}
\and S.~Ligori\orcid{0000-0003-4172-4606}\inst{\ref{aff34}}
\and P.~B.~Lilje\orcid{0000-0003-4324-7794}\inst{\ref{aff75}}
\and V.~Lindholm\orcid{0000-0003-2317-5471}\inst{\ref{aff88},\ref{aff89}}
\and I.~Lloro\orcid{0000-0001-5966-1434}\inst{\ref{aff92}}
\and G.~Mainetti\orcid{0000-0003-2384-2377}\inst{\ref{aff90}}
\and D.~Maino\inst{\ref{aff72},\ref{aff45},\ref{aff73}}
\and E.~Maiorano\orcid{0000-0003-2593-4355}\inst{\ref{aff25}}
\and O.~Mansutti\orcid{0000-0001-5758-4658}\inst{\ref{aff20}}
\and S.~Marcin\inst{\ref{aff93}}
\and O.~Marggraf\orcid{0000-0001-7242-3852}\inst{\ref{aff94}}
\and K.~Markovic\orcid{0000-0001-6764-073X}\inst{\ref{aff77}}
\and M.~Martinelli\orcid{0000-0002-6943-7732}\inst{\ref{aff46},\ref{aff47}}
\and N.~Martinet\orcid{0000-0003-2786-7790}\inst{\ref{aff64}}
\and F.~Marulli\orcid{0000-0002-8850-0303}\inst{\ref{aff95},\ref{aff25},\ref{aff30}}
\and R.~Massey\orcid{0000-0002-6085-3780}\inst{\ref{aff96}}
\and S.~Maurogordato\inst{\ref{aff97}}
\and H.~J.~McCracken\orcid{0000-0002-9489-7765}\inst{\ref{aff76}}
\and E.~Medinaceli\orcid{0000-0002-4040-7783}\inst{\ref{aff25}}
\and S.~Mei\orcid{0000-0002-2849-559X}\inst{\ref{aff98},\ref{aff99}}
\and M.~Melchior\inst{\ref{aff100}}
\and Y.~Mellier\inst{\ref{aff101},\ref{aff76}}
\and M.~Meneghetti\orcid{0000-0003-1225-7084}\inst{\ref{aff25},\ref{aff30}}
\and E.~Merlin\orcid{0000-0001-6870-8900}\inst{\ref{aff46}}
\and G.~Meylan\inst{\ref{aff102}}
\and A.~Mora\orcid{0000-0002-1922-8529}\inst{\ref{aff103}}
\and M.~Moresco\orcid{0000-0002-7616-7136}\inst{\ref{aff95},\ref{aff25}}
\and P.~W.~Morris\orcid{0000-0002-5186-4381}\inst{\ref{aff104}}
\and L.~Moscardini\orcid{0000-0002-3473-6716}\inst{\ref{aff95},\ref{aff25},\ref{aff30}}
\and S.~Mourre\orcid{0009-0005-9047-0691}\inst{\ref{aff97},\ref{aff105}}
\and R.~Nakajima\orcid{0009-0009-1213-7040}\inst{\ref{aff94}}
\and C.~Neissner\orcid{0000-0001-8524-4968}\inst{\ref{aff106},\ref{aff49}}
\and R.~C.~Nichol\orcid{0000-0003-0939-6518}\inst{\ref{aff23}}
\and S.-M.~Niemi\orcid{0009-0005-0247-0086}\inst{\ref{aff43}}
\and J.~W.~Nightingale\orcid{0000-0002-8987-7401}\inst{\ref{aff107}}
\and T.~Nutma\inst{\ref{aff7},\ref{aff8}}
\and C.~Padilla\orcid{0000-0001-7951-0166}\inst{\ref{aff106}}
\and S.~Paltani\orcid{0000-0002-8108-9179}\inst{\ref{aff66}}
\and F.~Pasian\orcid{0000-0002-4869-3227}\inst{\ref{aff20}}
\and J.~A.~Peacock\orcid{0000-0002-1168-8299}\inst{\ref{aff56}}
\and K.~Pedersen\inst{\ref{aff108}}
\and W.~J.~Percival\orcid{0000-0002-0644-5727}\inst{\ref{aff14},\ref{aff13},\ref{aff15}}
\and V.~Pettorino\inst{\ref{aff43}}
\and S.~Pires\orcid{0000-0002-0249-2104}\inst{\ref{aff1}}
\and G.~Polenta\orcid{0000-0003-4067-9196}\inst{\ref{aff33}}
\and J.~E.~Pollack\inst{\ref{aff109},\ref{aff98}}
\and M.~Poncet\inst{\ref{aff10}}
\and L.~A.~Popa\inst{\ref{aff110}}
\and L.~Pozzetti\orcid{0000-0001-7085-0412}\inst{\ref{aff25}}
\and G.~D.~Racca\inst{\ref{aff43},\ref{aff8}}
\and F.~Raison\orcid{0000-0002-7819-6918}\inst{\ref{aff16}}
\and R.~Rebolo\orcid{0000-0003-3767-7085}\inst{\ref{aff55},\ref{aff111},\ref{aff112}}
\and A.~Renzi\orcid{0000-0001-9856-1970}\inst{\ref{aff113},\ref{aff68}}
\and J.~Rhodes\orcid{0000-0002-4485-8549}\inst{\ref{aff77}}
\and G.~Riccio\inst{\ref{aff38}}
\and H.-W.~Rix\orcid{0000-0003-4996-9069}\inst{\ref{aff4}}
\and E.~Romelli\orcid{0000-0003-3069-9222}\inst{\ref{aff20}}
\and M.~Roncarelli\orcid{0000-0001-9587-7822}\inst{\ref{aff25}}
\and E.~Rossetti\orcid{0000-0003-0238-4047}\inst{\ref{aff29}}
\and B.~Rusholme\orcid{0000-0001-7648-4142}\inst{\ref{aff114}}
\and R.~Saglia\orcid{0000-0003-0378-7032}\inst{\ref{aff19},\ref{aff16}}
\and Z.~Sakr\orcid{0000-0002-4823-3757}\inst{\ref{aff115},\ref{aff83},\ref{aff116}}
\and A.~G.~S\'anchez\orcid{0000-0003-1198-831X}\inst{\ref{aff16}}
\and D.~Sapone\orcid{0000-0001-7089-4503}\inst{\ref{aff117}}
\and B.~Sartoris\orcid{0000-0003-1337-5269}\inst{\ref{aff19},\ref{aff20}}
\and M.~Sauvage\orcid{0000-0002-0809-2574}\inst{\ref{aff1}}
\and J.~A.~Schewtschenko\orcid{0000-0002-4913-6393}\inst{\ref{aff56}}
\and P.~Schneider\orcid{0000-0001-8561-2679}\inst{\ref{aff94}}
\and M.~Scodeggio\inst{\ref{aff45}}
\and A.~Secroun\orcid{0000-0003-0505-3710}\inst{\ref{aff69}}
\and E.~Sefusatti\orcid{0000-0003-0473-1567}\inst{\ref{aff20},\ref{aff21},\ref{aff27}}
\and G.~Seidel\orcid{0000-0003-2907-353X}\inst{\ref{aff4}}
\and M.~Seiffert\orcid{0000-0002-7536-9393}\inst{\ref{aff77}}
\and S.~Serrano\orcid{0000-0002-0211-2861}\inst{\ref{aff52},\ref{aff118},\ref{aff51}}
\and P.~Simon\inst{\ref{aff94}}
\and C.~Sirignano\orcid{0000-0002-0995-7146}\inst{\ref{aff113},\ref{aff68}}
\and G.~Sirri\orcid{0000-0003-2626-2853}\inst{\ref{aff30}}
\and J.~Skottfelt\orcid{0000-0003-1310-8283}\inst{\ref{aff119}}
\and A.~Spurio~Mancini\orcid{0000-0001-5698-0990}\inst{\ref{aff120}}
\and L.~Stanco\orcid{0000-0002-9706-5104}\inst{\ref{aff68}}
\and J.~Steinwagner\orcid{0000-0001-7443-1047}\inst{\ref{aff16}}
\and C.~Surace\orcid{0000-0003-2592-0113}\inst{\ref{aff64}}
\and P.~Tallada-Cresp\'{i}\orcid{0000-0002-1336-8328}\inst{\ref{aff48},\ref{aff49}}
\and D.~Tavagnacco\orcid{0000-0001-7475-9894}\inst{\ref{aff20}}
\and A.~N.~Taylor\inst{\ref{aff56}}
\and H.~I.~Teplitz\orcid{0000-0002-7064-5424}\inst{\ref{aff121}}
\and N.~Tessore\orcid{0000-0002-9696-7931}\inst{\ref{aff85}}
\and S.~Toft\orcid{0000-0003-3631-7176}\inst{\ref{aff122},\ref{aff123}}
\and R.~Toledo-Moreo\orcid{0000-0002-2997-4859}\inst{\ref{aff124}}
\and F.~Torradeflot\orcid{0000-0003-1160-1517}\inst{\ref{aff49},\ref{aff48}}
\and A.~Tsyganov\inst{\ref{aff125}}
\and I.~Tutusaus\orcid{0000-0002-3199-0399}\inst{\ref{aff83}}
\and E.~A.~Valentijn\inst{\ref{aff7}}
\and L.~Valenziano\orcid{0000-0002-1170-0104}\inst{\ref{aff25},\ref{aff126}}
\and J.~Valiviita\orcid{0000-0001-6225-3693}\inst{\ref{aff88},\ref{aff89}}
\and A.~Veropalumbo\orcid{0000-0003-2387-1194}\inst{\ref{aff24},\ref{aff36},\ref{aff35}}
\and Y.~Wang\orcid{0000-0002-4749-2984}\inst{\ref{aff121}}
\and J.~Weller\orcid{0000-0002-8282-2010}\inst{\ref{aff19},\ref{aff16}}
\and O.~R.~Williams\orcid{0000-0003-0274-1526}\inst{\ref{aff125}}
\and G.~Zamorani\orcid{0000-0002-2318-301X}\inst{\ref{aff25}}
\and F.~M.~Zerbi\inst{\ref{aff24}}
\and E.~Zucca\orcid{0000-0002-5845-8132}\inst{\ref{aff25}}
\and V.~Allevato\orcid{0000-0001-7232-5152}\inst{\ref{aff38}}
\and M.~Ballardini\orcid{0000-0003-4481-3559}\inst{\ref{aff127},\ref{aff128},\ref{aff25}}
\and R.~P.~Blake\inst{\ref{aff56}}
\and M.~Bolzonella\orcid{0000-0003-3278-4607}\inst{\ref{aff25}}
\and E.~Bozzo\orcid{0000-0002-8201-1525}\inst{\ref{aff66}}
\and C.~Burigana\orcid{0000-0002-3005-5796}\inst{\ref{aff129},\ref{aff126}}
\and R.~Cabanac\orcid{0000-0001-6679-2600}\inst{\ref{aff83}}
\and M.~Calabrese\orcid{0000-0002-2637-2422}\inst{\ref{aff130},\ref{aff45}}
\and A.~Cappi\inst{\ref{aff25},\ref{aff97}}
\and D.~Di~Ferdinando\inst{\ref{aff30}}
\and J.~A.~Escartin~Vigo\inst{\ref{aff16}}
\and L.~Gabarra\orcid{0000-0002-8486-8856}\inst{\ref{aff131}}
\and W.~G.~Hartley\inst{\ref{aff66}}
\and M.~Huertas-Company\orcid{0000-0002-1416-8483}\inst{\ref{aff55},\ref{aff132},\ref{aff133},\ref{aff134}}
\and J.~Mart\'{i}n-Fleitas\orcid{0000-0002-8594-569X}\inst{\ref{aff103}}
\and S.~Matthew\orcid{0000-0001-8448-1697}\inst{\ref{aff56}}
\and M.~Maturi\orcid{0000-0002-3517-2422}\inst{\ref{aff115},\ref{aff135}}
\and N.~Mauri\orcid{0000-0001-8196-1548}\inst{\ref{aff54},\ref{aff30}}
\and R.~B.~Metcalf\orcid{0000-0003-3167-2574}\inst{\ref{aff95},\ref{aff25}}
\and A.~Pezzotta\orcid{0000-0003-0726-2268}\inst{\ref{aff136},\ref{aff16}}
\and M.~P\"ontinen\orcid{0000-0001-5442-2530}\inst{\ref{aff88}}
\and C.~Porciani\orcid{0000-0002-7797-2508}\inst{\ref{aff94}}
\and I.~Risso\orcid{0000-0003-2525-7761}\inst{\ref{aff137}}
\and V.~Scottez\inst{\ref{aff101},\ref{aff138}}
\and M.~Sereno\orcid{0000-0003-0302-0325}\inst{\ref{aff25},\ref{aff30}}
\and M.~Tenti\orcid{0000-0002-4254-5901}\inst{\ref{aff30}}
\and M.~Viel\orcid{0000-0002-2642-5707}\inst{\ref{aff21},\ref{aff20},\ref{aff28},\ref{aff27},\ref{aff139}}
\and M.~Wiesmann\orcid{0009-0000-8199-5860}\inst{\ref{aff75}}
\and Y.~Akrami\orcid{0000-0002-2407-7956}\inst{\ref{aff140},\ref{aff141}}
\and S.~Alvi\orcid{0000-0001-5779-8568}\inst{\ref{aff127}}
\and I.~T.~Andika\orcid{0000-0001-6102-9526}\inst{\ref{aff142},\ref{aff143}}
\and S.~Anselmi\orcid{0000-0002-3579-9583}\inst{\ref{aff68},\ref{aff113},\ref{aff144}}
\and M.~Archidiacono\orcid{0000-0003-4952-9012}\inst{\ref{aff72},\ref{aff73}}
\and F.~Atrio-Barandela\orcid{0000-0002-2130-2513}\inst{\ref{aff145}}
\and S.~Avila\orcid{0000-0001-5043-3662}\inst{\ref{aff48}}
\and P.~Bergamini\orcid{0000-0003-1383-9414}\inst{\ref{aff72},\ref{aff25}}
\and D.~Bertacca\orcid{0000-0002-2490-7139}\inst{\ref{aff113},\ref{aff31},\ref{aff68}}
\and M.~Bethermin\orcid{0000-0002-3915-2015}\inst{\ref{aff146}}
\and L.~Bisigello\orcid{0000-0003-0492-4924}\inst{\ref{aff31}}
\and A.~Blanchard\orcid{0000-0001-8555-9003}\inst{\ref{aff83}}
\and L.~Blot\orcid{0000-0002-9622-7167}\inst{\ref{aff147},\ref{aff91}}
\and H.~B\"ohringer\orcid{0000-0001-8241-4204}\inst{\ref{aff16},\ref{aff148},\ref{aff149}}
\and S.~Borgani\orcid{0000-0001-6151-6439}\inst{\ref{aff150},\ref{aff21},\ref{aff20},\ref{aff27},\ref{aff139}}
\and A.~S.~Borlaff\orcid{0000-0003-3249-4431}\inst{\ref{aff151},\ref{aff152}}
\and M.~L.~Brown\orcid{0000-0002-0370-8077}\inst{\ref{aff57}}
\and S.~Bruton\orcid{0000-0002-6503-5218}\inst{\ref{aff104}}
\and F.~Buitrago\orcid{0000-0002-2861-9812}\inst{\ref{aff153},\ref{aff3}}
\and A.~Calabro\orcid{0000-0003-2536-1614}\inst{\ref{aff46}}
\and G.~Calderone\orcid{0000-0002-7738-5389}\inst{\ref{aff20}}
\and B.~Camacho~Quevedo\orcid{0000-0002-8789-4232}\inst{\ref{aff52},\ref{aff51}}
\and F.~Caro\inst{\ref{aff46}}
\and C.~S.~Carvalho\inst{\ref{aff3}}
\and T.~Castro\orcid{0000-0002-6292-3228}\inst{\ref{aff20},\ref{aff27},\ref{aff21},\ref{aff139}}
\and Y.~Charles\inst{\ref{aff64}}
\and F.~Cogato\orcid{0000-0003-4632-6113}\inst{\ref{aff95},\ref{aff25}}
\and S.~Conseil\orcid{0000-0002-3657-4191}\inst{\ref{aff59}}
\and A.~R.~Cooray\orcid{0000-0002-3892-0190}\inst{\ref{aff154}}
\and M.~Costanzi\orcid{0000-0001-8158-1449}\inst{\ref{aff150},\ref{aff20},\ref{aff21}}
\and O.~Cucciati\orcid{0000-0002-9336-7551}\inst{\ref{aff25}}
\and S.~Davini\orcid{0000-0003-3269-1718}\inst{\ref{aff36}}
\and F.~De~Paolis\orcid{0000-0001-6460-7563}\inst{\ref{aff155},\ref{aff156},\ref{aff157}}
\and G.~Desprez\orcid{0000-0001-8325-1742}\inst{\ref{aff7}}
\and A.~D\'iaz-S\'anchez\orcid{0000-0003-0748-4768}\inst{\ref{aff158}}
\and J.~J.~Diaz\inst{\ref{aff55}}
\and S.~Di~Domizio\orcid{0000-0003-2863-5895}\inst{\ref{aff35},\ref{aff36}}
\and J.~M.~Diego\orcid{0000-0001-9065-3926}\inst{\ref{aff159}}
\and P.~Dimauro\orcid{0000-0001-7399-2854}\inst{\ref{aff46},\ref{aff160}}
\and P.-A.~Duc\orcid{0000-0003-3343-6284}\inst{\ref{aff146}}
\and A.~Enia\orcid{0000-0002-0200-2857}\inst{\ref{aff29},\ref{aff25}}
\and Y.~Fang\inst{\ref{aff19}}
\and A.~M.~N.~Ferguson\inst{\ref{aff56}}
\and A.~G.~Ferrari\orcid{0009-0005-5266-4110}\inst{\ref{aff30}}
\and A.~Finoguenov\orcid{0000-0002-4606-5403}\inst{\ref{aff88}}
\and A.~Fontana\orcid{0000-0003-3820-2823}\inst{\ref{aff46}}
\and F.~Fontanot\orcid{0000-0003-4744-0188}\inst{\ref{aff20},\ref{aff21}}
\and A.~Franco\orcid{0000-0002-4761-366X}\inst{\ref{aff156},\ref{aff155},\ref{aff157}}
\and J.~Garc\'ia-Bellido\orcid{0000-0002-9370-8360}\inst{\ref{aff140}}
\and T.~Gasparetto\orcid{0000-0002-7913-4866}\inst{\ref{aff20}}
\and R.~Gavazzi\orcid{0000-0002-5540-6935}\inst{\ref{aff64},\ref{aff76}}
\and E.~Gaztanaga\orcid{0000-0001-9632-0815}\inst{\ref{aff51},\ref{aff52},\ref{aff161}}
\and F.~Giacomini\orcid{0000-0002-3129-2814}\inst{\ref{aff30}}
\and F.~Gianotti\orcid{0000-0003-4666-119X}\inst{\ref{aff25}}
\and A.~H.~Gonzalez\orcid{0000-0002-0933-8601}\inst{\ref{aff162}}
\and G.~Gozaliasl\orcid{0000-0002-0236-919X}\inst{\ref{aff163},\ref{aff88}}
\and A.~Gruppuso\orcid{0000-0001-9272-5292}\inst{\ref{aff25},\ref{aff30}}
\and M.~Guidi\orcid{0000-0001-9408-1101}\inst{\ref{aff29},\ref{aff25}}
\and C.~M.~Gutierrez\orcid{0000-0001-7854-783X}\inst{\ref{aff164}}
\and A.~Hall\orcid{0000-0002-3139-8651}\inst{\ref{aff56}}
\and C.~Hern\'andez-Monteagudo\orcid{0000-0001-5471-9166}\inst{\ref{aff112},\ref{aff55}}
\and H.~Hildebrandt\orcid{0000-0002-9814-3338}\inst{\ref{aff165}}
\and J.~Hjorth\orcid{0000-0002-4571-2306}\inst{\ref{aff108}}
\and J.~Jacobson\inst{\ref{aff114}}
\and S.~Joudaki\orcid{0000-0001-8820-673X}\inst{\ref{aff48}}
\and J.~J.~E.~Kajava\orcid{0000-0002-3010-8333}\inst{\ref{aff166},\ref{aff167}}
\and Y.~Kang\orcid{0009-0000-8588-7250}\inst{\ref{aff66}}
\and V.~Kansal\orcid{0000-0002-4008-6078}\inst{\ref{aff168},\ref{aff169}}
\and D.~Karagiannis\orcid{0000-0002-4927-0816}\inst{\ref{aff127},\ref{aff170}}
\and K.~Kiiveri\inst{\ref{aff86}}
\and C.~C.~Kirkpatrick\inst{\ref{aff86}}
\and S.~Kruk\orcid{0000-0001-8010-8879}\inst{\ref{aff6}}
\and F.~Lacasa\orcid{0000-0002-7268-3440}\inst{\ref{aff171},\ref{aff22}}
\and C.~Laigle\orcid{0009-0008-5926-818X}\inst{\ref{aff76}}
\and M.~Lattanzi\orcid{0000-0003-1059-2532}\inst{\ref{aff128}}
\and V.~Le~Brun\orcid{0000-0002-5027-1939}\inst{\ref{aff64}}
\and J.~Le~Graet\orcid{0000-0001-6523-7971}\inst{\ref{aff69}}
\and L.~Legrand\orcid{0000-0003-0610-5252}\inst{\ref{aff172},\ref{aff173}}
\and M.~Lembo\orcid{0000-0002-5271-5070}\inst{\ref{aff127},\ref{aff128}}
\and F.~Lepori\orcid{0009-0000-5061-7138}\inst{\ref{aff174}}
\and G.~Leroy\orcid{0009-0004-2523-4425}\inst{\ref{aff175},\ref{aff96}}
\and G.~F.~Lesci\orcid{0000-0002-4607-2830}\inst{\ref{aff95},\ref{aff25}}
\and J.~Lesgourgues\orcid{0000-0001-7627-353X}\inst{\ref{aff50}}
\and L.~Leuzzi\orcid{0009-0006-4479-7017}\inst{\ref{aff95},\ref{aff25}}
\and T.~I.~Liaudat\orcid{0000-0002-9104-314X}\inst{\ref{aff176}}
\and A.~Loureiro\orcid{0000-0002-4371-0876}\inst{\ref{aff177},\ref{aff178}}
\and M.~Magliocchetti\orcid{0000-0001-9158-4838}\inst{\ref{aff67}}
\and E.~A.~Magnier\orcid{0000-0002-7965-2815}\inst{\ref{aff9}}
\and C.~Mancini\orcid{0000-0002-4297-0561}\inst{\ref{aff45}}
\and F.~Mannucci\orcid{0000-0002-4803-2381}\inst{\ref{aff179}}
\and R.~Maoli\orcid{0000-0002-6065-3025}\inst{\ref{aff180},\ref{aff46}}
\and C.~J.~A.~P.~Martins\orcid{0000-0002-4886-9261}\inst{\ref{aff181},\ref{aff39}}
\and L.~Maurin\orcid{0000-0002-8406-0857}\inst{\ref{aff22}}
\and C.~J.~R.~McPartland\orcid{0000-0003-0639-025X}\inst{\ref{aff81},\ref{aff123}}
\and J.-B.~Melin\inst{\ref{aff182}}
\and M.~Migliaccio\inst{\ref{aff183},\ref{aff184}}
\and M.~Miluzio\inst{\ref{aff6},\ref{aff185}}
\and P.~Monaco\orcid{0000-0003-2083-7564}\inst{\ref{aff150},\ref{aff20},\ref{aff27},\ref{aff21}}
\and A.~Montoro\orcid{0000-0003-4730-8590}\inst{\ref{aff51},\ref{aff52}}
\and C.~Moretti\orcid{0000-0003-3314-8936}\inst{\ref{aff28},\ref{aff139},\ref{aff20},\ref{aff21},\ref{aff27}}
\and G.~Morgante\inst{\ref{aff25}}
\and C.~Murray\inst{\ref{aff98}}
\and S.~Nadathur\orcid{0000-0001-9070-3102}\inst{\ref{aff161}}
\and K.~Naidoo\orcid{0000-0002-9182-1802}\inst{\ref{aff161}}
\and A.~Navarro-Alsina\orcid{0000-0002-3173-2592}\inst{\ref{aff94}}
\and S.~Nesseris\orcid{0000-0002-0567-0324}\inst{\ref{aff140}}
\and L.~Nicastro\orcid{0000-0001-8534-6788}\inst{\ref{aff25}}
\and M.~Oguri\orcid{0000-0003-3484-399X}\inst{\ref{aff186},\ref{aff187}}
\and F.~Passalacqua\orcid{0000-0002-8606-4093}\inst{\ref{aff113},\ref{aff68}}
\and K.~Paterson\orcid{0000-0001-8340-3486}\inst{\ref{aff4}}
\and L.~Patrizii\inst{\ref{aff30}}
\and A.~Pisani\orcid{0000-0002-6146-4437}\inst{\ref{aff69},\ref{aff188}}
\and D.~Potter\orcid{0000-0002-0757-5195}\inst{\ref{aff174}}
\and S.~Quai\orcid{0000-0002-0449-8163}\inst{\ref{aff95},\ref{aff25}}
\and M.~Radovich\orcid{0000-0002-3585-866X}\inst{\ref{aff31}}
\and P.~Reimberg\orcid{0000-0003-3410-0280}\inst{\ref{aff101}}
\and P.-F.~Rocci\inst{\ref{aff22}}
\and G.~Rodighiero\orcid{0000-0002-9415-2296}\inst{\ref{aff113},\ref{aff31}}
\and R.~P.~Rollins\orcid{0000-0003-1291-1023}\inst{\ref{aff56}}
\and S.~Sacquegna\orcid{0000-0002-8433-6630}\inst{\ref{aff155},\ref{aff156},\ref{aff157}}
\and M.~Sahl\'en\orcid{0000-0003-0973-4804}\inst{\ref{aff189}}
\and D.~B.~Sanders\orcid{0000-0002-1233-9998}\inst{\ref{aff9}}
\and E.~Sarpa\orcid{0000-0002-1256-655X}\inst{\ref{aff28},\ref{aff139},\ref{aff27}}
\and C.~Scarlata\orcid{0000-0002-9136-8876}\inst{\ref{aff190}}
\and J.~Schaye\orcid{0000-0002-0668-5560}\inst{\ref{aff8}}
\and A.~Schneider\orcid{0000-0001-7055-8104}\inst{\ref{aff174}}
\and M.~Schultheis\inst{\ref{aff97}}
\and D.~Sciotti\orcid{0009-0008-4519-2620}\inst{\ref{aff46},\ref{aff47}}
\and D.~Scognamiglio\orcid{0000-0001-8450-7885}\inst{\ref{aff77}}
\and E.~Sellentin\inst{\ref{aff191},\ref{aff8}}
\and F.~Shankar\orcid{0000-0001-8973-5051}\inst{\ref{aff192}}
\and L.~C.~Smith\orcid{0000-0002-3259-2771}\inst{\ref{aff193}}
\and E.~Soubrie\orcid{0000-0001-9295-1863}\inst{\ref{aff22}}
\and S.~A.~Stanford\orcid{0000-0003-0122-0841}\inst{\ref{aff194}}
\and K.~Tanidis\orcid{0000-0001-9843-5130}\inst{\ref{aff131}}
\and C.~Tao\orcid{0000-0001-7961-8177}\inst{\ref{aff69}}
\and G.~Testera\inst{\ref{aff36}}
\and M.~Tewes\orcid{0000-0002-1155-8689}\inst{\ref{aff94}}
\and R.~Teyssier\orcid{0000-0001-7689-0933}\inst{\ref{aff188}}
\and S.~Tosi\orcid{0000-0002-7275-9193}\inst{\ref{aff35},\ref{aff36},\ref{aff24}}
\and A.~Troja\orcid{0000-0003-0239-4595}\inst{\ref{aff113},\ref{aff68}}
\and M.~Tucci\inst{\ref{aff66}}
\and C.~Valieri\inst{\ref{aff30}}
\and A.~Venhola\orcid{0000-0001-6071-4564}\inst{\ref{aff195}}
\and D.~Vergani\orcid{0000-0003-0898-2216}\inst{\ref{aff25}}
\and F.~Vernizzi\orcid{0000-0003-3426-2802}\inst{\ref{aff196}}
\and G.~Verza\orcid{0000-0002-1886-8348}\inst{\ref{aff197}}
\and P.~Vielzeuf\orcid{0000-0003-2035-9339}\inst{\ref{aff69}}
\and N.~A.~Walton\orcid{0000-0003-3983-8778}\inst{\ref{aff193}}
\and J.~R.~Weaver\orcid{0000-0003-1614-196X}\inst{\ref{aff198}}
\and J.~Wilde\orcid{0000-0002-4460-7379}\inst{\ref{aff60}}
\and L.~Zalesky\orcid{0000-0001-5680-2326}\inst{\ref{aff9}}}
										   
\institute{Universit\'e Paris-Saclay, Universit\'e Paris Cit\'e, CEA, CNRS, AIM, 91191, Gif-sur-Yvette, France\label{aff1}
\and
Departamento de F\'isica, Faculdade de Ci\^encias, Universidade de Lisboa, Edif\'icio C8, Campo Grande, PT1749-016 Lisboa, Portugal\label{aff2}
\and
Instituto de Astrof\'isica e Ci\^encias do Espa\c{c}o, Faculdade de Ci\^encias, Universidade de Lisboa, Tapada da Ajuda, 1349-018 Lisboa, Portugal\label{aff3}
\and
Max-Planck-Institut f\"ur Astronomie, K\"onigstuhl 17, 69117 Heidelberg, Germany\label{aff4}
\and
Univ. Grenoble Alpes, CNRS, Grenoble INP, LPSC-IN2P3, 53, Avenue des Martyrs, 38000, Grenoble, France\label{aff5}
\and
ESAC/ESA, Camino Bajo del Castillo, s/n., Urb. Villafranca del Castillo, 28692 Villanueva de la Ca\~nada, Madrid, Spain\label{aff6}
\and
Kapteyn Astronomical Institute, University of Groningen, PO Box 800, 9700 AV Groningen, The Netherlands\label{aff7}
\and
Leiden Observatory, Leiden University, Einsteinweg 55, 2333 CC Leiden, The Netherlands\label{aff8}
\and
Institute for Astronomy, University of Hawaii, 2680 Woodlawn Drive, Honolulu, HI 96822, USA\label{aff9}
\and
Centre National d'Etudes Spatiales -- Centre spatial de Toulouse, 18 avenue Edouard Belin, 31401 Toulouse Cedex 9, France\label{aff10}
\and
National Astronomical Observatory of Japan, 2-21-1 Osawa, Mitaka, Tokyo 181-8588, Japan\label{aff11}
\and
Graduate Institute for Advanced Studies, SOKENDAI, 2-21-1 Osawa, Mitaka, Tokyo 181-8588, Japan\label{aff12}
\and
Department of Physics and Astronomy, University of Waterloo, Waterloo, Ontario N2L 3G1, Canada\label{aff13}
\and
Waterloo Centre for Astrophysics, University of Waterloo, Waterloo, Ontario N2L 3G1, Canada\label{aff14}
\and
Perimeter Institute for Theoretical Physics, Waterloo, Ontario N2L 2Y5, Canada\label{aff15}
\and
Max Planck Institute for Extraterrestrial Physics, Giessenbachstr. 1, 85748 Garching, Germany\label{aff16}
\and
University Observatory, LMU Faculty of Physics, Scheinerstrasse 1, 81679 Munich, Germany\label{aff17}
\and
Department of Physics and Astronomy, University of British Columbia, Vancouver, BC V6T 1Z1, Canada\label{aff18}
\and
Universit\"ats-Sternwarte M\"unchen, Fakult\"at f\"ur Physik, Ludwig-Maximilians-Universit\"at M\"unchen, Scheinerstrasse 1, 81679 M\"unchen, Germany\label{aff19}
\and
INAF-Osservatorio Astronomico di Trieste, Via G. B. Tiepolo 11, 34143 Trieste, Italy\label{aff20}
\and
IFPU, Institute for Fundamental Physics of the Universe, via Beirut 2, 34151 Trieste, Italy\label{aff21}
\and
Universit\'e Paris-Saclay, CNRS, Institut d'astrophysique spatiale, 91405, Orsay, France\label{aff22}
\and
School of Mathematics and Physics, University of Surrey, Guildford, Surrey, GU2 7XH, UK\label{aff23}
\and
INAF-Osservatorio Astronomico di Brera, Via Brera 28, 20122 Milano, Italy\label{aff24}
\and
INAF-Osservatorio di Astrofisica e Scienza dello Spazio di Bologna, Via Piero Gobetti 93/3, 40129 Bologna, Italy\label{aff25}
\and
Mullard Space Science Laboratory, University College London, Holmbury St Mary, Dorking, Surrey RH5 6NT, UK\label{aff26}
\and
INFN, Sezione di Trieste, Via Valerio 2, 34127 Trieste TS, Italy\label{aff27}
\and
SISSA, International School for Advanced Studies, Via Bonomea 265, 34136 Trieste TS, Italy\label{aff28}
\and
Dipartimento di Fisica e Astronomia, Universit\`a di Bologna, Via Gobetti 93/2, 40129 Bologna, Italy\label{aff29}
\and
INFN-Sezione di Bologna, Viale Berti Pichat 6/2, 40127 Bologna, Italy\label{aff30}
\and
INAF-Osservatorio Astronomico di Padova, Via dell'Osservatorio 5, 35122 Padova, Italy\label{aff31}
\and
ATG Europe BV, Huygensstraat 34, 2201 DK Noordwijk, The Netherlands\label{aff32}
\and
Space Science Data Center, Italian Space Agency, via del Politecnico snc, 00133 Roma, Italy\label{aff33}
\and
INAF-Osservatorio Astrofisico di Torino, Via Osservatorio 20, 10025 Pino Torinese (TO), Italy\label{aff34}
\and
Dipartimento di Fisica, Universit\`a di Genova, Via Dodecaneso 33, 16146, Genova, Italy\label{aff35}
\and
INFN-Sezione di Genova, Via Dodecaneso 33, 16146, Genova, Italy\label{aff36}
\and
Department of Physics "E. Pancini", University Federico II, Via Cinthia 6, 80126, Napoli, Italy\label{aff37}
\and
INAF-Osservatorio Astronomico di Capodimonte, Via Moiariello 16, 80131 Napoli, Italy\label{aff38}
\and
Instituto de Astrof\'isica e Ci\^encias do Espa\c{c}o, Universidade do Porto, CAUP, Rua das Estrelas, PT4150-762 Porto, Portugal\label{aff39}
\and
Faculdade de Ci\^encias da Universidade do Porto, Rua do Campo de Alegre, 4150-007 Porto, Portugal\label{aff40}
\and
Dipartimento di Fisica, Universit\`a degli Studi di Torino, Via P. Giuria 1, 10125 Torino, Italy\label{aff41}
\and
INFN-Sezione di Torino, Via P. Giuria 1, 10125 Torino, Italy\label{aff42}
\and
European Space Agency/ESTEC, Keplerlaan 1, 2201 AZ Noordwijk, The Netherlands\label{aff43}
\and
Institute Lorentz, Leiden University, Niels Bohrweg 2, 2333 CA Leiden, The Netherlands\label{aff44}
\and
INAF-IASF Milano, Via Alfonso Corti 12, 20133 Milano, Italy\label{aff45}
\and
INAF-Osservatorio Astronomico di Roma, Via Frascati 33, 00078 Monteporzio Catone, Italy\label{aff46}
\and
INFN-Sezione di Roma, Piazzale Aldo Moro, 2 - c/o Dipartimento di Fisica, Edificio G. Marconi, 00185 Roma, Italy\label{aff47}
\and
Centro de Investigaciones Energ\'eticas, Medioambientales y Tecnol\'ogicas (CIEMAT), Avenida Complutense 40, 28040 Madrid, Spain\label{aff48}
\and
Port d'Informaci\'{o} Cient\'{i}fica, Campus UAB, C. Albareda s/n, 08193 Bellaterra (Barcelona), Spain\label{aff49}
\and
Institute for Theoretical Particle Physics and Cosmology (TTK), RWTH Aachen University, 52056 Aachen, Germany\label{aff50}
\and
Institute of Space Sciences (ICE, CSIC), Campus UAB, Carrer de Can Magrans, s/n, 08193 Barcelona, Spain\label{aff51}
\and
Institut d'Estudis Espacials de Catalunya (IEEC),  Edifici RDIT, Campus UPC, 08860 Castelldefels, Barcelona, Spain\label{aff52}
\and
INFN section of Naples, Via Cinthia 6, 80126, Napoli, Italy\label{aff53}
\and
Dipartimento di Fisica e Astronomia "Augusto Righi" - Alma Mater Studiorum Universit\`a di Bologna, Viale Berti Pichat 6/2, 40127 Bologna, Italy\label{aff54}
\and
Instituto de Astrof\'{\i}sica de Canarias, V\'{\i}a L\'actea, 38205 La Laguna, Tenerife, Spain\label{aff55}
\and
Institute for Astronomy, University of Edinburgh, Royal Observatory, Blackford Hill, Edinburgh EH9 3HJ, UK\label{aff56}
\and
Jodrell Bank Centre for Astrophysics, Department of Physics and Astronomy, University of Manchester, Oxford Road, Manchester M13 9PL, UK\label{aff57}
\and
European Space Agency/ESRIN, Largo Galileo Galilei 1, 00044 Frascati, Roma, Italy\label{aff58}
\and
Universit\'e Claude Bernard Lyon 1, CNRS/IN2P3, IP2I Lyon, UMR 5822, Villeurbanne, F-69100, France\label{aff59}
\and
Institut de Ci\`{e}ncies del Cosmos (ICCUB), Universitat de Barcelona (IEEC-UB), Mart\'{i} i Franqu\`{e}s 1, 08028 Barcelona, Spain\label{aff60}
\and
Instituci\'o Catalana de Recerca i Estudis Avan\c{c}ats (ICREA), Passeig de Llu\'{\i}s Companys 23, 08010 Barcelona, Spain\label{aff61}
\and
UCB Lyon 1, CNRS/IN2P3, IUF, IP2I Lyon, 4 rue Enrico Fermi, 69622 Villeurbanne, France\label{aff62}
\and
Canada-France-Hawaii Telescope, 65-1238 Mamalahoa Hwy, Kamuela, HI 96743, USA\label{aff63}
\and
Aix-Marseille Universit\'e, CNRS, CNES, LAM, Marseille, France\label{aff64}
\and
Instituto de Astrof\'isica e Ci\^encias do Espa\c{c}o, Faculdade de Ci\^encias, Universidade de Lisboa, Campo Grande, 1749-016 Lisboa, Portugal\label{aff65}
\and
Department of Astronomy, University of Geneva, ch. d'Ecogia 16, 1290 Versoix, Switzerland\label{aff66}
\and
INAF-Istituto di Astrofisica e Planetologia Spaziali, via del Fosso del Cavaliere, 100, 00100 Roma, Italy\label{aff67}
\and
INFN-Padova, Via Marzolo 8, 35131 Padova, Italy\label{aff68}
\and
Aix-Marseille Universit\'e, CNRS/IN2P3, CPPM, Marseille, France\label{aff69}
\and
School of Physics, HH Wills Physics Laboratory, University of Bristol, Tyndall Avenue, Bristol, BS8 1TL, UK\label{aff70}
\and
FRACTAL S.L.N.E., calle Tulip\'an 2, Portal 13 1A, 28231, Las Rozas de Madrid, Spain\label{aff71}
\and
Dipartimento di Fisica "Aldo Pontremoli", Universit\`a degli Studi di Milano, Via Celoria 16, 20133 Milano, Italy\label{aff72}
\and
INFN-Sezione di Milano, Via Celoria 16, 20133 Milano, Italy\label{aff73}
\and
NRC Herzberg, 5071 West Saanich Rd, Victoria, BC V9E 2E7, Canada\label{aff74}
\and
Institute of Theoretical Astrophysics, University of Oslo, P.O. Box 1029 Blindern, 0315 Oslo, Norway\label{aff75}
\and
Institut d'Astrophysique de Paris, UMR 7095, CNRS, and Sorbonne Universit\'e, 98 bis boulevard Arago, 75014 Paris, France\label{aff76}
\and
Jet Propulsion Laboratory, California Institute of Technology, 4800 Oak Grove Drive, Pasadena, CA, 91109, USA\label{aff77}
\and
Department of Physics, Lancaster University, Lancaster, LA1 4YB, UK\label{aff78}
\and
Felix Hormuth Engineering, Goethestr. 17, 69181 Leimen, Germany\label{aff79}
\and
Technical University of Denmark, Elektrovej 327, 2800 Kgs. Lyngby, Denmark\label{aff80}
\and
Cosmic Dawn Center (DAWN), Denmark\label{aff81}
\and
Universit\'e Paris-Saclay, CNRS/IN2P3, IJCLab, 91405 Orsay, France\label{aff82}
\and
Institut de Recherche en Astrophysique et Plan\'etologie (IRAP), Universit\'e de Toulouse, CNRS, UPS, CNES, 14 Av. Edouard Belin, 31400 Toulouse, France\label{aff83}
\and
NASA Goddard Space Flight Center, Greenbelt, MD 20771, USA\label{aff84}
\and
Department of Physics and Astronomy, University College London, Gower Street, London WC1E 6BT, UK\label{aff85}
\and
Department of Physics and Helsinki Institute of Physics, Gustaf H\"allstr\"omin katu 2, 00014 University of Helsinki, Finland\label{aff86}
\and
Universit\'e de Gen\`eve, D\'epartement de Physique Th\'eorique and Centre for Astroparticle Physics, 24 quai Ernest-Ansermet, CH-1211 Gen\`eve 4, Switzerland\label{aff87}
\and
Department of Physics, P.O. Box 64, 00014 University of Helsinki, Finland\label{aff88}
\and
Helsinki Institute of Physics, Gustaf H{\"a}llstr{\"o}min katu 2, University of Helsinki, Helsinki, Finland\label{aff89}
\and
Centre de Calcul de l'IN2P3/CNRS, 21 avenue Pierre de Coubertin 69627 Villeurbanne Cedex, France\label{aff90}
\and
Laboratoire d'etude de l'Univers et des phenomenes eXtremes, Observatoire de Paris, Universit\'e PSL, Sorbonne Universit\'e, CNRS, 92190 Meudon, France\label{aff91}
\and
SKA Observatory, Jodrell Bank, Lower Withington, Macclesfield, Cheshire SK11 9FT, UK\label{aff92}
\and
University of Applied Sciences and Arts of Northwestern Switzerland, School of Computer Science, 5210 Windisch, Switzerland\label{aff93}
\and
Universit\"at Bonn, Argelander-Institut f\"ur Astronomie, Auf dem H\"ugel 71, 53121 Bonn, Germany\label{aff94}
\and
Dipartimento di Fisica e Astronomia "Augusto Righi" - Alma Mater Studiorum Universit\`a di Bologna, via Piero Gobetti 93/2, 40129 Bologna, Italy\label{aff95}
\and
Department of Physics, Institute for Computational Cosmology, Durham University, South Road, Durham, DH1 3LE, UK\label{aff96}
\and
Universit\'e C\^{o}te d'Azur, Observatoire de la C\^{o}te d'Azur, CNRS, Laboratoire Lagrange, Bd de l'Observatoire, CS 34229, 06304 Nice cedex 4, France\label{aff97}
\and
Universit\'e Paris Cit\'e, CNRS, Astroparticule et Cosmologie, 75013 Paris, France\label{aff98}
\and
CNRS-UCB International Research Laboratory, Centre Pierre Binetruy, IRL2007, CPB-IN2P3, Berkeley, USA\label{aff99}
\and
University of Applied Sciences and Arts of Northwestern Switzerland, School of Engineering, 5210 Windisch, Switzerland\label{aff100}
\and
Institut d'Astrophysique de Paris, 98bis Boulevard Arago, 75014, Paris, France\label{aff101}
\and
Institute of Physics, Laboratory of Astrophysics, Ecole Polytechnique F\'ed\'erale de Lausanne (EPFL), Observatoire de Sauverny, 1290 Versoix, Switzerland\label{aff102}
\and
Aurora Technology for European Space Agency (ESA), Camino bajo del Castillo, s/n, Urbanizacion Villafranca del Castillo, Villanueva de la Ca\~nada, 28692 Madrid, Spain\label{aff103}
\and
California Institute of Technology, 1200 E California Blvd, Pasadena, CA 91125, USA\label{aff104}
\and
OCA, P.H.C Boulevard de l'Observatoire CS 34229, 06304 Nice Cedex 4, France\label{aff105}
\and
Institut de F\'{i}sica d'Altes Energies (IFAE), The Barcelona Institute of Science and Technology, Campus UAB, 08193 Bellaterra (Barcelona), Spain\label{aff106}
\and
School of Mathematics, Statistics and Physics, Newcastle University, Herschel Building, Newcastle-upon-Tyne, NE1 7RU, UK\label{aff107}
\and
DARK, Niels Bohr Institute, University of Copenhagen, Jagtvej 155, 2200 Copenhagen, Denmark\label{aff108}
\and
CEA Saclay, DFR/IRFU, Service d'Astrophysique, Bat. 709, 91191 Gif-sur-Yvette, France\label{aff109}
\and
Institute of Space Science, Str. Atomistilor, nr. 409 M\u{a}gurele, Ilfov, 077125, Romania\label{aff110}
\and
Consejo Superior de Investigaciones Cientificas, Calle Serrano 117, 28006 Madrid, Spain\label{aff111}
\and
Universidad de La Laguna, Departamento de Astrof\'{\i}sica, 38206 La Laguna, Tenerife, Spain\label{aff112}
\and
Dipartimento di Fisica e Astronomia "G. Galilei", Universit\`a di Padova, Via Marzolo 8, 35131 Padova, Italy\label{aff113}
\and
Caltech/IPAC, 1200 E. California Blvd., Pasadena, CA 91125, USA\label{aff114}
\and
Institut f\"ur Theoretische Physik, University of Heidelberg, Philosophenweg 16, 69120 Heidelberg, Germany\label{aff115}
\and
Universit\'e St Joseph; Faculty of Sciences, Beirut, Lebanon\label{aff116}
\and
Departamento de F\'isica, FCFM, Universidad de Chile, Blanco Encalada 2008, Santiago, Chile\label{aff117}
\and
Satlantis, University Science Park, Sede Bld 48940, Leioa-Bilbao, Spain\label{aff118}
\and
Centre for Electronic Imaging, Open University, Walton Hall, Milton Keynes, MK7~6AA, UK\label{aff119}
\and
Department of Physics, Royal Holloway, University of London, TW20 0EX, UK\label{aff120}
\and
Infrared Processing and Analysis Center, California Institute of Technology, Pasadena, CA 91125, USA\label{aff121}
\and
Cosmic Dawn Center (DAWN)\label{aff122}
\and
Niels Bohr Institute, University of Copenhagen, Jagtvej 128, 2200 Copenhagen, Denmark\label{aff123}
\and
Universidad Polit\'ecnica de Cartagena, Departamento de Electr\'onica y Tecnolog\'ia de Computadoras,  Plaza del Hospital 1, 30202 Cartagena, Spain\label{aff124}
\and
Centre for Information Technology, University of Groningen, P.O. Box 11044, 9700 CA Groningen, The Netherlands\label{aff125}
\and
INFN-Bologna, Via Irnerio 46, 40126 Bologna, Italy\label{aff126}
\and
Dipartimento di Fisica e Scienze della Terra, Universit\`a degli Studi di Ferrara, Via Giuseppe Saragat 1, 44122 Ferrara, Italy\label{aff127}
\and
Istituto Nazionale di Fisica Nucleare, Sezione di Ferrara, Via Giuseppe Saragat 1, 44122 Ferrara, Italy\label{aff128}
\and
INAF, Istituto di Radioastronomia, Via Piero Gobetti 101, 40129 Bologna, Italy\label{aff129}
\and
Astronomical Observatory of the Autonomous Region of the Aosta Valley (OAVdA), Loc. Lignan 39, I-11020, Nus (Aosta Valley), Italy\label{aff130}
\and
Department of Physics, Oxford University, Keble Road, Oxford OX1 3RH, UK\label{aff131}
\and
Instituto de Astrof\'isica de Canarias (IAC); Departamento de Astrof\'isica, Universidad de La Laguna (ULL), 38200, La Laguna, Tenerife, Spain\label{aff132}
\and
Universit\'e PSL, Observatoire de Paris, Sorbonne Universit\'e, CNRS, LERMA, 75014, Paris, France\label{aff133}
\and
Universit\'e Paris-Cit\'e, 5 Rue Thomas Mann, 75013, Paris, France\label{aff134}
\and
Zentrum f\"ur Astronomie, Universit\"at Heidelberg, Philosophenweg 12, 69120 Heidelberg, Germany\label{aff135}
\and
INAF - Osservatorio Astronomico di Brera, via Emilio Bianchi 46, 23807 Merate, Italy\label{aff136}
\and
INAF-Osservatorio Astronomico di Brera, Via Brera 28, 20122 Milano, Italy, and INFN-Sezione di Genova, Via Dodecaneso 33, 16146, Genova, Italy\label{aff137}
\and
ICL, Junia, Universit\'e Catholique de Lille, LITL, 59000 Lille, France\label{aff138}
\and
ICSC - Centro Nazionale di Ricerca in High Performance Computing, Big Data e Quantum Computing, Via Magnanelli 2, Bologna, Italy\label{aff139}
\and
Instituto de F\'isica Te\'orica UAM-CSIC, Campus de Cantoblanco, 28049 Madrid, Spain\label{aff140}
\and
CERCA/ISO, Department of Physics, Case Western Reserve University, 10900 Euclid Avenue, Cleveland, OH 44106, USA\label{aff141}
\and
Technical University of Munich, TUM School of Natural Sciences, Physics Department, James-Franck-Str.~1, 85748 Garching, Germany\label{aff142}
\and
Max-Planck-Institut f\"ur Astrophysik, Karl-Schwarzschild-Str.~1, 85748 Garching, Germany\label{aff143}
\and
Laboratoire Univers et Th\'eorie, Observatoire de Paris, Universit\'e PSL, Universit\'e Paris Cit\'e, CNRS, 92190 Meudon, France\label{aff144}
\and
Departamento de F{\'\i}sica Fundamental. Universidad de Salamanca. Plaza de la Merced s/n. 37008 Salamanca, Spain\label{aff145}
\and
Universit\'e de Strasbourg, CNRS, Observatoire astronomique de Strasbourg, UMR 7550, 67000 Strasbourg, France\label{aff146}
\and
Center for Data-Driven Discovery, Kavli IPMU (WPI), UTIAS, The University of Tokyo, Kashiwa, Chiba 277-8583, Japan\label{aff147}
\and
Ludwig-Maximilians-University, Schellingstrasse 4, 80799 Munich, Germany\label{aff148}
\and
Max-Planck-Institut f\"ur Physik, Boltzmannstr. 8, 85748 Garching, Germany\label{aff149}
\and
Dipartimento di Fisica - Sezione di Astronomia, Universit\`a di Trieste, Via Tiepolo 11, 34131 Trieste, Italy\label{aff150}
\and
NASA Ames Research Center, Moffett Field, CA 94035, USA\label{aff151}
\and
Bay Area Environmental Research Institute, Moffett Field, California 94035, USA\label{aff152}
\and
Departamento de F\'{i}sica Te\'{o}rica, At\'{o}mica y \'{O}ptica, Universidad de Valladolid, 47011 Valladolid, Spain\label{aff153}
\and
Department of Physics \& Astronomy, University of California Irvine, Irvine CA 92697, USA\label{aff154}
\and
Department of Mathematics and Physics E. De Giorgi, University of Salento, Via per Arnesano, CP-I93, 73100, Lecce, Italy\label{aff155}
\and
INFN, Sezione di Lecce, Via per Arnesano, CP-193, 73100, Lecce, Italy\label{aff156}
\and
INAF-Sezione di Lecce, c/o Dipartimento Matematica e Fisica, Via per Arnesano, 73100, Lecce, Italy\label{aff157}
\and
Departamento F\'isica Aplicada, Universidad Polit\'ecnica de Cartagena, Campus Muralla del Mar, 30202 Cartagena, Murcia, Spain\label{aff158}
\and
Instituto de F\'isica de Cantabria, Edificio Juan Jord\'a, Avenida de los Castros, 39005 Santander, Spain\label{aff159}
\and
Observatorio Nacional, Rua General Jose Cristino, 77-Bairro Imperial de Sao Cristovao, Rio de Janeiro, 20921-400, Brazil\label{aff160}
\and
Institute of Cosmology and Gravitation, University of Portsmouth, Portsmouth PO1 3FX, UK\label{aff161}
\and
Department of Astronomy, University of Florida, Bryant Space Science Center, Gainesville, FL 32611, USA\label{aff162}
\and
Department of Computer Science, Aalto University, PO Box 15400, Espoo, FI-00 076, Finland\label{aff163}
\and
Instituto de Astrof\'\i sica de Canarias, c/ Via Lactea s/n, La Laguna 38200, Spain. Departamento de Astrof\'\i sica de la Universidad de La Laguna, Avda. Francisco Sanchez, La Laguna, 38200, Spain\label{aff164}
\and
Ruhr University Bochum, Faculty of Physics and Astronomy, Astronomical Institute (AIRUB), German Centre for Cosmological Lensing (GCCL), 44780 Bochum, Germany\label{aff165}
\and
Department of Physics and Astronomy, Vesilinnantie 5, 20014 University of Turku, Finland\label{aff166}
\and
Serco for European Space Agency (ESA), Camino bajo del Castillo, s/n, Urbanizacion Villafranca del Castillo, Villanueva de la Ca\~nada, 28692 Madrid, Spain\label{aff167}
\and
ARC Centre of Excellence for Dark Matter Particle Physics, Melbourne, Australia\label{aff168}
\and
Centre for Astrophysics \& Supercomputing, Swinburne University of Technology,  Hawthorn, Victoria 3122, Australia\label{aff169}
\and
Department of Physics and Astronomy, University of the Western Cape, Bellville, Cape Town, 7535, South Africa\label{aff170}
\and
Universit\'e Libre de Bruxelles (ULB), Service de Physique Th\'eorique CP225, Boulevard du Triophe, 1050 Bruxelles, Belgium\label{aff171}
\and
DAMTP, Centre for Mathematical Sciences, Wilberforce Road, Cambridge CB3 0WA, UK\label{aff172}
\and
Kavli Institute for Cosmology Cambridge, Madingley Road, Cambridge, CB3 0HA, UK\label{aff173}
\and
Department of Astrophysics, University of Zurich, Winterthurerstrasse 190, 8057 Zurich, Switzerland\label{aff174}
\and
Department of Physics, Centre for Extragalactic Astronomy, Durham University, South Road, Durham, DH1 3LE, UK\label{aff175}
\and
IRFU, CEA, Universit\'e Paris-Saclay 91191 Gif-sur-Yvette Cedex, France\label{aff176}
\and
Oskar Klein Centre for Cosmoparticle Physics, Department of Physics, Stockholm University, Stockholm, SE-106 91, Sweden\label{aff177}
\and
Astrophysics Group, Blackett Laboratory, Imperial College London, London SW7 2AZ, UK\label{aff178}
\and
INAF-Osservatorio Astrofisico di Arcetri, Largo E. Fermi 5, 50125, Firenze, Italy\label{aff179}
\and
Dipartimento di Fisica, Sapienza Universit\`a di Roma, Piazzale Aldo Moro 2, 00185 Roma, Italy\label{aff180}
\and
Centro de Astrof\'{\i}sica da Universidade do Porto, Rua das Estrelas, 4150-762 Porto, Portugal\label{aff181}
\and
Universit\'e Paris-Saclay, CEA, D\'epartement de Physique des Particules, 91191, Gif-sur-Yvette, France\label{aff182}
\and
Dipartimento di Fisica, Universit\`a di Roma Tor Vergata, Via della Ricerca Scientifica 1, Roma, Italy\label{aff183}
\and
INFN, Sezione di Roma 2, Via della Ricerca Scientifica 1, Roma, Italy\label{aff184}
\and
HE Space for European Space Agency (ESA), Camino bajo del Castillo, s/n, Urbanizacion Villafranca del Castillo, Villanueva de la Ca\~nada, 28692 Madrid, Spain\label{aff185}
\and
Center for Frontier Science, Chiba University, 1-33 Yayoi-cho, Inage-ku, Chiba 263-8522, Japan\label{aff186}
\and
Department of Physics, Graduate School of Science, Chiba University, 1-33 Yayoi-Cho, Inage-Ku, Chiba 263-8522, Japan\label{aff187}
\and
Department of Astrophysical Sciences, Peyton Hall, Princeton University, Princeton, NJ 08544, USA\label{aff188}
\and
Theoretical astrophysics, Department of Physics and Astronomy, Uppsala University, Box 515, 751 20 Uppsala, Sweden\label{aff189}
\and
Minnesota Institute for Astrophysics, University of Minnesota, 116 Church St SE, Minneapolis, MN 55455, USA\label{aff190}
\and
Mathematical Institute, University of Leiden, Einsteinweg 55, 2333 CA Leiden, The Netherlands\label{aff191}
\and
School of Physics \& Astronomy, University of Southampton, Highfield Campus, Southampton SO17 1BJ, UK\label{aff192}
\and
Institute of Astronomy, University of Cambridge, Madingley Road, Cambridge CB3 0HA, UK\label{aff193}
\and
Department of Physics and Astronomy, University of California, Davis, CA 95616, USA\label{aff194}
\and
Space physics and astronomy research unit, University of Oulu, Pentti Kaiteran katu 1, FI-90014 Oulu, Finland\label{aff195}
\and
Institut de Physique Th\'eorique, CEA, CNRS, Universit\'e Paris-Saclay 91191 Gif-sur-Yvette Cedex, France\label{aff196}
\and
Center for Computational Astrophysics, Flatiron Institute, 162 5th Avenue, 10010, New York, NY, USA\label{aff197}
\and
Department of Astronomy, University of Massachusetts, Amherst, MA 01003, USA\label{aff198}}


\abstract{
The first Euclid Quick Data Release, Q1, comprises 63.1\,deg$^2$ of the Euclid Deep Fields (EDFs) to nominal wide-survey depth. It encompasses visible and near-infrared space-based imaging and spectroscopic data, ground-based photometry in the $u$, $g$, $r$, $i$, and $z$ bands, as well as corresponding masks. Overall, Q1 contains about 30 million objects in three areas near the ecliptic poles around the EDF-North and EDF-South, as well as the EDF-Fornax field in the constellation of the same name. The purpose of this data release -- and its associated technical papers -- is twofold. First, it is meant to inform the community of the enormous potential of the \Euclid survey data, to describe what is contained in these data, and to help prepare expectations for the forthcoming first major data release DR1. Second, it
enables a wide range of initial scientific projects with wide-survey \Euclid data, ranging from the early Universe to the Solar System. The Q1 data were processed with early versions of the processing pipelines, which already demonstrate good performance, with numerous improvements in implementation compared to pre-launch development. In this paper, we describe the sky areas released in Q1, the observations, a top-level view of the data processing of \Euclid and associated external data, the Q1 photometric masks, and how to access the data. We also give an overview of initial scientific results obtained using the Q1 data set by Euclid Consortium scientists, and conclude with important caveats when using the data. As a complementary product, Q1 also contains observations of a star-forming area in Lynd's Dark Nebula 1641 in the Orion~A Cloud, observed for technical purposes during \Euclid's performance-verification phase. This is a unique target, of a type not commonly found in \Euclid's nominal sky survey. 
    }

   \titlerunning{\Euclid Quick Data Release (Q1): Overview}
   \authorrunning{Euclid Collaboration: H.~Aussel et al.}
   
   \maketitle
%
%
%
%


\section{\label{sc:Intro}Introduction}

\Euclid is a space mission of the \ac{ESA} with the primary goal of studying dark matter and dark energy using two main probes, weak gravitational lensing and galaxy clustering \citep{EuclidSkyOverview}. \Euclid uses a 1.2-m diameter Korsch telescope with a field of view of 0.54\,deg$^2$, imaged by two instruments, VIS \citep{EuclidSkyVIS} and the Near-Infrared Spectrometer and Photometer \citep[NISP;][]{EuclidSkyNISP}, with the mission of conducting the \ac{EWS}, covering 14\,000\,deg$^2$ of the extragalactic sky \citep{Scaramella-EP1}. VIS is a broad-band optical imager with a spatial resolution of \ang{;;0.18}, designed to measure the distortion of galaxy shapes with $\IE \lesssim 24.5$. NISP combines the capabilities of an imager in the \ac{NIR} bands \YE, \JE, and \HE \citep{Schirmer-EP18} to derive the photometric redshifts of the galaxies whose shapes are measured with VIS, together with a near-infrared slitless spectrograph to measure accurate redshifts of galaxies with bright emission lines.

The VIS single \IE band is too wide to allow for the determination of the photometric redshifts of the objects whose shapes are being measured. To this end, the \Euclid space data are combined with ground-based photometry in the \textit{u}, \textit{g}, \textit{r}, \textit{i}, and \textit{z} bands from large-area surveys. In the southern sky, the \acl{DES} \citep[\acs{DES};][]{Abbott:2021} is currently used until deeper data from the Vera C.\ Rubin Observatory \citep{Ivezic:2019} become available. In the northern sky, a new collaboration has been set up, the \acl{UNIONS} (\acs{UNIONS}; Gwyn et al., in prep.), with the aim to survey the sky in the {\it ugriz} bands. This is a joint effort between the \acl{CFIS} \citep[\acs{CFIS};][]{Ibata:2017} for the {\it u} and {\it r }bands, the \acl{Pan-STARRS} \citep[\acs{Pan-STARRS};][]{Chambers:2016} for the {\it i} band, and the \acl{HSC} \citep[\acs{HSC};][]{Miyazaki2018} for both the {\it g} band, through the Waterloo-Hawaii-IfA {\it g}-band Survey (WHIGS, PIs K.~C. Chambers and M.~J. Hudson), and the {\it z} band, through the Wide Imaging with Subaru-Hyper Suprime-Cam Euclid Sky survey (WISHES, PI M.~Oguri).
In the \Euclid project, we refer to these external data as `EXT'. They are ingested and recalibrated to a flux scale in common with the VIS and NISP data. Together, the space- and ground-based data form the \Euclid mission data set.

In addition to the main survey, a significant fraction of observational time is spent on specific fields that, thanks to repeated visits, will accumulate greater depth than the \ac{EWS}, up to a gain of about two magnitudes \citep{EuclidSkyOverview}. These are the \acp{EDF} and the \acp{EAF}, 
supporting our instrument calibration and  characterising the source population (Scaramella et al., in prep.). 

Nominal \ac{EWS} observations started on 14 February 2024. It will take \Euclid 6 years to collect all of its 14\,000\,deg$^2$ and associated \ac{EDS}. The project foresees three major data releases of the survey data, DR1 to DR3, with DR1 using the first year of data collected, DR2 the first three years, and DR3 occurring after all survey observations have ended. The internal and public releases of DR1 are scheduled for October 2025 and 2026, respectively. A special data release, Q1, aimed at giving a taste of the capacities of the \Euclid mission to the astronomical community was planned to take place 14 months after the start of the survey. The fields to be included in the release were to be the three \acp{EDF} and other areas of interests. The area of Q1 is not large enough to allow meaningful derivation of cosmological parameters, but it is large enough for a slew of non-cosmological studies, as testified by the more than 30 publications from consortium members based on this release.  

In \cref{sc:skycontent} we present the sky fields that comprise Q1, and \cref{sc:Obs} contains a summary of the associated \Euclid and ground-based EXT observations. Overviews of the \Euclid mission data processing are given in \cref{sc:Processing}, with details expanded in separate papers, and in \cref{sc:ext} for the EXT data. The survey masks are discussed in \cref{sec:vmpzid}, and data access is outlined in \cref{sc:dataaccess}. We conclude with a presentation of various scientific results enabled by the Q1 data release in \cref{sc:scienceandcaveat} and discuss a few important caveats to bear in mind while using the data in \cref{sc:caveats}. 

All \Euclid magnitudes are in the physical AB system \citep{oke1983}, the astrometric calibration is against \textit{Gaia} DR3 \citep{vallenari2023}, and the photometric calibration against the \ac{HST} CALSPEC database \citep[see e.g.,][]{bohlin2020}.

\section{\label{sc:skycontent} Q1 sky content}

\subsection{\label{sc:EDF}Euclid Deep Fields}

The 63.1\,deg$^2$ of Q1 comprise observations of the \ac{EDF-N}, \ac{EDF-S}, and \ac{EDF-F} -- see \cref{tab:EDFs} -- to the single-visit depth of the \ac{EWS}. They provide a preview of the typical depth expected across most of the \Euclid survey. By DR3 the \acp{EDF} will have been observed multiple times,  reaching 2\,mag deeper than the \ac{EWS} over an area of 53\,deg$^2$ \citep{EuclidSkyOverview}. The total number of visits to each \ac{EDF} is adjusted to the different levels of zodiacal background present at the location of each field to reach a uniform depth at DR3.
Each \ac{EDF} visit consists of a so-called `patch' of \acp{ROS}, the building block of the \ac{EWS} \citep{Scaramella-EP1}. The tiling of a patch places the \acp{ROS} side-by-side with some overlap. Each visit covers the full deep field counting toward 53\,deg$^2$. A margin is needed because of the varying position angle from the orbit progression and the tiling itself, resulting in the additional 10\,deg$^2$ for Q1.

The \ac{EWS} and \ac{EDS} differ concerning the NISP spectroscopic observations. While the \ac{EWS} is exclusively observed with the `red grism' (\RGE, 1206--1892\,\micron), the \acp{EDF} are also observed with the `blue grism' (\BGE, 926--1366\,\micron), maintaining an approximate blue-to-red exposure-time ratio of 5:3. This enables the construction of a pure and complete spectroscopic reference sample of galaxies. The \ac{EDF} observations selected for Q1 are all made with the red grism.

\begin{figure*}[htbp!]
\centering
\includegraphics[width=1.0\textwidth]
{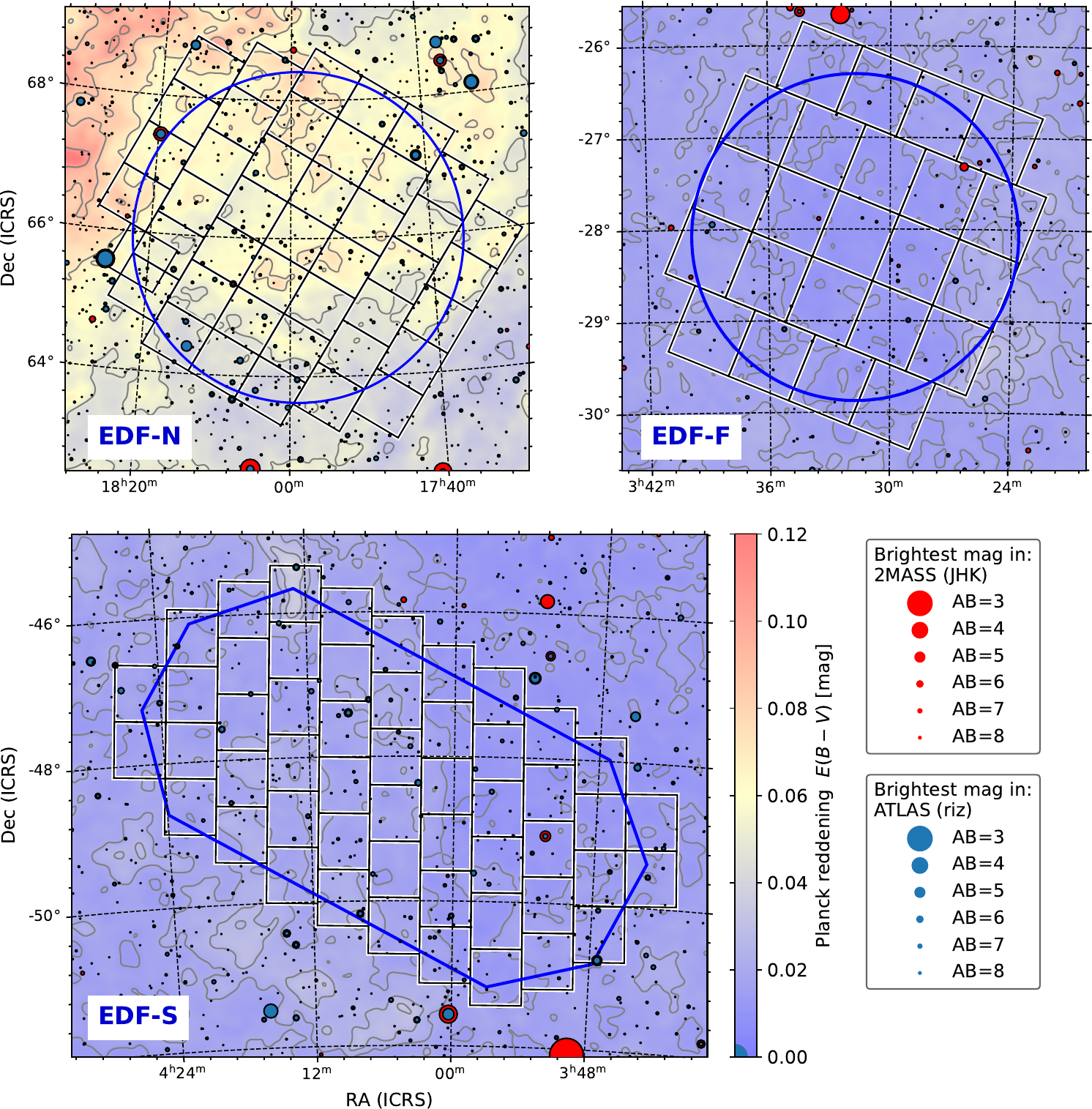}
\caption{Layout of the three \ac{EDF} tilings  comprising Q1 (black squares), overlaid on the reddening map from \cite{Planck2013dust}, with bright stars from 2MASS \citep{Skrutskie2006} and ATLAS-Refcat2 \citep{tonry2018}. The thick blue lines show the approximate areas that will be covered to full depth by DR3.}
\label{fig:3fieldsEBV}
\end{figure*}

The main \ac{EDF} properties are summarised in \cref{tab:EDFs}. \ac{EDF-N} is an ecliptic-polar field with a circular shape covering 20\,deg$^2$ to full depth by DR3. This field has the lowest zodiacal background, but it also has a lower Galactic latitude and thus a somewhat higher stellar density, extinction, and reddening. Because this field is always visible, it can be observed with a wide spread of position angles throughout the year. This strategy yields the diversity of spectral directions needed to build the complete and pure spectroscopic reference sample; the \ac{EDF-N} Q1 data correspond to one particular orientation. \ac{EDF-S} is at a lower ecliptic latitude with two relatively long visibility windows per year, meaning that the range of position angles is restricted compared to that of \ac{EDF-N}. 
It has a stadium shape, encompassing two tangent circles each covering 10\,deg$^2$, with a total coverage of 23\,deg$^2$. The \ac{EDF-F} covers a 10\,deg$^2$ circular region centred on the \ac{CDFS}, a location with a considerable amount of multi-wavelength data that is easily accessible from ground-based observatories; for \Euclid it has two short visibility windows per year. \Cref{fig:EDFs} shows the \ac{EDF} sky footprints of the Q1 visits. The footprints are reproduced in \cref{sc:footprint} as standard-format region files. \Cref{fig:3fieldsEBV} zooms in the three EDF areas showing information on sky quality and coverage by other surveys.

\begin{table*}[htbp!]
 	\caption{Summary of the properties of the \acp{EDF}. We list the Q1 area, and the smaller area in DR3 that reaches full target depth. Here $\lambda$ and $\beta$ refer to ecliptic longitude and latitude, respectively.} 
 	\centering           
 		\begin{tabular}{ l  c  c  c  c  c  c c } 
 	\hline\hline
 \noalign{\vskip 1pt}
	Field & Q1 area & DR3-depth area & RA & Dec & $\lambda$ & $\beta$ & DR3 visits\\
 	\hline 
 \noalign{\vskip 1pt}
 	EDF-N	& 22.9\,${\rm deg^2}$ & 20\,${\rm deg^2}$ & \ang{269.733;;}   & \ps\ang{66.018;;}   &	 \ang{258.690;;}  & \ps\ang{89.446;;} & 40\\
	EDF-S	& 28.1\,${\rm deg^2}$ & 23\,${\rm deg^2}$ & \pz\ang{61.241;;} & \ang{-48.423;;} & \pz\ang{36.493;;} & \ang{-66.599;;} & 45 \\
	EDF-F	& 12.1\,${\rm deg^2}$ & 10\,${\rm deg^2}$ & \pz\ang{52.932;;} & \ang{-28.088;;} & \pz\ang{40.772;;} & \ang{-45.397;;} & 52\\
 	\hline
 		\end{tabular}
 	\label{tab:EDFs}  
 \end{table*}

\subsection{\label{sc:darkcloud}\texorpdfstring{LDN\,1641}{LDN 1641} in the Orion A Cloud}
\begin{figure}[htbp!]
\centering
\includegraphics[width=1.0\hsize]{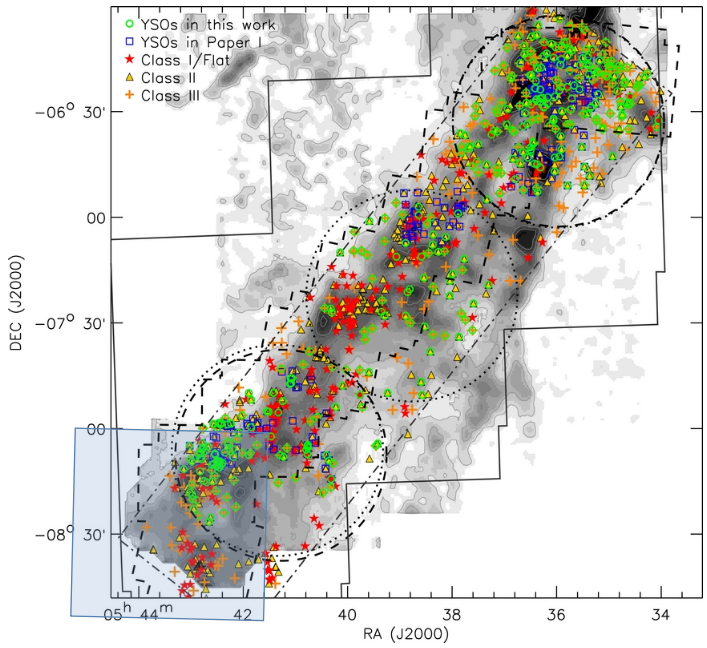}
    \caption{Map of \acp{YSO} in the Orion A Cloud \citep[adapted from][]{fang2013}. The markers show the various object classes, superimposed on a grey-scale \ce{^{13}CO} emission map. The large blue square in the bottom-left corner marks the \Euclid observations, comprising about 100 \acp{YSO}.}
\label{fig:ldn1641_ysomap}
\end{figure}

\Euclid's \ac{FGS} comprises four \acp{CCD} adjacent to the VIS detectors, observing in the same 530--920\,nm \IE passband as VIS itself. To test and further optimise the performance of the \ac{FGS} in September 2023, two months after launch, we needed to observe an area that had a particularly low number density of suitable guide stars. Because of thermal constraints, \Euclid can observe within a narrow meridian circle only \citep{Scaramella-EP1}. The only suitable area visible at that time, where all \ac{FGS} detectors would see very low guide-star densities, was a part of \ac{LDN}\,1641 \citep{Lynds1962}, an extended dust-obscured part of the Orion~A Cloud. We refer to this area  observed by \Euclid as the `dark cloud', centred near ${\rm RA}=\ra{5;43;0}$ and ${\rm Dec}=\ang{-8;22;0}$ (see \cref{fig:ldn1641_ysomap}). \Euclid observed a single field of view in the dark cloud area, i.e., approximately 0.5\,deg$^2$.

This star-forming area is known for its \aclp{YSO} \citep[\acsp{YSO}, e.g.][]{fang2009,fang2013,roquette2025}, predominantly becoming visible at wavelengths above 1\,\micron. There are targeted near-infrared \ac{HST} observations of protostars \citep[programme ID 11548,][]{megeath2007} that were used, for example, by \cite{habel2021}. The area was catalogued as part of the \acl{HOPS} \citep[\acs{HOPS},][]{furlan2016}, facilitating far-infrared studies \citep[e.g.][]{fischer2020}, and was mapped at sub-millimetre wavelengths by the \acl{ALMA} \citep{grant2021}. 

The dust extinction in the area ranges from 2 to at least 19\,mag in $R$-band \citep{Schlafly:2011}, with a typical value of about 7--8\,mag that reduces to about 2 to 1\,mag at wavelengths of 1 and 2\,\micron, respectively. Accordingly, the VIS data reveal comparatively little, whereas NISP shows an unprecedented, wide and contiguous view of the embedded objects in their larger environment.

\subsection{Target populations\label{sc:targetpops}}
With 63.1\,deg$^2$, the Q1 area coverage is likely to be comparable within a factor of a few to that of \ac{HST} since its launch\footnote{The actual \ac{HST} sky coverage is difficult to determine, due to overlapping observations and parallel fields. We did not attempt to make an accurate estimate.}. While the range of astrophysical targets in Q1 is naturally smaller than \ac{HST}'s, due to the restricted instrument suite onboard \Euclid and because of \Euclid's preferred sky areas, it is nonetheless considerable. Aside from tens of millions of galaxies the data comprise, for example, three extended planetary nebulae: PN K 1-16 \citep{kohoutek1963,montez2013} and the well-known Cat's Eye Nebula (NGC\,6543), both in the \ac{EDF-N}; and the little-studied Robin's Egg Nebula \citep[NGC\,1360; see also][]{goldman2004,garciadiaz2008} in \ac{EDF-F}. The Q1 data of NGC\,1360 are arguably the best images ever taken of it, resolving numerous cometary globules and showing the two bipolar jets in great detail. However, we note that our pipelines have an active background subtraction geared toward cosmological science. Therefore, extended nebulosity and low-surface brightness features such as tidal tails, galactic haloes, and intra-cluster light are either suppressed, removed, or over-corrected in the stacked Q1 images \citep[see also][]{Q1-TP002,Q1-TP003,Q1-TP004}. Notably, the \ac{EDF-N} contains the \Euclid self-calibration field, which will eventually become the deepest \Euclid field, with many hundreds of visits. The self-calibration field contains our recently published Einstein ring around the nucleus of NGC\,6505 \citep{oriordan2025}.


\section{\label{sc:Obs} Observations}
\subsection{Selection of the Q1 observations of the EDF areas}

The criteria for the selection of the single pass of each \ac{EDF} for Q1 were as follows. First, 
the pass had to be observed using the red grism to reproduce the observations of the \ac{EWS}. Second, 
the data set had to be as complete as possible, meaning no instrument failure during observations, and absence of high Solar activity for optimal data quality. Third, the pass had to be observed shortly after \Euclid's second decontamination campaign for maximum photometric stability. Lastly, the visit had to occur in time for the planned reprocessing campaign in October 2024.

For the \ac{EDF-F}, only two passes were available, one with the blue grism and one with the red (patch 59). The latter was obtained between 2024-08-05T20:56Z and 2024-08-07T00:55Z, and matched all our criteria. 
For the \ac{EDF-N}, out of six available passes with the red grism, one was lost to high Solar activity and another was acquired with heavy ice contamination on one of the mirrors \citep{Schirmer-EP29}. Among the remaining four, we selected the one with the least data loss to cosmic rays, patch 49, obtained between 2024-07-17T16:06Z and 2024-07-19T20:43Z.
For the \ac{EDF-S}, we used patch 71, observed between 2024-09-05TT13:52Z and 2024-09-08T05:34Z, which was preferred over two other passes because of low Solar activity. 

All selected passes were executed after \Euclid's second thermal decontamination on 8 June 2024 to remove water ice, restoring a largely stable throughput since then. The associated self-calibration observations to infer uniform relative photometric calibration at the level of a few millimag, and absolute calibration at the level of 1\%, were taken on 19 June 2024, 19 July 2024, 23 August 2024, and 29 September 2024. 

Further calibration products, such as biases, darks, lamp flats, nonlinearity, distortion, spectral traces, and wavelength calibration, are based on a large number of observations with varying cadences from \Euclid's routine calibration plan and \ac{PV} phase. They were selected to provide the best match for the Q1 data. It is beyond the scope of this paper to describe this process in detail. More information about the calibration is given in the technical papers complementing Q1 (see also \cref{sc:Processing}).

\begin{figure*}[htbp!]

\includegraphics[width=\textwidth]{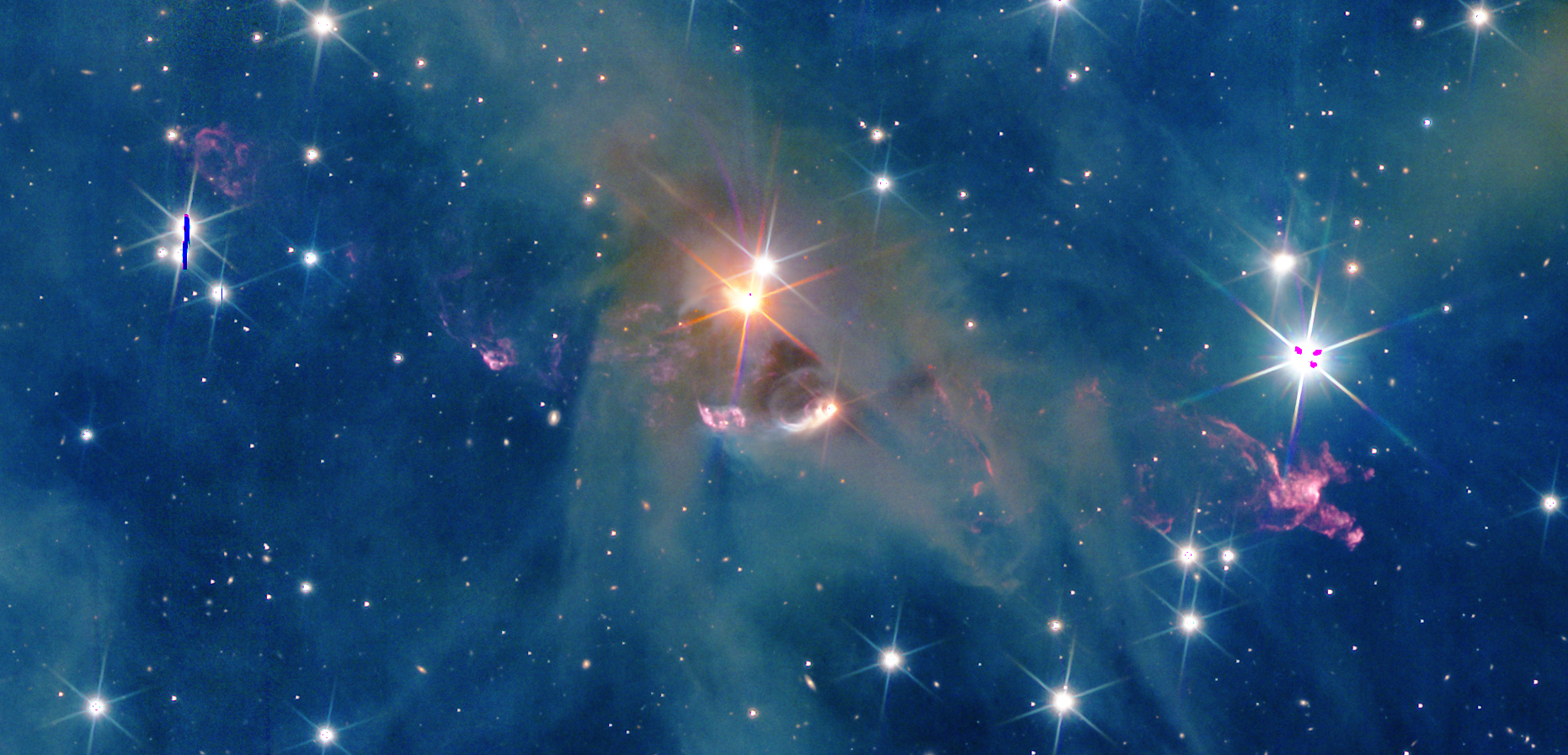}
\caption{Small cutout of the NISP dark-cloud image. Shown near the centre is HOPS\,221, a \ac{YSO} with prominent outflows that can be traced across the image and beyond. North is up, east is to the left, and the field is \ang{;7;} wide.}
\label{fig:darkcloud_jet}
\end{figure*}

\subsection{Dark-cloud observations}
The dark-cloud data included in Q1 were observed using the \ac{ROS} on September 24 and 27, 2023, using three different roll angles of ${\rm AA}=-5.5$, $-6.5$, and $-7.5$ \citep[see][for spacecraft angles]{EuclidSkyOverview}. The slitless spectra were not processed by the pipeline due to the extended, rich nebulosity, and are therefore not available for this particular field.

In total, 68 and 34 nominal and short-science exposures were taken with VIS, for a total of 41\,140\,s or 17 times the \ac{EWS} exposure time. However, due to the heavy dust extinction along this line of sight (\cref{sc:darkcloud}), the VIS data do not reveal much. 68 exposures were also taken in each of the \YE, \JE, and \HE bands, for a total of 5930\,s per band using the standard NISP imaging mode. The NISP data also comprise 17 times the exposure time of the \ac{EWS} and thus -- in the absence of shot-noise from foreground emission -- the $5\,\sigma$ point-source depth would be expected to be about 1.5\,mag deeper than that of the \ac{EWS}, or about 25.9\,mag in each band.
\Cref{fig:darkcloud_jet} shows a small cut-out of the NISP data centred on HOPS 221.

\subsection{Ground-based observations}

The \acp{EDF} are the target of dedicated deep ground-based observations. The \ac{EDF-N} is one of the targets of the Cosmic Dawn Survey \citep{EP-McPartland}, gathering MegaCam $u$ and \ac{HSC} $g$, $r$, $i$, $z$, and $y$ imaging together with Spitzer/IRAC 3.6\,\micron~and 4.5\,\micron\ data. The \ac{EDF-F} and \ac{EDF-S} sit in Vera C.\ Rubin Observatory's \ac{LSST} footprint  \citep{Ivezic:2019} and will accumulate deep data over the course of its mission. Moreover, \ac{EDF-F} is one of the Vera C.\ Rubin Observatory's deep-drilling fields\footnote{\url{https://www.lsst.org/scientists/survey-design/ddf}}. All these will be at least two magnitudes deeper than the \ac{EWS} EXT data, and will be ideal for fully exploiting the final \acp{EDF}
 data accumulated in the course of the mission. 

For Q1 our goal is to demonstrate the potential of the \ac{EWS}, and therefore we use observations from the EXT data set that overlap with the \acp{EDF} locations and are representative of the \ac{EWS} depth over the \Euclid \ac{ROI}. 
For \ac{EDF-N} we use data collected by the \ac{UNIONS} survey (Gwyn et al., in prep.), a collaboration of wide-field imaging surveys of the northern sky obtained using facilities in Hawai'i. They provide $u$ and $r$ band imaging using MegaCam \citep{MegaCam} at the 3.6-m Canada--France--Hawaii Telescope (CFHT) on Maunakea, $g$ and $z$ band imaging with \ac{HSC} \citep{Miyazaki2018} mounted on the Subaru 8.2-m telescope and $i$ band imaging from Pan-STARRS on Haleakala. All these data are delivered fully reduced, astrometrically calibrated and with an initial photometric calibration.
For \ac{EDF-F} and \ac{EDF-S}, we use data taken largely by the \ac{DES} \citep{Abbott:2021} but supplemented with additional DECam \citep{Flaugher2015AJ....150..150F} observations obtained by a variety of projects.  These raw exposures in the $g$, $r$, $i$, and $z$ bands are then detrended and astrometrically and photometrically calibrated by the \Euclid collaboration as described below in \cref{sc:ext}.  As part of that `Euclidisation' of these heterogeneous datasets, the \ac{UNIONS} and \ac{DES} data sets are photometrically calibrated as described below using the exquisite Gaia spectrophotometry.


\section{Data processing}
\label{sc:Processing}

\begin{figure*}
    \includegraphics[width=\textwidth]{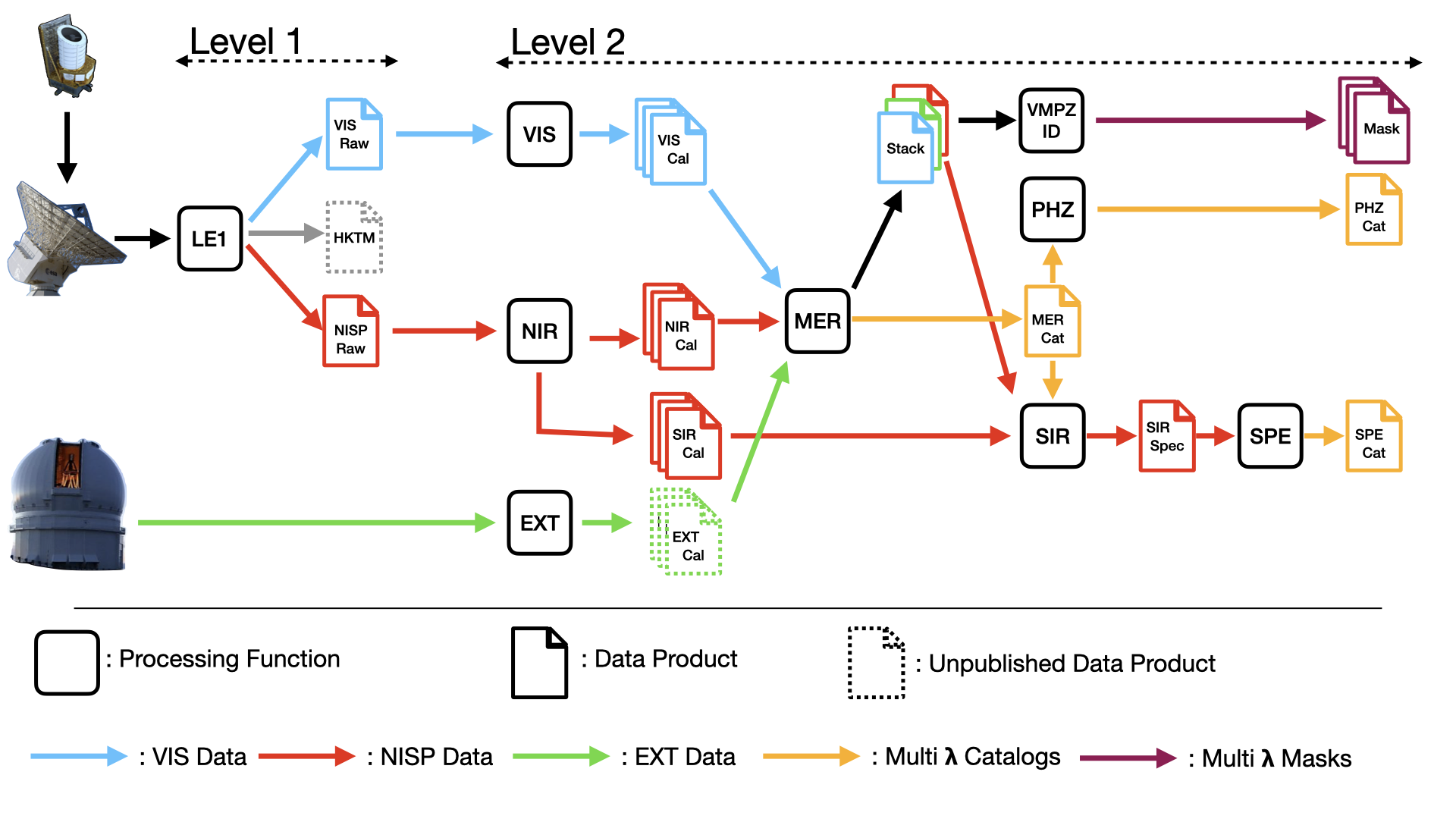}
    \caption{Overview of the Q1 data processing, showing the main data products making up the Q1 data release, together with the \ac{SGS} \acp{PF} creating them, and outlining the flow of the various data sets. `Raw' products refer to products created from  the decommuted telemetry without any further processing. `Cal' products refer to calibrated data, with instrument signatures being removed and data put on a physical scale. `Stack' refers to image combinations, `Spec' to combined 1D spectra, and `Cat' to catalogues.}\label{fig:dp}
\end{figure*}

The \Euclid mission data are processed by the \ac{SGS}, a distributed system across ten \acp{SDC} in Europe and the United States. The \ac{SGS} executes a pipeline chaining together different \acp{PF} and producing various products. The Q1 pipeline is outlined in \cref{fig:dp}, where the various data products and \acp{PF} are identified. In this section, we present a very top-level view of the \ac{SGS} pipeline only, and for details refer to the individual papers that describe the various \acp{PF}.

\subsection{Telemetry decommutation}
Decommutation is the process of extracting and reconstructing individual data streams from a time-multiplexed signal, typically in telemetry systems used for spacecraft communication. First, the telemetry from the {\it Euclid} satellite is received daily at \ac{ESA}'s \ac{MOC} located at the \ac{ESOC} in Darmstadt, Germany. It is then transferred to the \ac{SOC} at the \ac{ESAC} in Spain. Telemetry consists of  
VIS and NISP detector data, together with VIS, NISP, and spacecraft housekeeping data, as well as \ac{MOC} auxiliary products. It is transformed by the `LE1' \ac{PF} into the three \ac{LE1} products that are the inputs to the \ac{SGS} pipeline: the VIS 
raw-frame product, the NISP raw-frame product, and the \ac{HKTM} product. The Q1 processing of the VIS data uses only VIS raw frames, and the NISP processing of both images and spectra uses only the NISP raw-frame products.

These raw-frame products are distributed in Q1  and are described in detail in the \ac{DPDD}\footnote{\url{https://euclid.esac.esa.int/dr/q1/dpdd}}. They are the starting point of the \ac{SGS} data processing. Information from the \ac{HKTM} product is used for the precise reconstruction of the VIS \ac{PSF}, taking into account the satellite's attitude variation during an observation. This will be used in the \ac{DR1} for accurate weak-lensing measurements, which are not part of Q1. Hence the \ac{HKTM} product is not distributed in Q1. For completeness, we could not include the numerous calibration products with accurate descriptions of their validity ranges and the way they have to be applied to the data. This is well outside the scope and purpose of Q1. 

\subsection{Calibration and stacking\label{sec:LE2processing}}
The \ac{LE1} products are then distributed among the remaining nine \acp{SDC}, according to their position on the celestial sphere, in order to be fully processed into LE2 products. This separation is required in order to minimise the amount of data transfer between \acp{SDC} during the processing. 

The VIS data are processed by the VIS \ac{PF} \citep{Q1-TP002}, delivering single-frame calibrated images in the \IE band, together with associated catalogues. Moreover, VIS \ac{PF} produces stacked images of the six frames collected during a \ac{ROS}, that is four nominal science exposures of 560\,s integration time and two short ones with 90\,s; these stacks are not included in Q1. The NISP imaging data are processed by the \ac{NIR} \ac{PF}, delivering single calibrated frames in the \YE, \JE, and \HE bands, together with associated catalogues \citep{Q1-TP003}. The associated ground-based imaging data are processed by the \ac{EXT} \ac{PF} detailed in \cref{sc:ext}. 

The VIS, NIR, and EXT data available on a survey tile are then collected by the \ac{MER} \ac{PF} that proceeds to build combined stacks per band, and creates a master catalogue selected in both the \IE-band and in a combined \YE + \JE + \HE detection stack. \ac{MER} \ac{PF} provides a large number of photometric and morphology measurements for each source, presented in detail in \citet{Q1-TP004}. The S{\'e}rsic morphology measurements are further discussed in \citet{Q1-SP040}. The MER stacks contain, for each tile, \ac{RMS} and flag maps at the scale of 0\farcs1 for VIS and NIR \citep{Q1-TP003}. These are combined into coarser masks over the entire data release to allow for statistical quantities to be derived by the VMPZ-ID \ac{PF} (see \cref{sec:vmpzid}).

\subsection{Further catalogue and spectra extraction\label{sec:phz_spe_extraction}}

The MER photometric information is used by the \ac{PHZ} \ac{PF} to determine the type, photometric redshift if applicable, and physical properties of each source. This step is described in \citet{Q1-TP005}.

Spectra of all sources detected at $\HE \le 22.5$ are extracted from the NISP spectroscopic exposures by the SIR-\ac{PF} to provide calibrated spectra, as described in \citet{Q1-TP006}. These are used by the \ac{SPE} \ac{PF} to measure redshifts and lines fluxes \citep{Q1-TP007}.

\subsection{Complementary dark-cloud stacks\label{sc:complementary_darkcloud_stacks}}
The MER processing divides the sky into $\ang{;32;}\times\ang{;32;}$ wide tiles with a \ang{;2;} wide overlap. This also applies to the dark-cloud data. The default stacks produced by \ac{MER} \ac{PF} for the dark cloud, however, are not ideal for two reasons.

First, the $\ang{;46;}\times\ang{;51;}$ area covered  resulted in five separate tiles, some of which contain comparatively little data. While special tiles will be available in future data releases centred on dedicated objects, such as larger galaxies, they would still be restricted in size. Second, the background computed by the VIS and \ac{NIR} \acp{PF} is subtracted by \ac{MER} \ac{PF} and not restored as a separate data product in Q1. In case of the dark cloud, the true background variations have much higher frequency than the background smoothing length in the \acp{PF}, resulting in a highly uneven result that complicates photometric measurements of objects in their local environment.  

To mitigate this limitation of Q1, we provide complementary NISP stacks of the dark cloud. They were created from the released background-preserving LE2 frames with the {\tt THELI} software \citep[v3.2.0,][]{schirmer2013}, outside the context of the SGS. Using the software's {\tt NISP\_LE2@EUCLID} instrument configuration, we corrected -- as in \ac{MER} \ac{PF} -- for the individual detector-response offsets ({\tt PHRELDT} LE2 keyword), and used the astrometric solution to create single stacks per band that encompass the entire area. As in the MER PF, the stacks were created with {\tt SWarp} \citep{bertin2010}. Notable differences are that we chose to preserve the native NISP resolution of \ang{;;0.3}\,pixel$^{-1}$, and normalised the stacks to a flux of 1\,e$^-$\,s$^{-1}$. Also, due to the particularly strong undersampling of the NISP PSF in \YE band, the bilinear resampling kernel was used for \YE band, and the Lanczos2 kernel for \JE and \HE. The LE2 photometric zero points were propagated to the {\tt ZPAB} header keyword, and are independent of the dark-cloud observations \citep[see also][]{Q1-TP003}. 

At the time of Q1, curation of these complementary stacks has not been fully completed. We expect them to become available at an ESA server within a few weeks after publication of the Q1 data.

\section{External data processing}
\label{sc:ext}

\Euclid relies on ground-based optical imaging to complement VIS and NISP data for photometric-redshift estimation \citep{Abdalla08} and to derive galaxy and star \acp{SED}, ensuring the correct \ac{SED}-weighted \ac{PSF} assignment in the VIS lensing analysis \citep{Eriksen18}. A previous publication \citep{EuclidSkyOverview} contains an overview of the plans for the preparation and use of the external data in the \Euclid mission. Future publications prepared for \Euclid DR1 will provide detailed descriptions of each step of the processing and calibration.  Here we describe the specifics of the Q1 external data, which have been observed as part of \ac{UNIONS} and \ac{DES}, the calibration and processing applied to these data and the resulting data quality of the released data set.

Because of the heterogeneity of the external data set, we enforce a common data model that contains the information required for the calibration and processing.  This common data product -- termed a \ac{SEF} -- consists of a detrended and astrometrically and photometrically calibrated single CCD image, the associated 
position-dependent \ac{PSF} model and an associated catalogue that includes -- at a minimum -- the sky positions and PSF-fitting magnitudes of the brighter, unresolved sources.  In the case of \ac{UNIONS}, these \acp{SEF} are created using output data products from the external surveys. In particular, the ensemble of $i$ band \acp{SEF} from \ac{Pan-STARRS} is prepared using \ac{Pan-STARRS} collaboration specific software \citep{magnier2017.datasystem,waters2016}.  Similarly, the \ac{HSC} data in the $g$ and $z$ bands from  \acs{WHIGS} and \acs{WISHES} are produced using output data products from {\tt HSCpipe}  \citep{HSCpipe2018PASJ...70S...5B,LSST2019ASPC..523..521B}.  The $r$ band \acp{SEF} from \ac{CFIS} are created using the {\tt MegaPipe} software \citep{Gwyn2008PASP..120..212G}. 

\begin{figure*}[htbp!]
    \includegraphics[width=\textwidth]{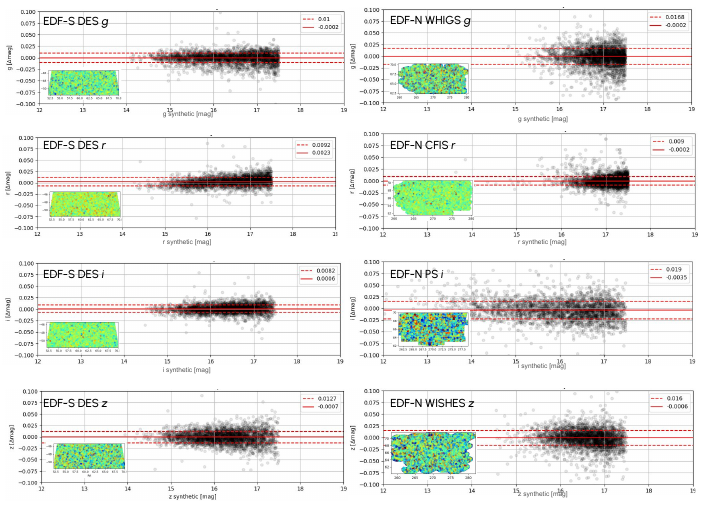}
    \caption{Photometric quality of the DES (left) and UNIONS (right) \acp{SEF} after calibration using \textit{Gaia} spectrophotometry and the appropriate survey bandpass, assumed to be independent of focal plane location for Q1.  Shown in each panel is the median offset (solid line) with respect to the \textit{Gaia} calibrators, as well as the \ac{NMAD} scatter (dashed line). }
    \label{fig:SEFphotometry}
\end{figure*}

\begin{figure*}[htbp!]
    \includegraphics[width=\textwidth]{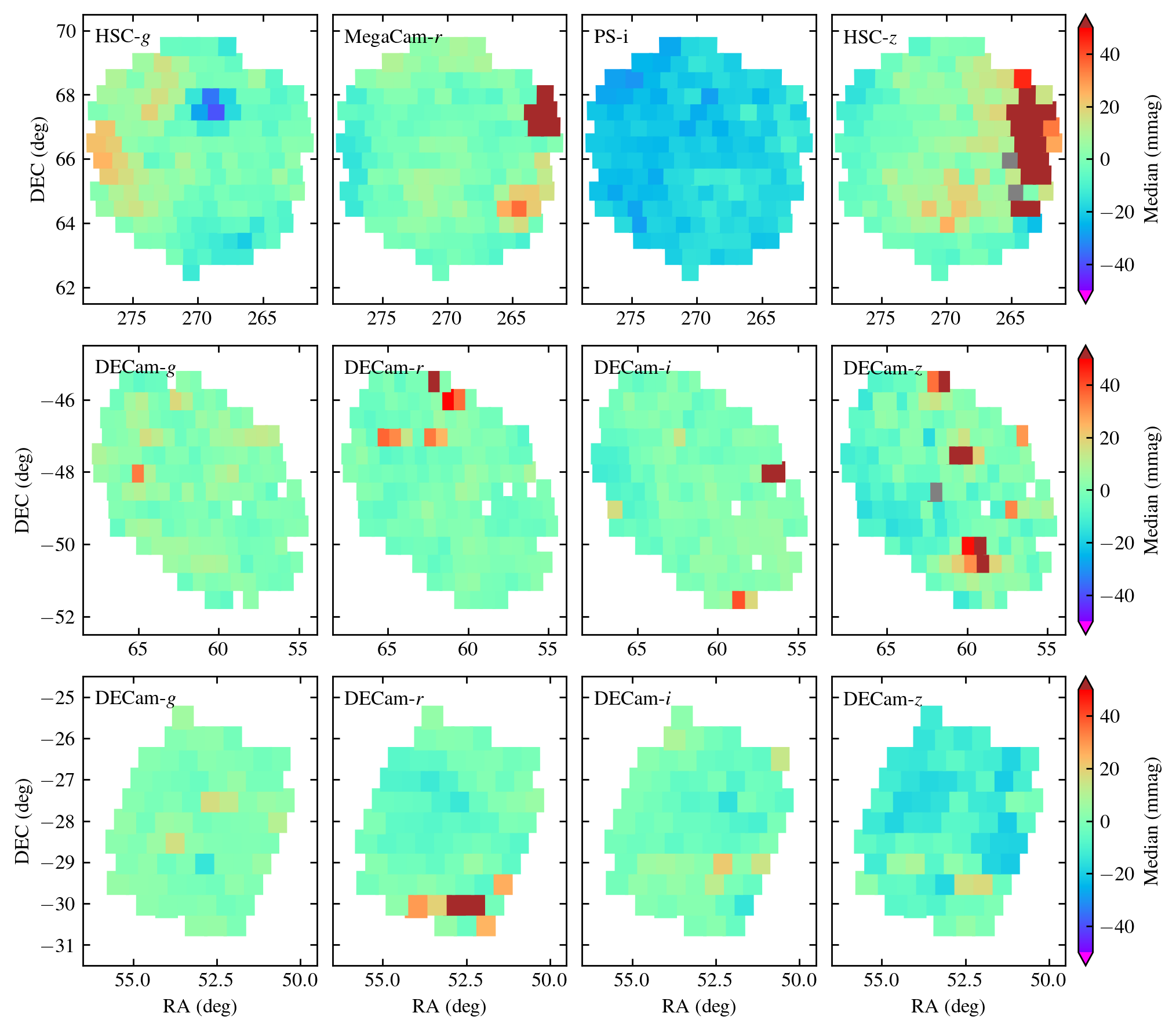}
    \caption{Per tile median magnitude residuals between the coadd \ac{PSF} photometry and \textit{Gaia} synthetic total magnitude for stars in the magnitude range 17--19. {\em Top to bottom:} \ac{EDF-N}, \ac{EDF-S}, and \ac{EDF-F}. {\em Left to right:} $g$, $r$, $i$, and $z$ bands. }
    \label{fig:ext_median_offset}
\end{figure*}

\begin{figure*}[htbp!]
    \includegraphics[width=\textwidth]{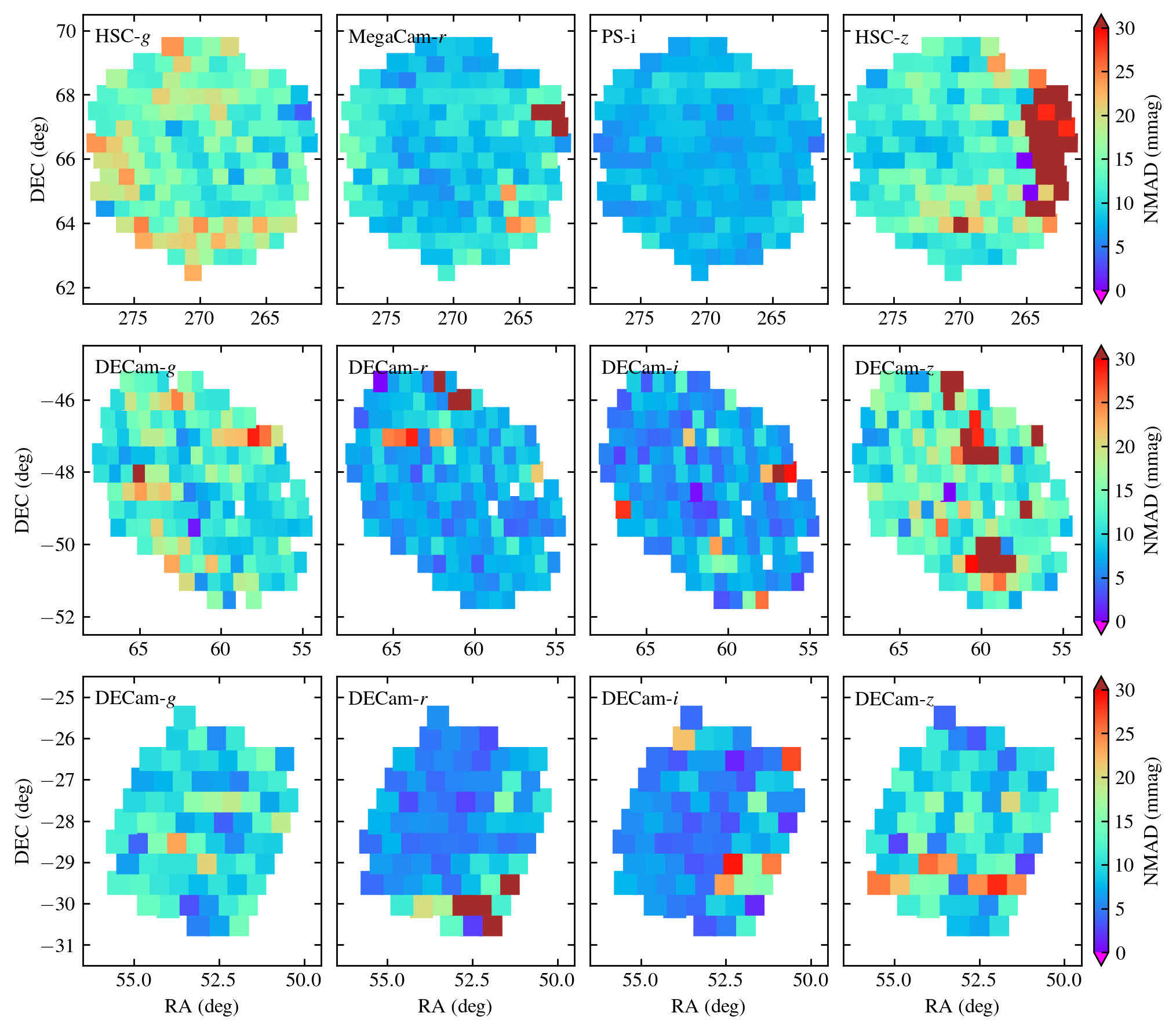}
    \caption{Intra-tile normalised \ac{NMAD} scatter of the \ac{PSF} fitting coadd photometry about the \textit{Gaia} synthetic magnitudes for stars in the magnitude range 17--19. {\em Top to bottom:} \ac{EDF-N}, \ac{EDF-S}, and \ac{EDF-F}. {\em Left to right:} $g$, $r$, $i$, and $z$ bands.}\label{fig:ext_nmad_offset}
\end{figure*}

\begin{figure*}[htbp!]
    \includegraphics[width=\textwidth]{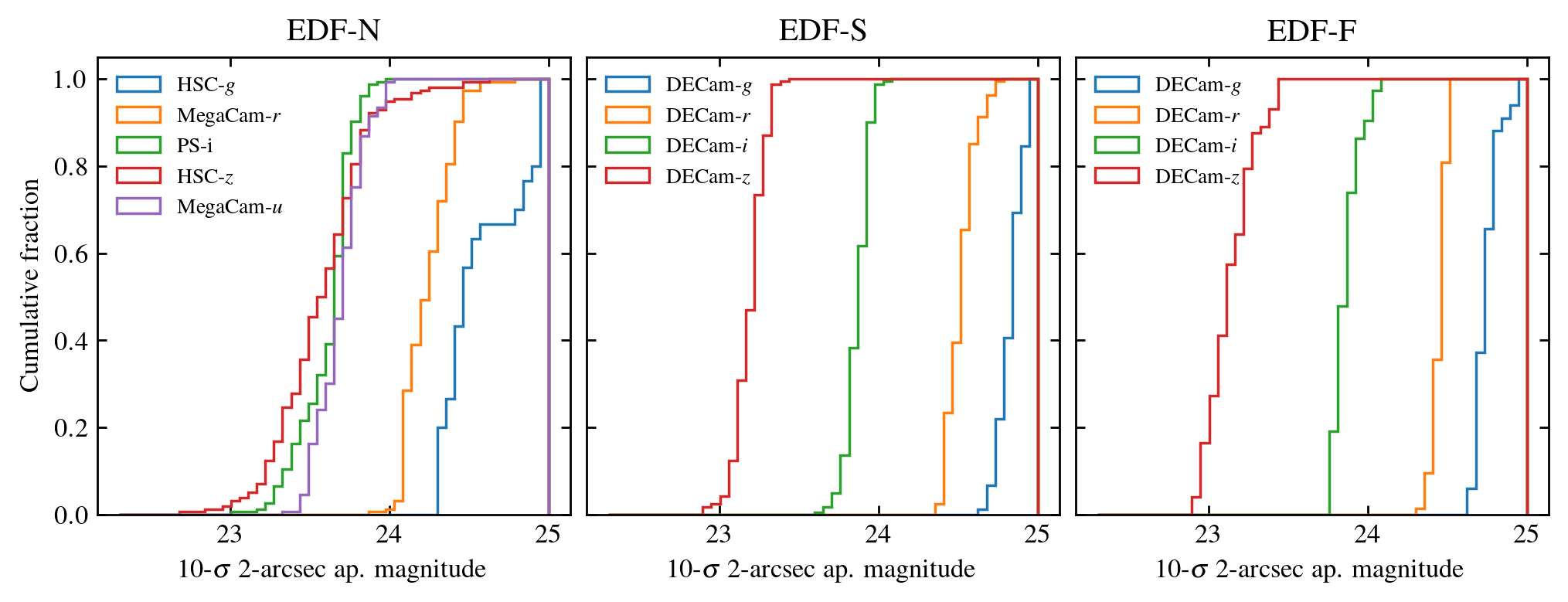}
    \caption{Cumulative fraction of the achieved depth per tile and band across the \acp{EDF}. Depth is defined as the 2\arcsec\ diameter aperture magnitude for which the signal-to-noise ratio is 10.}\label{fig:Q1_EXT_Depth}
\end{figure*}

\begin{figure*}[htbp!]
    \includegraphics[width=\textwidth]{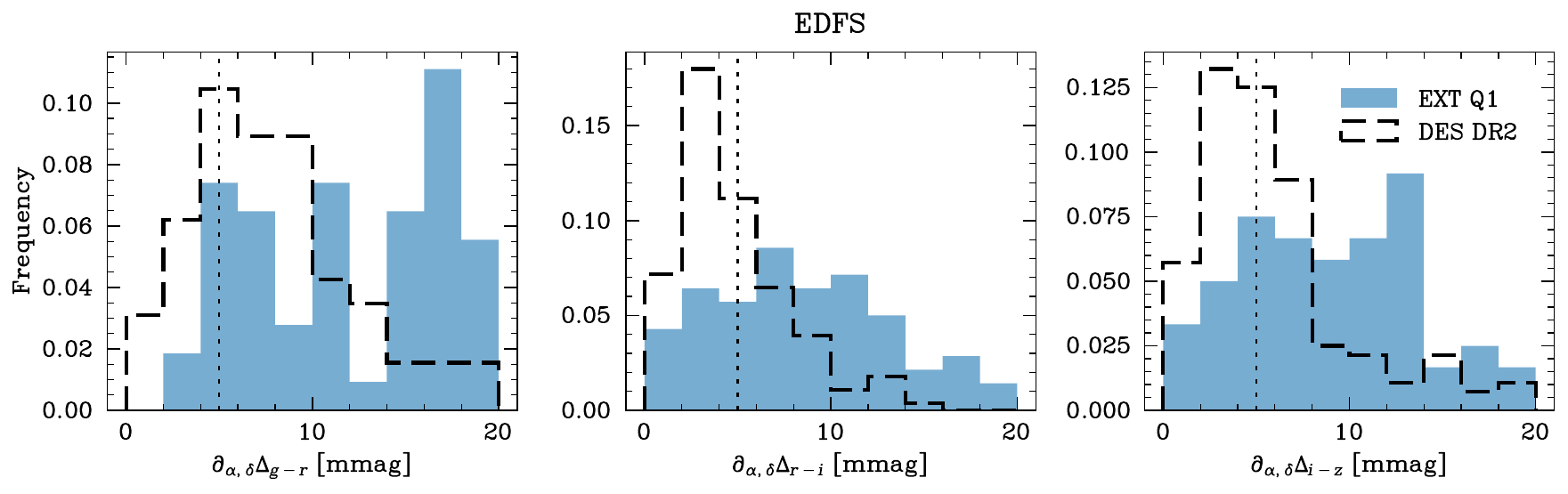}
    \includegraphics[width=\textwidth]{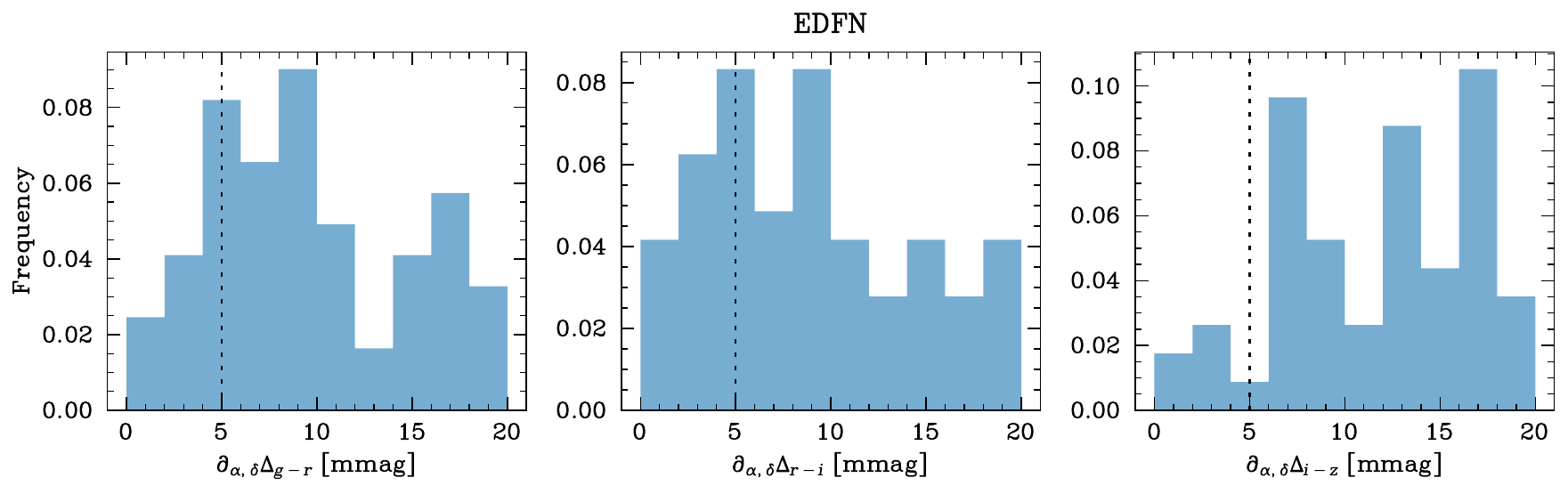}
    \caption{Colour offsets.  Each panel here shows the spatial variation ($\partial_{\alpha,
    \delta}$) on scales of \ang{0.3;;} of the best-fit colour offset
    ($\Delta_{\rm color}$) for two of the \acp{EDF}. This
    statistic was determined for $g-r$ (left), $r-i$ (middle), and $i-z$ (right),
    relative to the average stellar colour distribution in the \ac{EDF-S} area
    (top), and \ac{EDF-N} (bottom). In the case of \ac{EDF-S} we also show this
    result for DES DR2 data as the dashed dark histogram. The vertical black
    dotted line indicates the end-of-mission requirement.}
   \label{fig:Q1colorHomogeneityDECAMhistograms}
\end{figure*}

On the other hand, the \ac{DES} data \citep{DESDR12018ApJS..239...18A} and other \ac{DECam} \citep{Flaugher2015AJ....150..150F} observations are processed and calibrated within the Euclid Collaboration using extended versions of pipelines originally developed for \ac{DES} \citep{Mohr2008SPIE.7016E..0LM,Mohr2012SPIE.8451E..0DM,Desai2012ApJ...757...83D} and extended for the CosmoDM data management system \citep{Desai2015JInst..10C6014D}. These pipelines carry out detrending, astrometric calibration using \texttt{SCAMP} \citep{Bertin2006ASPC..351..112B}, position-dependent \ac{PSF} modelling using \texttt{PSFEx} \citep{Bertin2011ASPC..442..435B}, and an initial photometric calibration using a statistical method that relies on the \textit{Gaia} $G$, $BP$, and $RP$ photometry \citep{George2020ASPC..527..701G}.  In a final step, the \ac{DECam} images are masked using a tool developed for \Euclid \citep{Desai2016A&C....16...67D}, which employs \ac{PSF} homogenisation \citep{Darnell2009ASPC..411...18D} in the creation of surface-brightness templates for each \ac{SEF}. 

The \acp{SEF} produced through the methods described above are then subjected to a homogeneous photometric calibration using the \textit{Gaia} spectrophotometric data set and the corresponding survey passbands. For the Q1 calibration, these passbands are assumed to be constant across the different instrument focal planes. The \textit{Gaia} data set \citep{Gaia_2016A&A...595A...1G} offers a tremendous resource for producing stable, well-understood, ground-based photometry, because the \textit{Gaia} photometry and spectroscopy are stable across the sky with a systematic uncertainty of around 2\,mmag \citep{Gaia-DR3}. As previously reported, photometric calibration of the \Euclid \ac{SEF} collection using the statistical transformations from \textit{Gaia} $G$, $BP$, and $RP$ \citep{George2020ASPC..527..701G} to each of the external $griz$ bands demonstrated good consistency with the \ac{DES} calibration, and improved photometric stability and internal consistency in comparison to the \ac{UNIONS} calibration. 

For the Q1 calibration, \textit{Gaia} spectrophotometry is employed by using the survey passbands to create \textit{Gaia} synthetic photometry for each of the \Euclid external bands. Those calibrators are adopted for direct zero-point constraints and then combined with relative zero-point constraints from overlapping \acp{SEF} to solve for the \ac{SEF} zero points and zero-point uncertainties within collections of \acp{SEF} that lie within overlapping patches of sky.  The calibration tiles vary in size from 
1--5\,deg$^2$, depending on the density of \acp{SEF} in each of the \acp{EDF}.

\Cref{fig:SEFphotometry} shows -- for a randomly selected sample of stellar sources -- the scatter of the PSF-fitting stellar photometry versus the \textit{Gaia} synthetic magnitude calibrators for \ac{EDF-S} and \ac{EDF-N}.  The \ac{EDF-F} data-set performance is similar to that for \ac{EDF-S}.  Each panel corresponds to a different band $griz$, and in the inset the median offset and \acf{NMAD} scatter of the stellar PSF magnitudes are presented with respect to the calibrators.  The inset sky maps plot the positional variation of the offsets and no spatial trends can be seen.  In all cases, this photometric validation shows high quality and uniform \ac{SEF} photometry.  There are some differences from band to band in the \ac{NMAD} scatter, which is reflective of the photometric flatness of the detrended \acp{SEF} and the quality of the derived zero points, which are impacted to some degree by the assumption that a single passband describes each survey band, independent of the location of the \ac{SEF} within the focal plane of each instrument.
 
These homogeneously calibrated single-epoch images and their associated \ac{PSF} models are processed into coadded images and per-object coadd-\ac{PSF} models that are then used for photometry extraction alongside the VIS and NIR coadded images.  The coaddition pipeline is a version of the coadd pipeline originally developed for the \ac{DES} data management system \citep{Mohr2008SPIE.7016E..0LM,Mohr2012SPIE.8451E..0DM} and then later further developed within the CosmoDM data management system \citep{Desai2015JInst..10C6014D}.  The coadd pipeline uses a coadd tile definition to automatically select the relevant \acp{SEF} needed for coaddition, and then these images, their RMS noise maps and the metadata describing the zero point and \ac{WCS} of each \ac{SEF} are then coadded using calls to the {\tt SWarp} code \citep{bertin2002,bertin2010}.  Similarly, RMS and flag maps are produced for each coadd image.  This coadd pipeline has been applied to create high-quality, science ready, multi-band coadd imaging for a range of previous projects \citep[e.g.,][]{Desai2012ApJ...757...83D,Liu2015MNRAS.449.3370L,Zenteno2016MNRAS.462..830Z,Hennig2017MNRAS.467.4015H}.

The coadd-\ac{PSF} modelling code has been created and validated as part of the \Euclid development program.  It draws upon the position-dependent \ac{PSF} models extracted from every contributing \ac{SEF}, scaling these models to have the measured flux and position appropriate for each \ac{SEF}, and then assigning the appropriate RMS sky noise to each model. This collection of \ac{SEF}-based PSF models are then coadded in the same manner as the \ac{SEF} images themselves.  These PSF models are created individually for every detected object, and they provide 
high-quality models of the resulting unresolved sources in the coadded images.  PSF models for resolved objects are modelled under the assumption that the full flux of the source in each contributing \ac{SEF} is assigned to the PSF model.  A more complete description of this method along with validation tests will be presented in a future DR1 publication. 

These coadd and \ac{PSF} modelling codes have been integrated within the \Euclid \ac{SGS} software base for use in the external data processing.  
These data are validated using \texttt{SourceXtractor++} to extract object catalogues from the images while using the prepared \ac{PSF} models. The astrometry is cross-checked against \textit{Gaia}, and the \ac{PSF}-fitting magnitudes of unresolved (stellar) sources are compared to the synthetic magnitudes that have been extracted from the \textit{Gaia} spectrophotometry.

To illustrate the photometric stability of the $griz$ coadd photometry in \ac{EDF-N} (UNIONS: CFIS/MegaCam, HSC, PS), \ac{EDF-S}, and \ac{EDF-F} (DES/DECam), we show in \cref{fig:ext_median_offset} the median photometric offset between coadd PSF-fitting photometry and synthetic photometry computed from \textit{Gaia} XP spectra.
These figures show the relatively small tile-to-tile zero-point variations across most of each Q1 field, while also indicating that some tiles are outliers at the level of up to 0.04\,mag from the photometric system of the calibrated \acp{SEF} used to create the coadds.

\Cref{fig:ext_nmad_offset} shows the normalised absolute deviation of the PSF magnitudes extracted from the coadds around the median offset described above.  This measure serves as an indicator of the quality of the photometry of individual objects and can be compared to the \ac{NMAD} scatter reported for the input \acp{SEF} presented in \cref{fig:SEFphotometry}.  In general, one would expect the photometric scatter in the coadds to always decrease relative to that seen in the input \ac{SEF} population, but that is not the case for all tiles in the Q1 coadd data set.  In addition, one can see that particular tiles have poorer performance than the others, suggesting imperfect performance of the coaddition and coadd PSF-modelling process. These metrics from the Q1 processing of the external data allowed us to identify and resolve these minor issues, which are being corrected in preparation for the \Euclid DR1 processing.

\begin{table}[htbp!]
	\centering
	\caption{
    External data coadd and SEF data summary statistics.  Photometry offsets and scatter are given in mmag, and scatter is estimated using \ac{NMAD}.}
    \label{tab:ext_stats}
	\begin{tabular}{lcccc}
    \hline\hline
    \noalign{\vskip 2pt}
    Bandpasses & $g$ & $r$ & $i$ & $z$ \\
    \hline
    \noalign{\vskip 2pt}
    \multicolumn{5}{c}{EDF-N} \\
    \hline
    \noalign{\vskip 2pt}
    Median offset & $-1^{+9}_{-7}$ & $-1^{+8}_{-4}$ & $-19^{+4}_{-4}$ & $\phantom{1}3^{+14}_{-8}$ \\
    Coadd scatter & $14^{+6}_{-3}$ & $\phantom{-}9^{+3}_{-2}$ & $\phantom{-1}8^{+1}_{-1}$ & $12^{+9\phantom{0}}_{-3}$ \\
    SEF scatter & $17\phantom{^{+0}}$ & $\phantom{-}9\phantom{^{+0}}$ & $\phantom{-}19\phantom{^{+0}}$ & $16\phantom{^{+00}}$ \\
    Depth & $25.3^{+0.3}_{-0.3}$ & $24.3^{+0.2}_{-0.1}$ & $23.7^{+0.1}_{-0.2}$ & $23.6^{+0.2}_{-0.3}$ \\
    \noalign{\vskip 3pt}
    \hline
    \noalign{\vskip 2pt}
    \multicolumn{5}{c}{EDF-S} \\
    \hline
    \noalign{\vskip 2pt}    Median offset & $-1^{+6}_{-4}$ & $-1^{+3}_{-3}$ & $0^{+4}_{-4}$ & \phantom{1}$-1^{+7}_{-8}$ \\
    Coadd scatter & $12^{+7}_{-3}$ &$\phantom{-}7^{+2}_{-2}$ & $7^{+3}_{-2}$ & $\phantom{-}13^{+6}_{-4}$ \\
    SEF scatter & $10\phantom{^{+0}}$ & $\phantom{-}9\phantom{^{+0}}$ & $8\phantom{^{+0}}$ & $\phantom{-}13\phantom{^{+0}}$ \\
    Depth  & $23.9^{+1.2}_{-0.4}$ & $24.0^{+1.2}_{-0.4}$ & $23.9^{+1.3}_{-0.4}$ & $24.1^{+1.1}_{-0.5}$ \\
        \noalign{\vskip 3pt}
    \hline
    \noalign{\vskip 2pt}
    \multicolumn{5}{c}{EDF-F} \\
    \hline
    \noalign{\vskip 2pt}
    Median offset & $-1^{+4}_{-2}$ & $-3^{+5}_{-5}$ & $-2^{+5}_{-4}$ & $-10^{+7}_{-7}$ \\
    Coadd scatter & $10^{+4}_{-4}$ &$\phantom{-}6^{+5}_{-1}$ & $\phantom{-}6^{+4}_{-2}$ & $\phantom{-}10^{+5}_{-4}$ \\
    SEF scatter & $10\phantom{^{+0}}$ & $\phantom{-}9\phantom{^{+0}}$ & $\phantom{-}8\phantom{^{+0}}$ & $\phantom{-}13\phantom{^{+0}}$ \\
    Depth  & $24.1^{+1.1}_{-0.5}$ & $23.9^{+1.2}_{-0.4}$ & $23.9^{+0.6}_{-0.3}$ & $23.9^{+1.3}_{-0.4}$ \\
    \noalign{\vskip 3pt}
    \hline
	\end{tabular}
\end{table}

\Cref{tab:ext_stats} contains summary statistics of the coadd and SEF photometry within each Q1 field.  For \ac{EDF-N}, \ac{EDF-S}, and \ac{EDF-F}, we find typical median offsets that are consistent with zero, with the exception of the PS data set.  The \ac{NMAD} scatter of the coadd photometry about the \textit{Gaia} synthetic magnitudes is approximately 1\%, and the depths are consistent with the depth requirements set by the \Euclid mission.  The photometric scatter within the \acp{SEF} is smaller in \ac{DES} than in \ac{UNIONS}, and in most cases the photometric scatter in the coadds is smaller, as expected.
\Cref{fig:Q1_EXT_Depth} shows the depth distributions of each band within each of the Q1 fields.  These depths are summarised in \cref{tab:ext_stats}.

To assess the colour stability of the Q1 data set, we utilize the stellar locus (SL). We
compute a fiducial SL for \ac{EDF-N} and \ac{EDF-S} and assume that intrinsic variations
(for example, due to the spatial variations of the underlying stellar populations in the
Galaxy) on the SL are minimal in these areas. We run the SL analysis in
\ac{EDF-N} and \ac{EDF-S}. In the former case, we use the bands HSC $g$, MegaCam $r$,
PS $i$, and HSC $z$, and in the latter DECam $g$, $r$, $i$, and $z$. The fiducial SL is built
using high-confidence non-saturated stars with low photometric uncertainties.
The magnitudes and colours adopted are measured using \texttt{SourceXtractor++} on
the coadd images as part of \Euclid's internal validation pipeline. 
For the purpose of this colour validation test we must first remove the variations due to Galactic extinction.  Therefore, all
magnitudes are extinction corrected using \citet{Schlegel1998} dust maps, with \citet{Schlafly:2011} corrections and adopting the \citet{Cardelli:1989} extinction law. 

Given a list of excellent photometric quality objects, we build the SL model by
fitting a Gaussian mixture model, taking into account all the photometric
uncertainties. To avoid overfitting, this model was tuned to represent the
data set with the least amount of components. We then partition the sky into \ang{0.3;;} bins, each containing at least 200 stars. For each bin, we measure a colour
offset that maximizes the likelihood of its stars being drawn from the fiducial
SL model. The likelihood maximisation is done through a Nelder--Mead gradient
descent method. This effectively produces a map of best-fit colour offsets. 

To assess the stability of the colour in the Q1 EXT data we need to take the
spatial derivative of the colour offset maps described above. In
\cref{fig:Q1colorHomogeneityDECAMhistograms} we show this quantity for
\ac{EDF-S} and \ac{EDF-N}. In the case of \ac{EDF-S} we also compare it to what
is obtained using public \ac{DES} DR2 data \citep{Abbott:2021}. On all panels
the dotted vertical line marks the end-of-mission requirement of 5\,mmag.

We note that the colour stability for the Q1 data set is generally below the
end-of-mission requirements.  One can see in the upper panels of
\cref{fig:Q1colorHomogeneityDECAMhistograms} that the \ac{DES} DR2 colour
stability is better, but it also does not meet the \Euclid requirements. The Q1
data quality is affected by a combination of factors that mostly have affected
the integrated version of the PSF-modelling code and have since been identified
and corrected.  With these code fixes, the incidence of tiles with strongly
outlying photometry has been reduced, but has not been completely resolved.
Further work to improve the integrated versions of the coadd and \ac{PSF}
modelling software is ongoing in preparation for the \Euclid DR1 processing.

\section{Photometric masks}\label{sec:vmpzid}

\begin{figure*}
	\begin{centering}
	\includegraphics[
    scale=0.33]{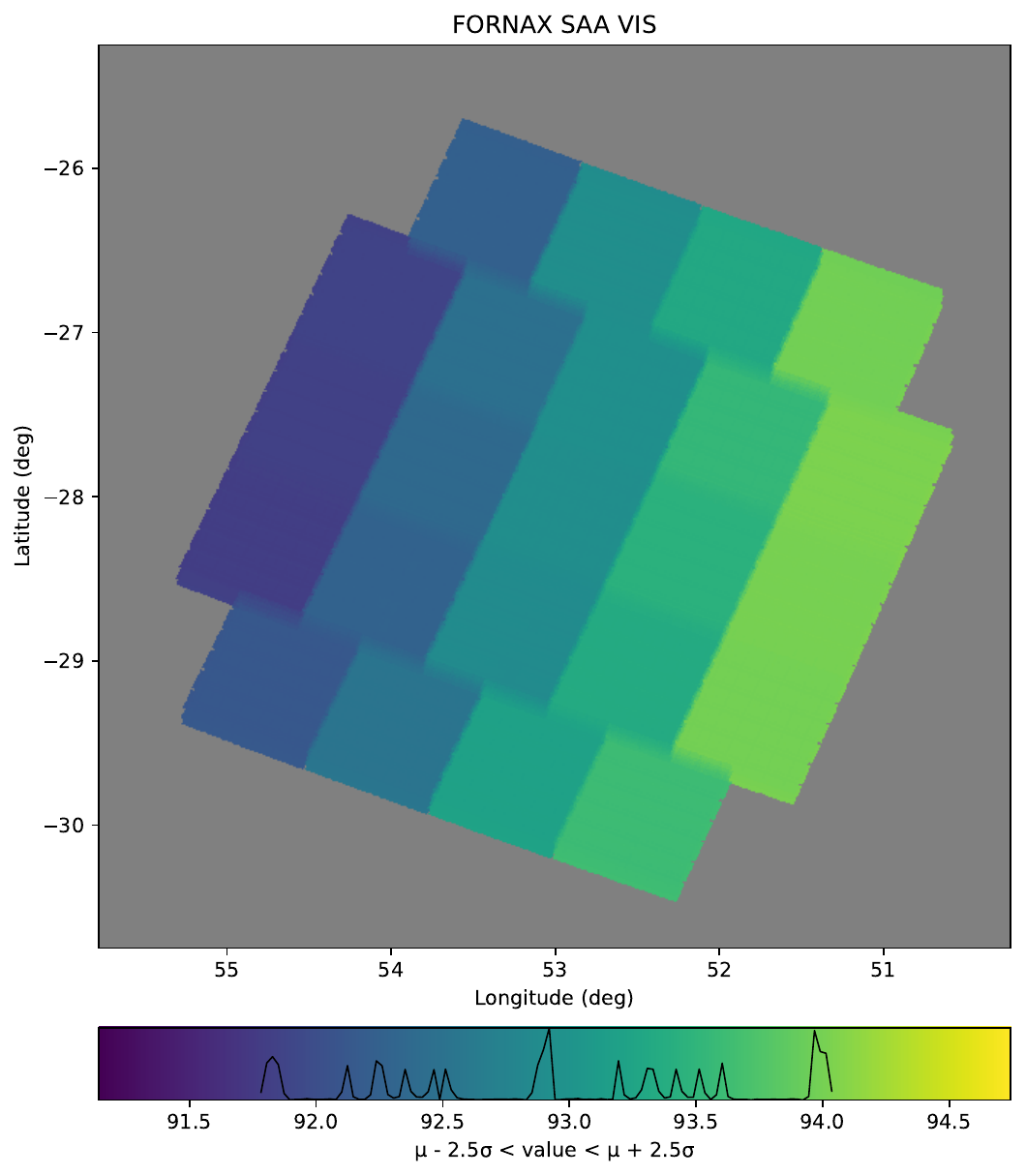} 
    \includegraphics[trim=0 0 0 20, clip, scale=0.33]{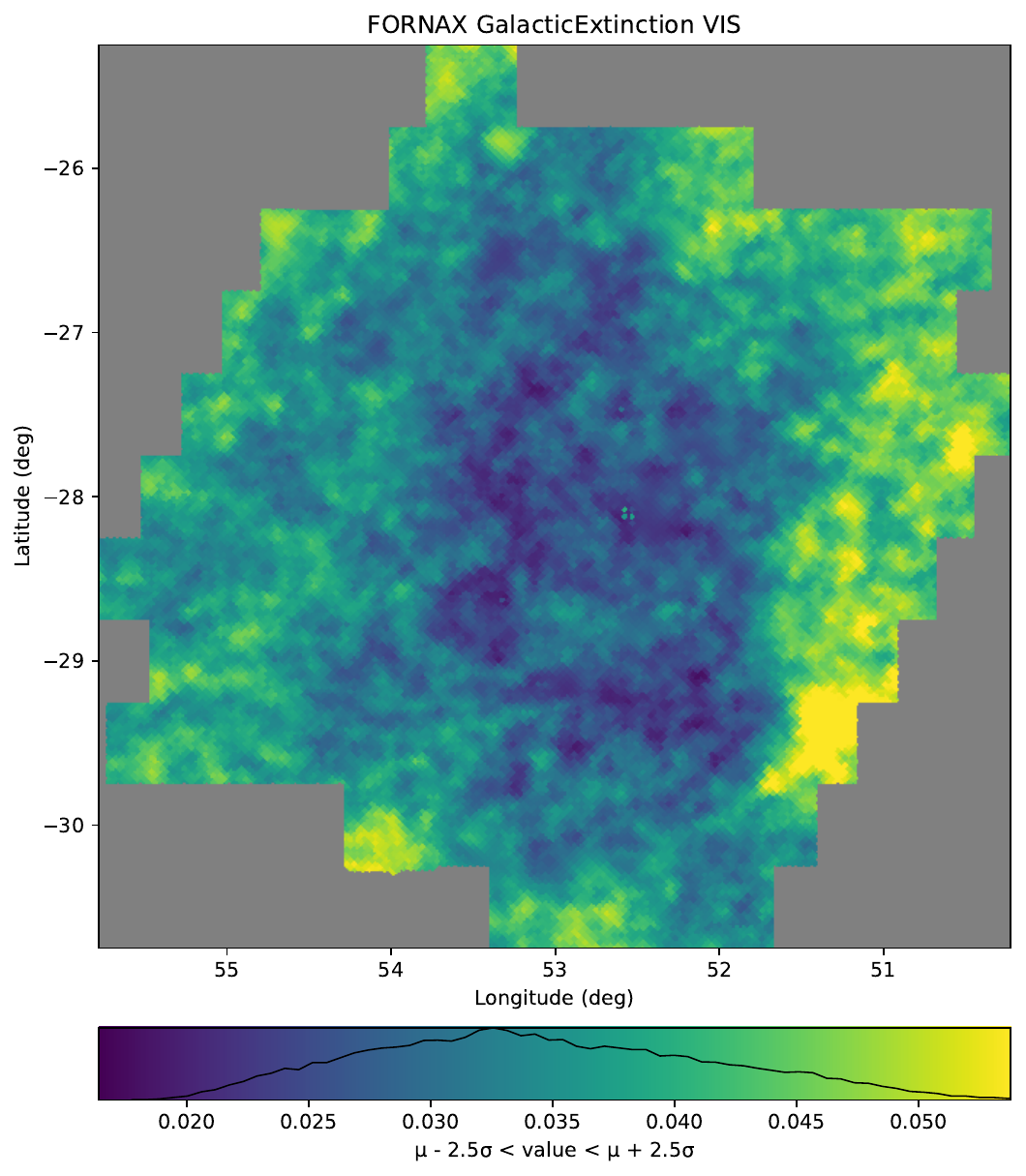} 
    \includegraphics[trim=0 0 0 20, clip,scale=0.33]{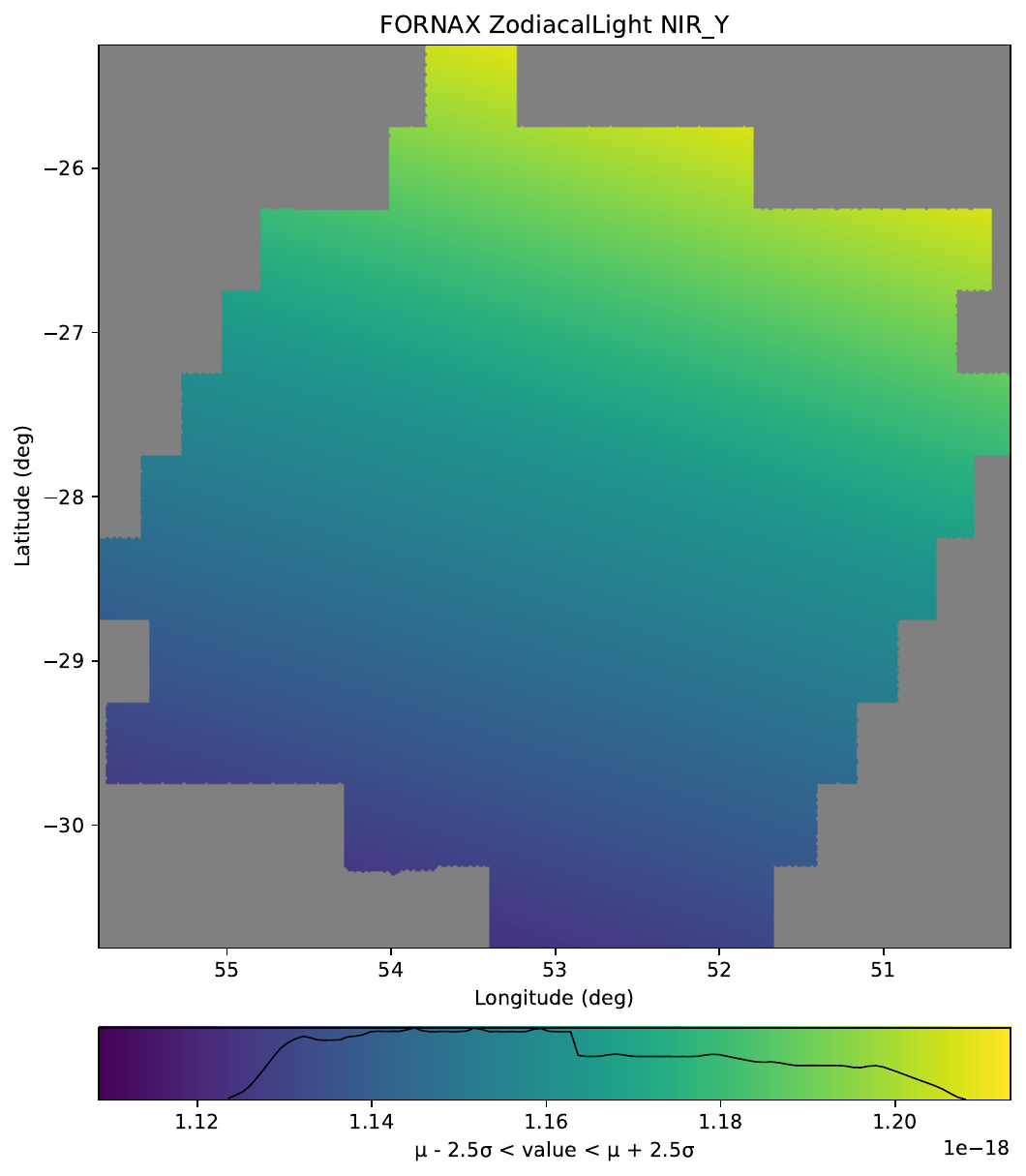} 
    \includegraphics[trim=0 0 0 20, clip,scale=0.33]{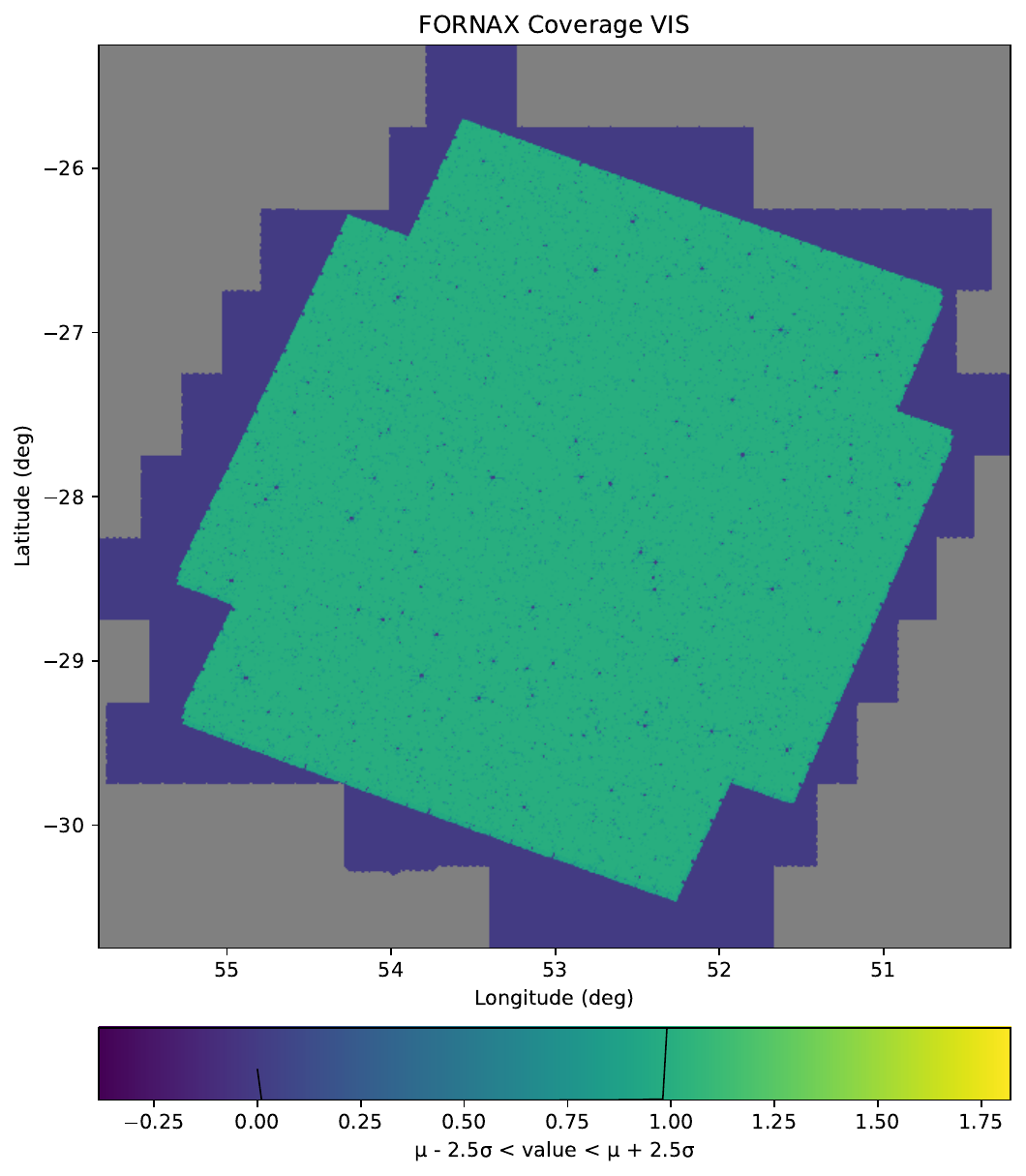} 
    \includegraphics[trim=0 0 0 20, clip,scale=0.33]{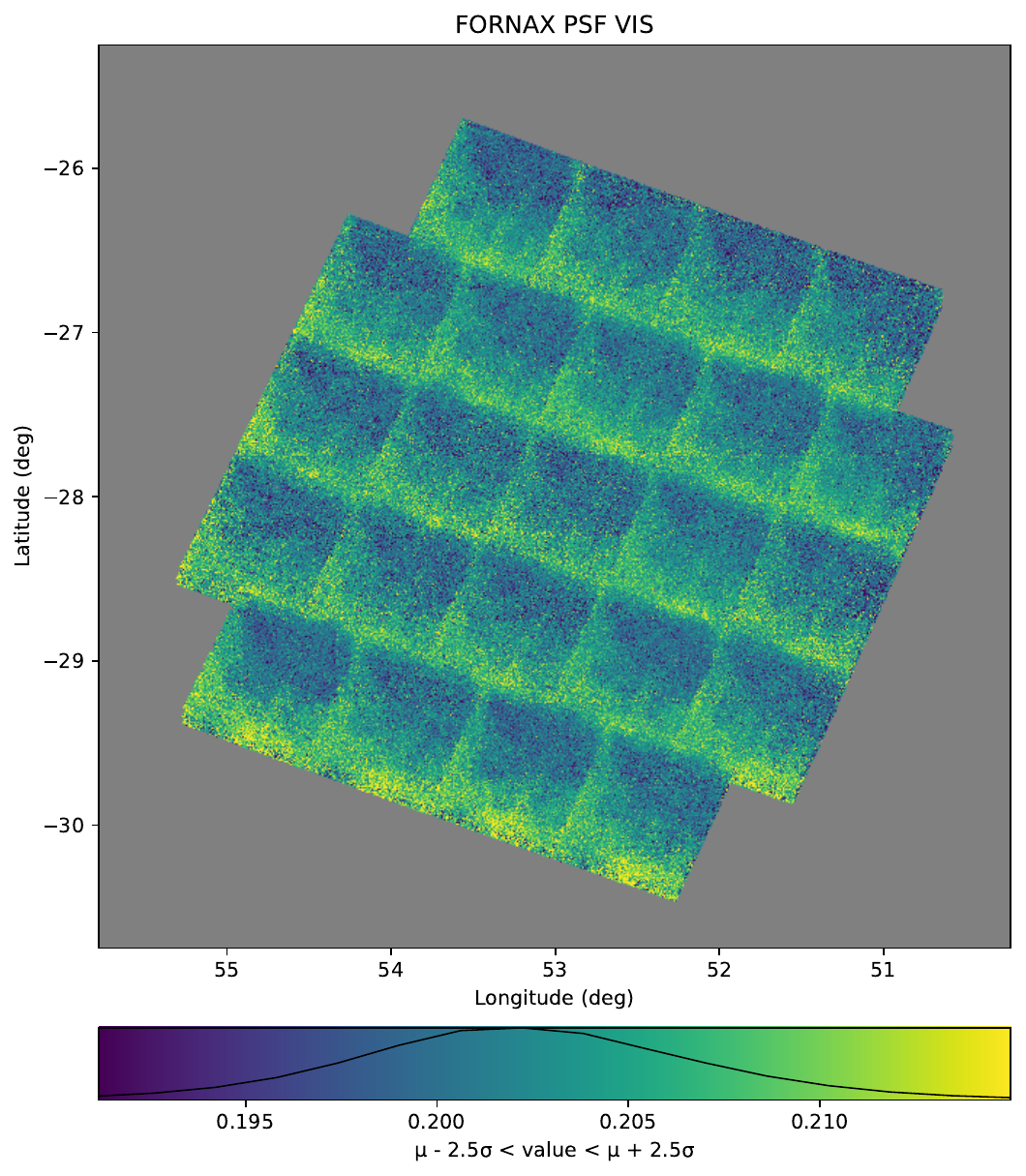} 
    \includegraphics[trim=0 0 0 20, clip,scale=0.33]{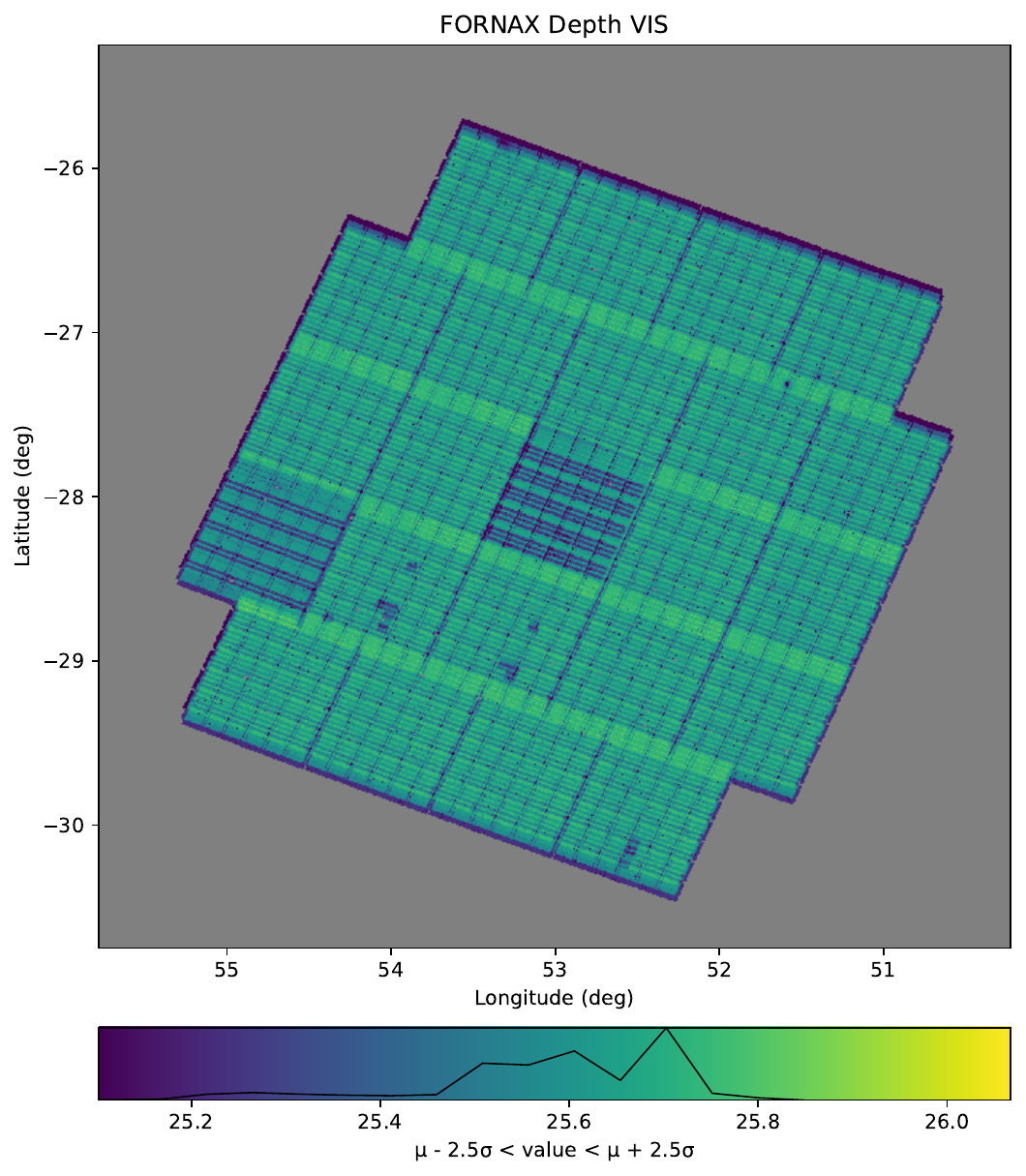} 
	\caption{Examples of photometric masks for the \ac{EDF-F} field. \textit{Top}: \ac{SAA} and VIS Galactic extinction and zodiacal light masks. \textit{Bottom}: VIS coverage, \ac{PSF}, and depth masks. Histograms of the represented signals are shown within the colour bars.}
	\label{fig:VMPZmasks}
	\end{centering}
\end{figure*}

A set of masks allowing for the assessment of the data quality over each of the Q1 \acp{EDF} were produced by the VMPZ-ID \ac{PF}. While this \ac{PF} is formally a \ac{LE3} \ac{PF} in the \ac{SGS}, it is at the interface between LE2, where data are at the pixel level, and LE3 that consider the whole \Euclid mission data set on the celestial sphere. The information from the VIS, NIR, EXT, and MER masks at the exposure and tile level is combined into a series of masks in the  \texttt{HEALPix}\footnote{Hierarchical Equal Area isoLatitude Pixelation of a sphere, see\\
\url{https://healpix.jpl.nasa.gov/}} \citep{2005ApJ...622..759G} format using a $N_{\mathrm{side}}=16\,384$, which corresponds to a resolution of approximately \ang{;;12.88}. VMPZ-ID also produce so--called {\tt DpdHealpixInfoMapVMPZ} that enable us to track satellite and instrumental properties and project them on the sky, allowing for easy cross-check between spatial information (for example the depth at a given point of the survey) with temporal information (for example the value of a temperature sensor on the \Euclid focal plane) and look for possible effects affecting the performance of the \Euclid mission. The masks distributed with the Q1 data release are listed in \cref{tab:photomasks}.

\begin{table*}
 	\caption{Summary of the main characteristics of the photometric maps. For each mask we give the mask family (indicating if it is related to the survey properties, to the satellite status, to the instrument performance, or to the sky emission), the mask type as defined in the text, the InfoMap type (only for InfoMap masks) and the numerical type. We also indicate if masks per band are produced.} 
    \newlength{\tabcs}
\setlength{\tabcs}{0mm}
 	\centering           
 	\scalebox{0.98}{
 		\begin{tabular}{@{\hskip 0mm}c@{\hskip -2mm}c@{\hskip 2mm}c@{\hskip -2mm}c@{\hskip\tabcs}c@{\hskip 2mm}c@{\hskip -2mm}c@{\hskip 0mm}c} 
 	\hline\hline
 \noalign{\vskip 1pt} 
    Mask  & Mask family & Mask product type & InfoMap type& Numerical&  Per-band& Comments \\
    & & &type&type&mask&\\
    \hline
 \noalign{\vskip 1pt}
    Tile index & Survey  &\texttt{DpdHealpixInfoMapVMPZ} & Tile & integer & no &   \\
    Solar aspect angle & Survey/Satellite  &\texttt{DpdHealpixInfoMapVMPZ} & SAA & float32 & no &   \\
    Alpha angle & Survey/Satellite  &\texttt{DpdHealpixInfoMapVMPZ} & AA & float32 & no &   \\
    Beta angle & Survey/Satellite  &\texttt{DpdHealpixInfoMapVMPZ} & BA & float32 & no &   \\
    \hline
 \noalign{\vskip 1pt}
    Exposure time & Survey  &\texttt{DpdHealpixInfoMapVMPZ} & Exposures & float32 & yes  &  \Euclid only \\
    Footprint & Survey/Instrument  &\texttt{DpdHealpixFootprintMaskVMPZ} & \dots & float32 & no &  all bands combined \\
    Coverage  & Survey/Instrument  &\texttt{DpdHealpixCoverageMaskVMPZ} & \dots & float32 & yes &   \\
    \hline 
 \noalign{\vskip 1pt} 
    Bit mask  & Instrument         &\texttt{DpdHealpixBitMaskVMPZ} & \dots & uint32 & yes &   \\
    Depth      & Instrument         &\texttt{DpdHealpixDepthMapVMPZ} & \dots & float32 & yes &   \\
    RMS Noise  & Instrument         &\texttt{DpdHealpixInfoMapVMPZ}  & Noise & float32 & yes &   \\
    Point-spread function        & Instrument  &\texttt{DpdHealpixInfoMapVMPZ}  & PSF & float32 & yes &   \\
    \hline
 \noalign{\vskip 1pt}
    Zodiacal light  & Sky emission         &\texttt{DpdHealpixInfoMapVMPZ}  & ZodiacalLight & float32 & yes & \Euclid only  \\
    Galactic extinction   & Instrument         &\texttt{DpdHealpixInfoMapVMPZ}  & GalacticExtinction & float32 & yes & \Euclid only \\
 	\hline
 		\end{tabular}
 		}
 	\label{tab:photomasks}  
 \end{table*}

In terms of \Euclid data products the photometric masks are constructed per MER data products\footnote{The description of the \Euclid MER products can be found in the \ac{DPDD} at 
\url{http://https://euclid.esac.esa.int/dr/q1/dpdd/}\,.} (Euclid Collaboration: Kang et al., in prep.) and are of five main types. \\
-- \texttt{DpdHealpixFootprintMaskVMPZ} mask has value 1 if the sky region has been observed and 0 otherwise in any band (\Euclid and EXT). \\
--  \texttt{DpdHealpixCoverageVMPZ} has values between 0 and 1, representing the coverage for the pixel each of the \Euclid and EXT bands. \\
- \texttt{DpdHealpixBitMaskVMPZ} is an unsigned integer mask with its bits associated with the MER flags. \\
-- \texttt{DpdHealpixDepthMapVMPZ} is a float map representing the depth as computed from the \ac{RMS} for each \Euclid and EXT band. \\
-- \texttt{DpdHealpixInfoMapVMPZ} is a generic mask to represent different quantities of interest.

In practice, 108 photometric masks per MER tile have been produced for the Q1 data release. In \cref{tab:photomasks} we summarise the main properties of those masks. A more detailed description of each of the masks is given below.
Examples of those masks for the EDF-F region are shown in \cref{fig:VMPZmasks}.

\subsection{Survey-definition masks}

-- {\bf Tile number}: This mask, of type \texttt{DpdHealpixInfoMapVMPZ}, is constructed so that for each \texttt{HEALPix} pixel the number of the MER tile  
on which it lies is stored in integer form. \\
-- {\bf Footprint}: A footprint mask defining the observed or valid sky area is stored in the form of a \texttt{DpdHealpixFootprintMaskVMPZ}. It is constructed per \Euclid and EXT bands. Starting from the \texttt{MerBksMosaic} maps we set the sky regions observed in all bands to one and the others to zero. MER tile pixels for which the \ac{RMS} is zero or larger than a given threshold are considered as not observed. All MER tile pixels within a \texttt{HEALPix} pixel are combined using a \texttt{logical and} operator.  \\
-- {\bf Coverage}: The coverage per band (\Euclid and EXT) is calculated from the MER background-subtracted mosaic (hereafter \texttt{MerBksMosaic} maps. The footprint MER tile pixels, for which the \ac{RMS} value is zero or larger than a given threshold, are considered as not observed. Furthermore, for the \Euclid bands polygon masks are used to exclude sky regions affected by bright star emission. For each \texttt{HEALPix} pixel, the coverage is given by the fraction of valid MER tile pixels.

\subsection{Satellite related masks}

-- {\bf Satellite orientation angles}: for each \texttt{HEALPix} pixel the average \ac{SAA}, \ac{AA}, and \ac{BA} in degrees \citep[see][]{Scaramella-EP1} are calculated from the MER Layering files. They are stored in the form of a \texttt{DpdHealpixInfoMapVMPZ}. \\
-- {\bf Exposure Time}: for each of the \Euclid bands the average exposure time per \texttt{HEALPix} pixel is calculated and stored as a \texttt{DpdHealpixInfoMapVMPZ}.
\subsection{Instrumental properties masks}
For each of the \Euclid and EXT bands, we obtain masks of the main instrumental properties from the \texttt{MerBksMosaic} maps. For each \texttt{HEALPix} pixel, the weighted average of the instrumental properties is computed. We assign zero weight to MER tile pixels with \ac{RMS} equal to zero or above a given threshold. \\
-- {\bf \ac{RMS} noise}: \texttt{DpdHealpixInfoMapVMPZ} that monitors the average \ac{RMS} noise in the \texttt{MerBksMosaic} maps. \\
-- {\bf Depth}:  \texttt{DpdHealpixDepthMapVMPZ} that monitors the depth obtained from the \ac{RMS} moise:\\
\begin{equation}
    m=-2.5\log10\left(\mathrm{RMS}\cdot d_{\mathrm{th}} \cdot\sqrt{\frac{A_{\mathrm{aper}}}{A_{\mathrm{pixel}}}}\,\right)+\textrm{ZP},
\end{equation}
with ZP the zero point as found in the \texttt{MerBksMosaic} maps. The detection threshold, $d_{\mathrm{th}}$, is fixed to 5 corresponding to a 5$\,\sigma$ detection. $A_{\mathrm{aper}}$ and and $A_{\mathrm{pixel}}$ represent the detection aperture and pixel areas, respectively. For the Q1 data release we have chosen a 2$\arcsec$ diameter aperture to compute the depth. \\
-- {\bf PSF}: \texttt{DpdHealpixInfoMapVMPZ} that monitors the FWHM of the \ac{PSF}. Because the \ac{FWHM} can be measured only on bright stars, 2D interpolation is performed across the tile area. \\
-- {\bf Bit Mask:} \texttt{DpdHealpixBitMaskVMPZ} obtained from a bit-by-bit \texttt{bitwise or} of the Bit Mask of the \texttt{MerBksMosaic} maps. 

\subsection{Sky emission masks}
Following \cite{Scaramella-EP1} we construct per-tile masks of the zodiacal light emission and Galactic extinction at the \Euclid bands. These masks are stored as \texttt{DpdHealpixInfoMapVMPZ}.

\section{Data access}
\label{sc:dataaccess}

The public \Euclid science archive at the ESAC Science Data Centre (ESDC) opened on 19  March 2025 offering the curated Q1 data products at the following address: \url{https://eas.esac.esa.int/sas/}. Information about the Q1 data release is presented at \url{https://www.cosmos.esa.int/en/web/euclid/euclid-q1-data-release}.

The MER mosaics in the four \Euclid bands and four EXT bands, as well as VIS/NIR calibrated frames (and their auxiliary data sets) can be searched and downloaded from the web interface. An image cutout service on the mosaics is also offered. All MER, PHZ, and SPE catalogues are ingested into the database and can be queried, through the IVOA Table Access Protocol, with \ac{ADQL}, for which tutorials are provided. The results of queries can be downloaded or overlaid on the different Hierarchical Progressive Survey (HiPS) maps. Spectra can also be queried and
retrieved using the IVOA DataLink protocol. A subset of the data set can be accessed through ESASky at \url{https://sky.esa.int/}.

Data can also be accessed through Python with the \Euclid \texttt{astroquery} toolkit to query catalogues or retrieve files \citep{Astroquery2019}. However, since the FITS files are very large, we promote the use of the ESA Datalabs science platform \citep{Navarro2024}, accessible at \url{https://datalabs.esa.int/}. Access to Datalabs is limited to users with an ESA Cosmos account, which can be obtained through self-registration with invitation code `EUCLIDQ1'. This provides direct access to the \Euclid Q1 data repository with tutorial \texttt{Jupyter} notebooks to access, visualise, manipulate, and analyse data.

As explained in \cref{sc:complementary_darkcloud_stacks}, the complementary background-preserving stacks of the dark cloud will be fully curated soon. Their location and access will be communicated on the Euclid consortium's website at \url{https://www.euclid-ec.org}.


\section{Scientific exploitation of the Q1 data}\label{sc:scienceandcaveat}

The very large range of non-cosmological science enabled by \Euclid is evidenced by already $\sim$\,30 papers submitted as an initial batch of results at the time of Q1. They range from nearby galaxies to strong lenses and quasars, including the detection of rare objects. The publications also place a strong emphasis on automated and scalable search and modelling methods, in preparation for efficient exploitation of the vast amount of \Euclid data to come in DR1 to DR3.

\subsection{Nearby galaxies}

\citet{Q1-SP001} takes advantage of the unprecedented depth, spatial resolution, and field of view of Q1 data to detect and characterise dwarf galaxies. We have identified 2674 candidates, corresponding to 188 dwarfs per deg$^2$, in a 14.25\,deg$^2$ area of EDF-N. Candidates were selected using a semi-automatic method based on \Euclid pipeline measurements, followed by cuts in surface brightness, magnitude, morphology, and $\IE-\HE$ colour. A final visual classification assigned morphology, number of nuclei, globular cluster richness, and blue compact centres.

\subsection{Galaxy morphology}

Q1 data are used as a vast training data set for `foundation' deep-learning models. The goal is to produce catalogues of galaxy properties, in particular morphology-related features.

\cite{Q1-SP047} presents a detailed visual morphology catalogue created by fine-tuning the \texttt{Zoobot} foundation deep-learning model on annotations from an intensive 1-month campaign by Galaxy Zoo \citep{lintott2008} volunteers. Placing a trained deep-learning model within the survey image processing pipeline allows immediate morphology measurements, producing a detailed visual morphology catalogue for Q1 in weeks instead of years.
Detailed visual morphology refers to the recognisable features that comprise a galaxy, such as bars, spiral arms, and tidal tails. Our catalogue includes such galaxy features for the 378\,000 bright ($\IE < 20.5$) or extended (area ${\ge}$\,700 pixels) galaxies in Q1. This catalogue is the first 0.4\% of the approximately 100 million galaxies where \Euclid will ultimately resolve detailed morphology. Our measurements have already proven useful for exploring the relative abundance of stellar bars in disc galaxies from $z=1$ to 0 in \cite{Q1-SP043}. Stellar bars are key structures in disc galaxies, driving angular momentum redistribution and influencing processes such as bulge growth and star formation. Therefore, tracing their abundance over time serves as a proxy for disc assembly. We identified 7711 barred galaxies, which is an order of magnitude more than previous \ac{HST} and \ac{JWST} based results. At fixed redshift, massive galaxies exhibit higher bar fractions, while lower-mass systems show a steeper decline with redshift, suggesting earlier disc assembly in massive galaxies. 

\cite{Q1-SP049} presents the first application of \texttt{AstroPT}, a multi-modal autoregressive foundation model, to \Euclid's Q1 data release. Trained on around 300\,000 optical and infrared images along with \acp{SED}, \texttt{AstroPT} enables efficient self-supervised learning for key astrophysical tasks. We demonstrate its effectiveness in galaxy-morphology classification, redshift estimation, similarity searches, and anomaly detection. These findings showcase the potential of foundation models for scalable, data-driven astrophysical analysis in future larger \Euclid data releases.

\cite{Q1-SP040} also characterises the morphology of Q1 galaxies, but uses more conventional S\'ersic profiles. \Euclid's exquisite image resolution and the large survey area of Q1 data release enable a robust morphological description for more than a million galaxies. The analysis confirms the bimodality of galaxy populations between late- and early-type galaxies, which reflects further differences in their physical properties.

\subsection{Star-forming galaxies}

\cite{Q1-SP031} provides a first view of the \ac{SFMS} in the \acp{EDF}. The \ac{SFMS} is a fundamental relation linking together the galaxies' budget of cold gas, the efficiency in converting it into stars, and its stellar content. It manifests itself as a tight relation between galaxy stellar masses and star-formation rates across different epochs. We investigated the \ac{SFMS} in the redshift range $0.2 < z < 3.0$  and recovered more than  30\,000 galaxies with $\log_{10}(M_*/M_\odot) > 11$, giving a precise constraint of the \ac{SFMS} at the high-mass end. These results highlight the potential of \Euclid in studying the fundamental scaling relations that regulate galaxy formation and evolution.

At higher redshifts $z > 3$, our understanding of cosmic star formation mostly relies on rest-frame UV observations. However, these observations overlook massive, dust-obscured sources, and indeed recently infrared data from the \Spitzer and \ac{JWST} have revealed a hidden population at $z \simeq 3$--6 with extreme red colours. Taking advantage of the overlap between imaging in the \acp{EDF} and ancillary \Spitzer observations, \cite{Q1-SP016} identified 3900 extremely red objects with $\HE - \textrm{IRAC2} > 2.25$, dubbed \acp{HIERO}. Our results confirm that \ac{HIERO} galaxies may populate the high-mass end of the stellar mass function at $z > 3$, with some sources reaching extreme stellar masses $(M_* > 10^{11} M_\odot)$ and exhibiting high dust attenuation values $(A_V >3)$, contributing to a more complete census of early star-forming galaxies. Given the extreme nature of this population, characterising these sources is crucial for building a comprehensive picture of galaxy evolution and stellar mass assembly across most of the history of the Universe. This work demonstrates \Euclid’s potential to provide statistical samples of rare objects.

The first years of observations with \ac{JWST} have revealed a
novel population of compact red sources, the so-called \acp{LRD}. They are characterised
by a peculiar ‘v-shaped’ \ac{SED}, namely a blue rest-frame UV continuum and a red rest-frame optical continuum, and were mainly observed at $z>4$. The nature of these \acp{LRD} is debated, their emission being consistent with either hosting an \ac{AGN} or emission from dusty star formation.
In \citet{Q1-SP011} these sources are identified by combining a slope selection, a criterion for compactness and visual inspection. We found over $3000$ \ac{LRD} candidates at  $z<4$, while previous \ac{JWST} results mostly found objects at higher redshifts. We also show that \acp{LRD} are not the dominant \ac{AGN} population in this redshift range.

\subsection{Passive galaxies and galaxy quenching}

Galaxy quenching is the sudden cessation of star formation in galaxies. Investigating the drivers of the quenching of star formation in galaxies is key to understanding their evolution. Using Q1 data, \cite{Q1-SP044} develops a probabilistic classification framework, that combines the average specific star-formation rate inferred over two timescales ($10^8$ and $10^9$ years), to categorise galaxies as `Ageing' (secularly evolving), `Quenched' (recently halted star formation), or `Retired' (dominated by old stars). At $z < 0.1$ and $M_* \ge 3\,\times\,10^8 M_\odot$, we obtain  fractions of 68--72\%, 8--17\%, and 14--19\% for Ageing, Quenched, and Retired populations, respectively, consistent with previous studies.  We also explore how these fractions vary with different factors, including redshift, stellar mass, morphology, and chemical composition, finding, for example, that Ageing and Retired galaxies dominate at the low and high-mass ends, respectively, while Quenched galaxies surpass the Retired fraction for $M_* \le 3\,\times\,10^{10} M_\odot$.  Additionally, the evolution with redshift shows increasing/decreasing fraction of Ageing/Retired galaxies and Ageing galaxies generally exhibit disc morphologies and low metallicities.

\subsection{AGN evolution}

Several papers have used Q1 data to explore \ac{AGN} science.
In \citet{Q1-SP027},
three multi-wavelength catalogues of \ac{AGN} candidates from the Q1 \acp{EDF} are introduced. Traditional photometric selections, involving optical, \ac{NIR}, mid-IR, and spectroscopic diagnostics are employed to analyse the \ac{AGN} populations using \Euclid data.  Additionally, we explore new colour-colour criteria to identify \ac{AGN}. 
This catalogue of \ac{AGN} candidates is complemented by identification of X-ray \ac{AGN} in the Q1 footprint in \citet{Q1-SP003}. Here, the most likely X-ray emitters among the \Euclid sources are first identified and then, using machine learning, X-ray objects are classified as Galactic or extragalactic. Finally, for the extragalactic sources, photometric redshifts, their luminosity, and their basic SED properties are presented. 

This multiwavelength \ac{AGN} candidate catalogue is then used to understand the performance of new machine-learning methods presented in \citet{Q1-SP009} and \citet{Q1-SP015}. The first paper explores a novel application of diffusion-based inpainting for the identification of \ac{AGN} using VIS images alone. By exploiting the reconstruction error in regenerated galaxy cores, this method achieves high recall rates with candidates from the traditional multi-wavelength methods presented in \citet{Q1-SP027} and \citet{Q1-SP003}. Using only VIS images for training and inference, no prior knowledge about the presence of an \ac{AGN} component is required, making it applicable for use with future \Euclid data releases. 
The second paper \citep{Q1-SP015} 
presents a deep learning-based method to identify and quantify \ac{AGN} in \Euclid galaxy images by estimating the central point source contribution. Trained on `Euclidised' mock images with injected \ac{AGN}, the model accurately recovers the \ac{AGN} contribution fraction. Applied to \Euclid data, 8\% of galaxies show an \ac{AGN} contribution fraction above 20\%. 
We also find that \ac{AGN} luminosity correlates with host galaxy mass, suggesting faster \ac{SMBH} growth in massive galaxies. \ac{AGN} are more common in quiescent galaxies and most luminous in massive and starburst systems, supporting a link between \acp{SMBH} and galaxy assembly. 

The machine-learning approach is further explored in \citet{Q1-SP013} to study the role of major mergers in triggering \ac{AGN} using a classical binary classification of galaxies into `active' and `non-active'. The paper also examines \ac{AGN} properties, such as the point-source contribution, as well as luminosity, with four different \ac{AGN}-selection techniques
explored. The main results are that mergers influence all types of \ac{AGN} candidates, but overall they do not seem to be the main triggering mechanism. However, major mergers become more and more important in the fuelling of the most luminous \ac{AGN}, indicating that they might be the dominant triggering mechanism of the most powerful \ac{AGN} in the Universe. 

Q1 data also allowed us to explore the dust-obscured red \acp{QSO}. 
In \citet{Q1-SP023} a selection method based on machine-learning and multidimensional colour analysis is developed, identifying over 150\,000 candidate red \acp{QSO}. Compared to VISTA+DECAm-based colour selection criteria, \Euclid's superior depth, resolution, and optical-to-NIR coverage improves the identification of the reddest, most obscured sources. To refine the selection function, probabilistic random-forest classification is combined with \ac{UMAP} visualisation, achieving 98\% completeness and 88\% purity. This work provides a first census of candidate red \acp{QSO} in Q1 and sets the groundwork for future data releases, including spectral and host morphology analyses.

\subsection{Cosmic environment}

The impact of the environment in the properties of galaxies and galaxy clusters is addressed in several papers.
\cite{Q1-SP017} studies how the local environment of a galaxy affects its evolution. To do this, we measure the local environmental density for each galaxy in the Q1 sample. We then calculate the fractions of passive galaxies and early-type galaxies as function of stellar mass, environment, and redshift. We find that, up to $z \simeq 0.7$, the environment plays a significant role in transforming galaxies from star-forming to passive. At $z > 0.7$, the passive fraction and early-type-galaxy fraction are mostly determined by the stellar mass, and as such the environment only has a weak effect on these properties.

Galaxy morphologies and shape orientations are also expected to correlate with their large-scale environment, as they grow by accreting matter from the cosmic web and are subject to interactions with other galaxies. \cite{Q1-SP028} extracts cosmic filaments from the Q1 data at $0.5 < z < 0.9$, and analyses the 2D alignment of galaxy shapes with large-scale filaments as a function of S\'ersic indices and masses. We confirm the known trend that more massive galaxies are closer to filament spines. At fixed masses, morphologies correlate with both density and distances to large-scale filaments. In addition, the large volume of this data set allows us to detect a signal indicating that there is a preferential alignment of the major axis of massive early-type galaxies along cosmic filaments.

Clusters are also shaped by their connection to the cosmic web. In particular, since they are nodes in the large-scale cosmic web, a relevant property is the number of filaments connected to a cluster, known as its `connectivity'. \cite{Q1-SP005} uses Q1 data to investigate the connectivity of galaxy clusters and how it correlates with their own and their galaxy-member properties. Around 220 clusters in the redshift range $0.2 < z < 0.7$ were analysed. In agreement with previous measurements, we find that the most massive clusters are, on average, connected to a larger number of cosmic filaments, which is consistent with hierarchical structure formation models.  We also explored possible correlations between connectivity and the fraction of early-type galaxies and the S\'ersic index of galaxy members. Our result suggests that the clusters populated by early-type galaxies exhibit higher connectivity compared to clusters dominated by late-type galaxies. 
 
\subsection{Strong gravitational lensing}

The combination of a wide field of view with the capability of resolving small Einstein radii makes \Euclid the most efficient instrument for finding strong lenses ever built. Most galaxy-scale strong lenses are expected to have Einstein radii smaller than \ang{;;1.0}, below the resolution limits of ground-based surveys, while \Euclid's space-based \ac{PSF} can resolve lenses down to an Einstein radii of around \ang{;;0.6} in large numbers. \cite{Q1-SP048} presents the `strong lensing discovery engine', which combines the strengths of AI, citizen scientists, and experts into a system that efficiently searches for strong lenses in \Euclid data. In particular, the lenses are found through an initial sweep by deep-learning models, followed by Space Warps citizen-science inspection, expert vetting, and detailed lens modelling. A catalogue of 497 galaxy-galaxy strong lenses was constructed from Q1 data, which already doubles the total number of known lens candidates with space-based imaging. Extrapolating to the complete \ac{EWS} implies a likely yield of over 100\,000 high-confidence candidates, transforming strong lensing science. 

The Q1 search with the strong-lensing discovery engine is detailed in a series of four additional papers. \cite{Q1-SP052} uses Q1 images containing galaxies with high velocity dispersion, spectroscopically identified in the \ac{SDSS} and the \ac{DESI}, to search for lenses and build an initial training set for our machine-learning models. \cite{Q1-SP053} analyses five machine learning models and compares their performance on real \Euclid images. \cite{Q1-SP054} reports the discovery of four new \acp{DSPL} in Q1. Strong gravitational lensing systems with multiple source planes are powerful tools for probing the density profiles and dark matter substructure of the galaxies, and the ratio of Einstein radii is related to the dark energy equation of state. \acp{DSPL} are extremely rare, but \Euclid is expected to discover over 1000 such systems. Finally,  \cite{Q1-SP059} discusses lessons learned for future Euclid data releases, presenting a Bayesian ensemble method that combines lens classifiers to further optimise lens discovery within our discovery engine for DR1.

In addition to searching methods, also modelling techniques need to be optimised for handling \Euclid's vast data set. \cite{Q1-SP063} presents a Bayesian neural network, dubbed \ac{LEMON} designed to model \Euclid's gravitational lenses efficiently. \ac{LEMON} estimates key parameters of the mass and light profiles and is shown to perform well on simulated \Euclid lenses, real Euclidised \ac{HST} lenses, and real Q1 lenses. 

\begin{figure}[htbp!]
\centering
\includegraphics[width=1.0\hsize]{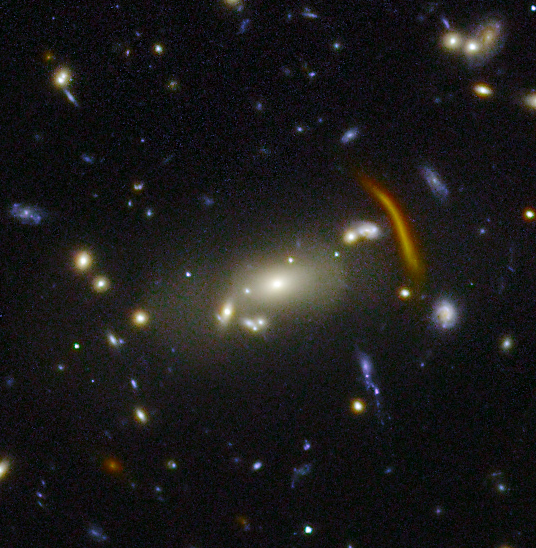}
\caption{The strong-lensing cluster Abell\,2280 in the \ac{EDF-N} Q1 area. This image is at the full VIS resolution, $\ang{;;52}\times\ang{;;52}$ wide, using VIS and all NISP photometric images for colour information.}
\label{fig:a2280}
\end{figure}

\subsection{Galaxy clusters}

The algorithms developed and implemented in the \ac{SGS} \ac{LE3} pipeline for cluster detection have already being run on the Q1 data (Euclid Collaboration: Bhargava et al., in prep.). They found several hundred high-confidence clusters in the Q1 area, stretching across in the redshift range $0.2<z<1.5$, validating the pipelines ahead of DR1 and allowing for improvements. 

Also using Q1 data, \cite{Q1-SP057} constructs the first catalogue of strong lensing galaxy clusters observed by \Euclid, based on the visual inspection of over a thousand richness-selected known galaxy clusters. Most of these galaxy clusters had never been observed from space before (such as Abell\,2280, \cref{fig:a2280}), and only a few were previously known to host strong lensing features. Specifically, we identified 83 strong gravitational lenses, including 14 exhibiting secure strong lensing features such as tangential and radial arcs and multiple images. These galaxy clusters will be re-observed by \Euclid multiple times during the mission as part of the \ac{EDS}. Based on the number density of detected lensing galaxy clusters, we expect that \Euclid will observe more than 6000 strong lensing galaxy clusters in its wide survey. This estimate is consistent with the forecasts from simulations in the $\Lambda$CDM cosmological model. Our work demonstrates the huge potential of the \Euclid mission for discovering new strong lensing galaxy clusters. 

 \cite{Q1-SP022} deals with the detection of galaxy clusters and protoclusters at high redshift $(z > 1.3)$. \Euclid is expected to detect tens of thousands of such objects over the course of its mission, and Q1 is large enough to detect tens of clusters and hundreds of protoclusters at these early epochs. To extend the detection to this redshift range, we have combined \Euclid and \Spitzer observations of the Q1 \acp{EDF}. We compute local projected densities of Spitzer-selected galaxies with two methods, and use high values of the computed surface density as signposts for cluster and protocluster candidates. We found that 2--3\% of the surface densities measured in \ac{EDF-N} and \ac{EDF-F} are $3\,\sigma$ above the mean density. This is a clear indication
that a large part of the less massive galaxies in these two \acp{EDF} have larger densities than galaxies with the same mass in the field, and hence there is high potential that they belong to groups or clusters.

\subsection{Transients}

While Q1, in distinction to future data releases, does not include multiple epoch observations of the target fields, \citet{Q1-SP002} reports on serendipitous \Euclid observations of previously known transients reported to the Transient Name Server in the \acp{EDF}. We were able to make photometric measurements at the location of 161 transients, obtaining deep photometry or upper limits in \IE, \YE, \JE, and \HE at various phases of the transient light-curves. These observations include one of the earliest NIR detection of a Type~Ia supernova, 15 days prior to peak brightness and the late-phase (435.9\,days post peak) observations of the core-collapse supernova 2023aew. In addition to this the hosts galaxies of several transients were detected that previously had been classified as hostless.

\section{Caveats}\label{sc:caveats}

Q1 is the first release of \Euclid data by the \ac{SGS} in a sequence ultimately gearing up towards DR3 in the early 2030s. The data processing is not yet fully mature, and many of the processing functions have identified pathways towards improvements, some that will be implemented for DR1 and some whose development will be longer and appear in subsequent data releases. We take the opportunity to remind the users of Q1 data of a few important caveats to keep in mind when working with the data. 

One important caveat discussed in \citet{Q1-TP007} is to be careful when using redshifts from the catalogues. On the one hand, by design, the \ac{SGS} pipelines produce one spectrum for each source of the MER catalogue with $\HE < 22.5$. This corresponds to about 4.3 million spectra. On the other hand, the expected number of sources with H$\alpha$ flux above the nominal flux limit of the \ac{EWS} (predicted from luminosity functions for an area of 63.1\,deg$^2$) is of the order of 100\,000. This corresponds to about 2.5\% of the spectra.  These data also include further spectra for stars and galaxies/\ac{AGN} outside of the target redshift range, measured from other lines.  These will still be a small part of the total, so that the fraction of spectra allowing for the measurement of a redshift is below 10\%. Yet, the \ac{SGS} pipeline provides a redshift estimate for everyone of them. The challenge the \ac{SGS} is confronted with is to correctly assess the reliability of the automated redshift measurements. The issue will be even more complex for DR1 when attempting to increase the limiting magnitude of the extracted sample. Meanwhile, we recommend to work with the highest quality of spectra, and to check if possible the redshift by examining the 1D spectra of the sources of interest. This can only be done for small samples of sources.

Another caveat concerns the imaging data. The \ac{SGS} pipeline is optimised for the \Euclid mission core science goals, namely to measure the shape and photometry of galaxies at redshifts above $z = 0.5$, that is for objects with arcsecond size on the sky. While the VIS and NIR \acp{PF} do strive to preserve low surface brightness features as much as possible, these are mostly filtered out at the MER stage when the stacks are being built. Members of the Euclid Consortium working on the intracluster medium, on local galaxies, or on Galactic cirrus are devising alternative processing outputs derived from the VIS and NIR calibrated frames. These alternatives will become available in the future through the Euclid Datalabs interface discussed in~\cref{sc:dataaccess}, but not in time for Q1. We additionally caution users about the MER catalogue from the LDN\,1641 region, as the structure of this field is very different from typical \ac{EWS} fields for which the SGS PFs are optimised, and advise them to start their analysis from the VIS and NIR LE2 calibrated frames for this part of the data set.

Finally, we note that many \acp{PF} have issued release notes for their processing procedures, either in the papers submitted with Q1, or in the explanatory supplement published online from the release page at \url{https://www.cosmos.esa.int/en/web/euclid/euclid-q1-data-release}. We recommend all users of the Q1 data to read these descriptions carefully.


\section{Conclusions}
\label{sec:conclusions}

Q1 contains data from about 7 days of \Euclid observations. Over these 63\,deg$^2$ of the \ac{EWS} \Euclid already detects about 30 million sources, including galaxies, stars, quasars, brown dwarfs, and even solar system objects. 
This release is only the first part of \Euclid's survey data, meant as an initial batch of data for scientific use, but also to provide to the community an impression of what is to come: in the sense of types of data and data quality, to sharpen tools and define interesting science questions, but also to sensitise the community to the sheer size of the \Euclid surveys.
The first `cosmological' data release (DR1), likely to be in late 2026, will cover 30 times the Q1 area or about $1900\,\textrm{deg}^2$, and hence about 1.5 orders of magnitude more objects in every category.
Any science project that finds a handful objects of interest in Q1 will find hundreds in DR1; however, an easy project in Q1 involving human eyes as part of the analysis or classification cascade might become challenging in DR1, and any analysis tasks that due to sheer catalogue size are already hard in Q1 will require very different tools in DR1 to be feasible.

The first wave of scientific publications based on Q1, sketched in \cref{sc:scienceandcaveat}, illustrates the growing role that machine learning will play in the analysis of upcoming big data sets in astronomy -- including \Euclid. Approximately half of the scientific papers accompanying Q1 make use of `artificial intelligence': generative and classification models are used for finding and characterising \ac{AGN} in galaxies \citep{Q1-SP009,Q1-SP013,Q1-SP015,Q1-SP003,Q1-SP023},
neural-network-based morphology classification and strong gravitational lens classifiers are directly deployed in the \ac{SGS} pipeline \citep{Q1-SP047,Q1-SP048, Q1-SP059,Q1-SP053, Q1-SP043,Q1-SP054}, and simulation-based inference is also employed for characterising lenses \citep{Q1-SP063}. The Q1 data additionally serve as benchmarks for the development of large multimodal foundation models in astronomy \citep{Q1-SP049}, which will probably play a major role in future releases.

While the first set of astronomy results from the Euclid Consortium -- many more beyond those referenced here are in progress -- have a scientific goal and value in themselves, these projects were also essential for vetting the \Euclid data processing pipeline that was sketched in \cref{sc:Processing,sc:ext}.
\Euclid's data processing arguably had a maturity upon launch like no or very few other mission before, due to the core cosmological mission goals being very far down the data-analysis pipeline. Their feasibility and resulting data-processing needs had to be demonstrated very early on in mission development. However, even with more than a decade of developing the \Euclid \ac{SGS}, the confrontation of planning with post-launch reality \citep{EuclidSkyOverview} naturally led to pipeline modification, incorporating all the new knowledge directly arising from actual space-based data.
These Q1 science projects provided a detailed hands-on deep-dive into all aspects of \Euclid data and have directly provided valuable and essential feedback for survey, pipeline, and archiving updates. These lessons learned from Q1 data will also directly enter DR1 and future releases.

Now these billions of pixels and the catalogues of \Euclid space- and ground-based data are released to the world-wide astronomy community. They will form the basis for many more science studies than those the Euclid Consortium already carried out during its short, four-month long head start. Q1 is a beginning, both of \Euclid as a readily available astronomical resource, enabling more and more astronomy and cosmology studies, but also as a steadily growing and increasingly vast database of unprecedented astronomical information -- as a standard reference for decades to come.

\begin{acknowledgements}
\AckQone\\

\AckEC\\

Based on data from UNIONS, a scientific collaboration using three Hawaii-based telescopes: CFHT, Pan-STARRS and Subaru \url{www.skysurvey.cc}\,.
This work is also based on
data from the Dark Energy Camera (DECam) on the Blanco 4-m Telescope
at CTIO in Chile \url{https://www.darkenergysurvey.org}\,.
This work uses results from the ESA mission {\it Gaia},
whose data are being processed by the Gaia Data Processing and
Analysis Consortium \url{https://www.cosmos.esa.int/gaia}\,.
We are honoured and grateful for the opportunity of observing the Universe from Maunakea and Haleakala, which both have cultural, historical, and natural significance in Hawaii.
\end{acknowledgements}

\bibliography{q1overview, Euclid, Q1}

\begin{appendix}
  \onecolumn 

\section{Footprints}\label{sc:footprint}

\begin{figure*}[htbp!]
	\begin{centering}
	\scalebox{1.0}{ 
	\includegraphics{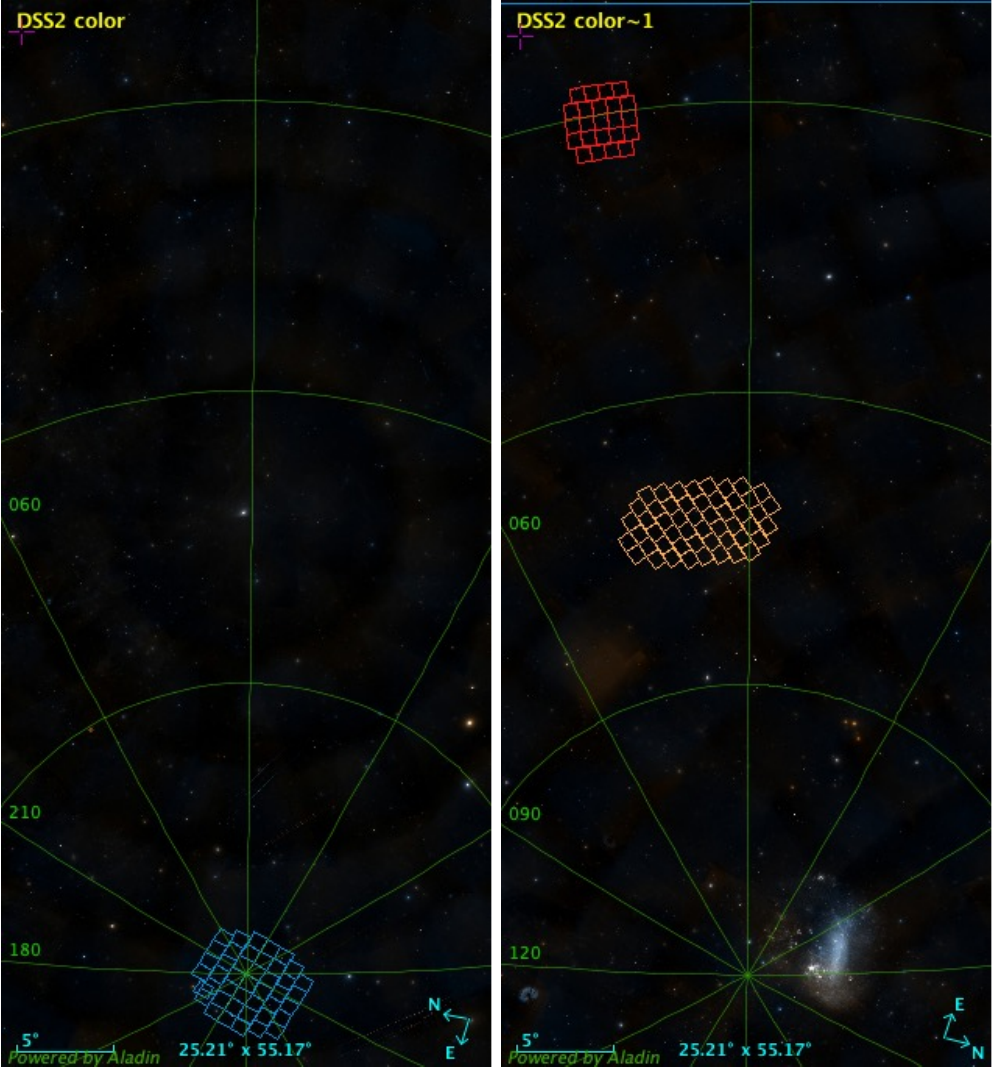} }
	\caption{Footprints of the Q1 visits to the \acp{EDF} shown on the sky in ecliptic coordinates overlaid on two 25$^\circ$\,$\times$\,55$^\circ$ cut-outs of a full-sky Digitized Sky Survey 2 optical image generated by the \ac{CDSdata}. 
    \textit{Left}: A strip of the northern ecliptic hemisphere down to $+45^\circ$, showing \ac{EDF-N} (blue) at the ecliptic pole in the Draco constellation; one of its fields is noticeably out of the regular pattern in order to avoid the bright 42~Dra star. \textit{Right}: A strip of the southern ecliptic hemisphere down to $-45^\circ$ showing \ac{EDF-F} (red) in the Fornax constellation and \ac{EDF-S} (orange) in the Dorado constellation.  The Large Magellanic Cloud is noticeable close to the ecliptic pole.}
	\label{fig:EDFs}
	\end{centering}
\end{figure*}

In order to facilitate visualisation and interaction with \Euclid's Q1 data, we reproduce below the DS9 region files around the EDF-N, EDF-S, and EDF-F fields, in an equatorial coordinate system. \Cref{fig:EDFs} shows the \acp{EDF} sky footprints of the Q1 visits.

\subsection{Euclid Deep Field North (Q1 visit)}

{\tt 
\# Region file format: DS9 version 4.1\\
global color=green dashlist=8 3 width=1 font="helvetica 10 normal roman" select=1 highlite=1 dash=0 fixed=0 edit=1 move=1 delete=1 include=1 source=1\\
icrs\\
polygon(264.893501,+66.876904, 266.545798,+67.314562, 265.507842,+67.901955, 263.834235,+67.453547)\\
polygon(266.544700,+67.314397, 268.257279,+67.734528, 267.243764,+68.332679, 265.506130,+67.901740)\\
polygon(268.256414,+67.734214, 270.029657,+68.135953, 269.045117,+68.744752, 267.242481,+68.332262)\\
polygon(270.502410,+67.829132, 272.320057,+68.206349, 271.394840,+68.828130, 269.543423,+68.440658)\\
polygon(269.219916,+67.130327, 270.963701,+67.521848, 270.029895,+68.136053, 268.256746,+67.734290)\\
polygon(267.533196,+66.720617, 269.220313,+67.130413, 268.257279,+67.734528, 266.544918,+67.314341)\\
polygon(265.902932,+66.293425, 267.533720,+66.720655, 266.545649,+67.314451, 264.892984,+66.876888)\\
polygon(263.820789,+66.137120, 265.405108,+66.585950, 264.371331,+67.166150, 262.769005,+66.706754)\\
polygon(262.808185,+65.390331, 264.329254,+65.849557, 263.301380,+66.422788, 261.765345,+65.953277)\\
polygon(263.805122,+64.820743, 265.310362,+65.269873, 264.328021,+65.849776, 262.806711,+65.390701)\\
polygon(264.824612,+65.560626, 266.390752,+65.999246, 265.404353,+66.585959, 263.819725,+66.137214)\\
polygon(266.866432,+65.703672, 268.475126,+66.120927, 267.533453,+66.720621, 265.902671,+66.293440)\\
polygon(268.474876,+66.120895, 270.136287,+66.520851, 269.220316,+67.130413, 267.533441,+66.720588)\\
polygon(270.136224,+66.520821, 271.850469,+66.902624, 270.963788,+67.521942, 269.220117,+67.130370)\\
polygon(271.412680,+67.212635, 273.196736,+67.580074, 272.320148,+68.206674, 270.502730,+67.829417)\\
polygon(272.764139,+67.893364, 274.621719,+68.245300, 273.759234,+68.879416, 271.863561,+68.517596)\\
polygon(273.617937,+67.265067, 275.437972,+67.607764, 274.621509,+68.245992, 272.764312,+67.893961)\\
polygon(274.427855,+66.632410, 276.211277,+66.966355, 275.437265,+67.608373, 273.617658,+67.265539)\\
polygon(275.197039,+65.995778, 276.944859,+66.321414, 276.210339,+66.966922, 274.427094,+66.632791)\\
polygon(273.099339,+65.965181, 274.817518,+66.314505, 274.027754,+66.949606, 272.276962,+66.591357)\\
polygon(272.277219,+66.591202, 274.028159,+66.949337, 273.196497,+67.580354, 271.412776,+67.212867)\\
polygon(271.008299,+65.906254, 272.693273,+66.278759, 271.850385,+66.902697, 270.136276,+66.520846)\\
polygon(269.372777,+65.515706, 271.008299,+65.906254, 270.136276,+66.520846, 268.474905,+66.120879)\\
polygon(267.786616,+65.108100, 269.372777,+65.515706, 268.474905,+66.120879, 266.866323,+65.703735)\\
polygon(265.784814,+64.977783, 267.331840,+65.406551, 266.390156,+65.999307, 264.824073,+65.560835)\\
polygon(264.760317,+64.244671, 266.249360,+64.684150, 265.309510,+65.270192, 263.803987,+64.821216)\\
polygon(265.676327,+63.662629, 267.148505,+64.092747, 266.248879,+64.684564, 264.759689,+64.245306)\\
polygon(266.555250,+63.074998, 268.010176,+63.496057, 267.148453,+64.093307, 265.676067,+63.663391)\\
polygon(268.427292,+63.195235, 269.915229,+63.596656, 269.091858,+64.205114, 267.583946,+63.795147)\\
polygon(267.583847,+63.794739, 269.091601,+64.204828, 268.231276,+64.808484, 266.703618,+64.389350)\\
polygon(266.703827,+64.389063, 268.231260,+64.808272, 267.331539,+65.406665, 265.784489,+64.978067)\\
polygon(268.666110,+64.507050, 270.229780,+64.905528, 269.372861,+65.515801, 267.786500,+65.108130)\\
polygon(270.229642,+64.905424, 271.839504,+65.286942, 271.008297,+65.906253, 269.372808,+65.515703)\\
polygon(271.839316,+65.286898, 273.495532,+65.650554, 272.693297,+66.278792, 271.008237,+65.906238)\\
polygon(273.881753,+65.334963, 275.568086,+65.675908, 274.816856,+66.314735, 273.098688,+65.965322)\\
polygon(274.776103,+65.125740, 276.460162,+65.456719, 275.737758,+66.099857, 274.021387,+65.760637)\\
polygon(272.631954,+64.663143, 274.259684,+65.018390, 273.495276,+65.650576, 271.839134,+65.286855)\\
polygon(271.048118,+64.290415, 272.632668,+64.663195, 271.839622,+65.286855, 270.229663,+64.905425)\\
polygon(269.507512,+63.900998, 271.048631,+64.290659, 270.229950,+64.905531, 268.666185,+64.507145)\\
polygon(270.313093,+63.290203, 271.831779,+63.671408, 271.048942,+64.290683, 269.507726,+63.901191)\\
polygon(271.830621,+63.671074, 273.390062,+64.035531, 272.632668,+64.663195, 271.048137,+64.290417)\\
polygon(273.388700,+64.035332, 274.988662,+64.382470, 274.259328,+65.018376, 272.631649,+64.663072)\\
polygon(274.627208,+64.700916, 276.282521,+65.033829, 275.567216,+65.676086, 273.880976,+65.335041)\\
}

\subsection{Euclid Deep Field South (Q1 visit)}
{\tt 
\# Region file format: DS9 version 4.1 1860\\
global color=green dashlist=8 3 width=1 font="helvetica 10 normal roman" select=1 highlite=1 dash=0 fixed=0 edit=1 move=1 delete=1 include=1 source=1\\
icrs\\
polygon(55.365656,-48.409852, 55.290812,-49.186124, 56.372854,-49.226311, 56.431187,-48.449491)\\
polygon(55.291518,-49.184519, 55.213699,-49.960700, 56.312971,-50.001827, 56.373586,-49.225062)\\
polygon(56.313552,-50.001223, 56.250422,-50.778036, 57.369497,-50.809646, 57.414565,-50.032375)\\
polygon(57.392655,-50.410546, 57.346300,-51.187750, 58.475886,-51.209195, 58.503568,-50.431609)\\
polygon(57.436152,-49.633845, 57.391803,-50.411142, 58.502655,-50.431983, 58.529267,-49.654403)\\
polygon(56.373323,-49.225408, 56.312652,-50.002320, 57.413660,-50.033225, 57.456916,-49.255876)\\
polygon(56.431157,-48.449620, 56.372858,-49.226546, 57.456422,-49.256664, 57.498134,-48.479363)\\
polygon(56.487136,-47.673775, 56.431049,-48.450718, 57.498030,-48.480220, 57.538165,-47.702809)\\
polygon(57.558185,-47.303640, 57.518907,-48.080971, 58.578657,-48.100366, 58.602272,-47.322776)\\
polygon(57.518854,-48.080412, 57.477918,-48.857712, 58.554018,-48.877576, 58.578519,-48.099957)\\
polygon(57.478188,-48.857135, 57.435615,-49.634420, 58.528756,-49.654843, 58.554207,-48.877152)\\
polygon(58.546849,-49.104042, 58.520982,-49.881765, 59.619863,-49.892023, 59.628506,-49.114226)\\
polygon(58.521281,-49.881409, 58.494198,-50.659020, 59.611274,-50.669641, 59.620271,-49.891863)\\
polygon(58.494853,-50.658750, 58.466564,-51.436428, 59.602462,-51.447240, 59.611908,-50.669504)\\
polygon(59.614373,-50.367278, 59.604859,-51.145118, 60.733611,-51.145253, 60.724482,-50.367411)\\
polygon(59.623145,-49.589565, 59.614122,-50.367378, 60.724283,-50.367489, 60.715546,-49.589714)\\
polygon(59.631601,-48.811896, 59.623070,-49.589711, 60.715365,-49.589667, 60.706904,-48.811874)\\
polygon(58.571568,-48.326777, 58.546638,-49.104359, 59.628323,-49.114413, 59.636614,-48.336618)\\
polygon(58.595350,-47.549380, 58.571559,-48.327012, 59.636582,-48.336745, 59.644464,-47.558945)\\
polygon(58.618388,-46.772027, 58.595513,-47.549662, 59.644733,-47.559187, 59.652307,-46.781375)\\
polygon(59.654821,-46.478762, 59.647478,-47.256514, 60.690842,-47.256413, 60.683054,-46.478547)\\
polygon(59.647353,-47.256446, 59.639703,-48.034217, 60.698685,-48.034136, 60.690719,-47.256344)\\
polygon(59.639630,-48.034130, 59.631496,-48.811936, 60.706852,-48.811893, 60.698632,-48.034156)\\
polygon(60.702587,-48.412256, 60.710933,-49.190081, 61.794182,-49.179992, 61.769221,-48.402358)\\
polygon(60.710933,-49.190081, 60.719369,-49.967856, 61.820041,-49.957640, 61.794182,-49.179992)\\
polygon(60.719369,-49.967856, 60.727976,-50.745645, 61.846813,-50.735253, 61.820041,-49.957640)\\
polygon(61.807086,-49.579576, 61.833189,-50.357189, 62.942370,-50.336534, 62.898522,-49.559221)\\
polygon(61.781893,-48.801951, 61.807187,-49.579538, 62.898647,-49.559290, 62.856345,-48.781942)\\
polygon(61.757450,-48.024337, 61.781944,-48.801932, 62.856366,-48.782048, 62.815520,-48.004683)\\
polygon(60.694429,-47.634438, 60.702587,-48.412256, 61.769221,-48.402358, 61.745133,-47.624745)\\
polygon(60.686443,-46.856672, 60.694429,-47.634438, 61.745133,-47.624745, 61.721804,-46.847025)\\
polygon(60.678587,-46.078835, 60.686443,-46.856672, 61.721804,-46.847025, 61.699282,-46.069404)\\
polygon(61.687600,-45.691436, 61.710067,-46.469112, 62.737627,-46.450246, 62.700753,-45.672814)\\
polygon(61.710286,-46.469036, 61.733348,-47.246678, 62.775971,-47.227521, 62.737900,-46.450155)\\
polygon(61.733577,-47.246711, 61.757397,-48.024356, 62.815537,-48.004790, 62.776065,-47.227377)\\
polygon(62.799976,-47.702391, 62.840325,-48.479760, 63.907321,-48.450208, 63.851063,-47.673247)\\
polygon(62.840358,-48.479635, 62.881969,-49.256962, 63.965585,-49.226745, 63.907269,-48.449888)\\
polygon(62.881771,-49.256805, 62.924845,-50.034120, 64.025843,-50.003267, 63.965370,-49.226480)\\
polygon(63.923974,-48.676258, 63.982844,-49.453143, 65.069941,-49.412751, 64.994248,-48.636457)\\
polygon(64.958555,-48.259253, 65.032887,-49.035559, 66.109770,-48.985518, 66.019141,-48.210022)\\
polygon(63.867244,-47.899836, 63.924169,-48.676755, 64.994451,-48.637060, 64.921478,-47.860769)\\
polygon(63.812115,-47.123350, 63.867145,-47.900316, 64.921335,-47.861371, 64.850946,-47.085019)\\
polygon(62.760840,-46.925292, 62.799900,-47.702642, 63.850879,-47.673531, 63.796574,-46.896601)\\
polygon(62.722630,-46.148011, 62.760463,-46.925420, 63.796218,-46.896939, 63.743710,-46.119907)\\
polygon(62.685424,-45.370797, 62.722143,-46.148283, 63.743230,-46.120281, 63.692455,-45.343247)\\
polygon(63.706269,-45.570338, 63.757795,-46.347301, 64.781862,-46.309945, 64.716056,-45.533418)\\
polygon(64.749278,-45.932675, 64.816383,-46.709097, 65.846335,-46.661972, 65.764790,-45.886200)\\
polygon(63.758487,-46.346856, 63.811721,-47.123813, 64.850569,-47.085691, 64.782542,-46.309293)\\
polygon(64.817087,-46.708214, 64.886559,-47.484630, 65.931512,-47.436610, 65.847033,-46.660883)\\
polygon(64.886848,-47.483762, 64.958623,-48.260123, 66.019223,-48.211103, 65.931792,-47.435532)\\
polygon(65.931750,-47.435656, 66.019236,-48.211209, 67.077617,-48.152500, 66.974519,-47.377796)\\
polygon(65.846731,-46.661304, 65.931332,-47.436996, 66.974211,-47.379418, 66.874679,-46.604582)\\
}

\subsection{Euclid Deep Field Fornax (Q1 visit)}
{\tt
\# Region file format: DS9 version 4.1\\
global color=black dashlist=8 3 width=1 font="helvetica 10 normal roman" select=1 highlite=1 dash=0 fixed=0 edit=1 move=1 delete=1 include=1 source=11920\\
icrs\\
polygon(50.585853,-27.644678, 50.902896,-28.370438, 51.651550,-28.112633, 51.330259,-27.388680)\\
polygon(50.902775,-28.370157, 51.223920,-29.095156, 51.977045,-28.835778, 51.651435,-28.112564)\\
polygon(51.223872,-29.094851, 51.549421,-29.819124, 52.307059,-29.558040, 51.977039,-28.835567)\\
polygon(51.928784,-29.688937, 52.261115,-30.411604, 53.021536,-30.146543, 52.684516,-29.425819)\\
polygon(51.977005,-28.835685, 52.307026,-29.558158, 53.060758,-29.292837, 52.726389,-28.572241)\\
polygon(51.651435,-28.112564, 51.977045,-28.835778, 52.726315,-28.572265, 52.396513,-27.850868)\\
polygon(51.330144,-27.388611, 51.651550,-28.112633, 52.396513,-27.850868, 52.071013,-27.128622)\\
polygon(50.643457,-26.791823, 50.958647,-27.517318, 51.701255,-27.259222, 51.382031,-26.535410)\\
polygon(51.381611,-26.535441, 51.701065,-27.259178, 52.440171,-26.997106, 52.116833,-26.275027)\\
polygon(52.070938,-27.128647, 52.396513,-27.850868, 53.137897,-27.585053, 52.808298,-26.864604)\\
polygon(52.396513,-27.850868, 52.726315,-28.572265, 53.471935,-28.304710, 53.137897,-27.585053)\\
polygon(52.726315,-28.572265, 53.060724,-29.292954, 53.810546,-29.023558, 53.471935,-28.304710)\\
polygon(52.684335,-29.425985, 53.021211,-30.146756, 53.777552,-29.877536, 53.436251,-29.158735)\\
polygon(53.436030,-29.158806, 53.777186,-29.877654, 54.529491,-29.604313, 54.183935,-28.887325)\\
polygon(53.471860,-28.304734, 53.810472,-29.023581, 54.556487,-28.750127, 54.213657,-28.033135)\\
polygon(53.137897,-27.585053, 53.471935,-28.304710, 54.213769,-28.033206, 53.875667,-27.315354)\\
polygon(52.808222,-26.864628, 53.137935,-27.585148, 53.875704,-27.315448, 53.542138,-26.596771)\\
polygon(52.116452,-26.275149, 52.439944,-26.997179, 53.175612,-26.731160, 52.848223,-26.010814)\\
polygon(52.847955,-26.011004, 53.175459,-26.731420, 53.907637,-26.461622, 53.576402,-25.742964)\\
polygon(53.542024,-26.596913, 53.875666,-27.315566, 54.609895,-27.042002, 54.272409,-26.325098)\\
polygon(53.875667,-27.315354, 54.213769,-28.033206, 54.951906,-27.757769, 54.609861,-27.041696)\\
polygon(54.213657,-28.033135, 54.556301,-28.750079, 55.298427,-28.472743, 54.951646,-27.757531)\\
polygon(54.183674,-28.887301, 54.529085,-29.604334, 55.277398,-29.326907, 54.927677,-28.611815)\\
}
  
\end{appendix}

\end{document}